\documentclass[12pt,a4paper]{article}
\pdfoutput=1
\usepackage{jheppub}
\usepackage{cite}

\usepackage{caption}
\usepackage{subcaption}
\usepackage{array,arydshln}
\usepackage{braket}
\usepackage{amsmath,amssymb}

	\newcommand{\be}{\begin{equation}}
		\newcommand{\ee}{\end{equation}}
	\newcommand{\bea}{\begin{eqnarray}}
		\newcommand{\eea}{\end{eqnarray}}
	\newcommand{\beas}{\begin{eqnarray*}}
		\newcommand{\eeas}{\end{eqnarray*}}
	
	\def\order{\ensuremath{\mathcal{O}}}

\newcommand{\pho}{F}

\title{
Krylov complexity in the IP matrix model}

\author[a]{Norihiro Iizuka} 
\author[b]{and Mitsuhiro Nishida}

\affiliation[a]{Department of Physics, Osaka University, Toyonaka, Osaka 560-0043, JAPAN}
\affiliation[b]{Department of Physics, Pohang University of Science and Technology, Pohang 37673, Korea}

\emailAdd{iizuka@phys.sci.osaka-u.ac.jp} 
\emailAdd{nishida124@postech.ac.kr}

\abstract{The IP matrix model is a simple large $N$ quantum mechanical model made up of an adjoint harmonic oscillator plus a fundamental harmonic oscillator. It is a model introduced previously as a toy model of the gauge theory dual of an AdS black hole. In the large $N$ limit, one can solve the Schwinger-Dyson equation for the fundamental correlator, and at sufficiently high temperature, this model shows key signatures of thermalization and information loss; the correlator decay exponentially in time, and the spectral density becomes continuous and gapless. We study the Lanczos coefficients $b_n$ in this model and at sufficiently high temperature, it grows linearly in $n$ with logarithmic corrections, which is one of the fastest growth under certain conditions. As a result, the Krylov complexity grows exponentially in time as $\sim \exp\left({\order{\left(\sqrt{t}\right) }}\right)$. These results indicate that the IP model at sufficiently high temperature is chaotic.}

\begin{document}

\begin{flushright}
{\small OU-HET-1189}
 \\
\end{flushright}

\maketitle

\section{Introduction}

The gauge gravity correspondence makes the large $N$ gauge theory dual to gravity and AdS black holes.  Famous examples include ${\cal{N}}=4$ SYM \cite{Maldacena:1997re} and D0-brane quantum mechanics \cite{Itzhaki:1998dd}, which is so-called the BFSS matrix model \cite{Banks:1996vh}. However, these large $N$ gauge theories are difficult to solve due to the difficulty of strong coupling. 
Therefore, one has to choose several approaches, such as full numerics like Monte-Carlo simulation, 
or simplifying it into a toy model.

From the viewpoint of $N$ D-brane dynamics, one can add a probe D-brane, and 
this ends up a theory with both matrices and vectors. 
Vectors correspond to open strings between $N$ D-branes and a probe D-brane, and matrices correspond to open strings between $N$ D-branes. Vectors transform in the $U(N)$ fundamental representation and matrices transform in the $U(N)$ adjoint representation. 
Especially for the D0-brane case, the theory is quantum mechanics instead of quantum field theory. 
The IP matrix model \cite{Iizuka:2008hg} was introduced as a toy model for that quantum mechanical model. It is supposed to be a toy model of the gauge theory dual of an AdS black hole. 

The IP model \cite{Iizuka:2008hg} is a simple large $N$ quantum mechanical model made up of a harmonic oscillator in the $U(N)$ adjoint representation, $X_{ij}$, plus a harmonic oscillator in the $U(N)$ fundamental 
representation $\phi_i$. They couple through a trilinear interaction, and through that, $X_{ij}$ acts as a heat bath for the fundamental $\phi_i$. 
This model has similarity to the 't Hooft's two-dimensional QCD model \cite{tHooft:1974pnl} and it is motivated by \cite{Festuccia:2006sa}. Although the IP model has a simple Hamiltonian, it shows a non-trivial result such that 
the two-point correlator for the fundamental decays exponentially in time at high enough temperature. The spectrum changes from discrete at zero temperature to continuous at high temperature. 
The exponential decay of the correlator indicates an information loss to a large $N$ D0-brane black hole and thermalization.
In this paper, we investigate the aspects of quantum chaos of the IP model. In particular, we study the operator growth at high enough temperature in the IP model.

Quantum chaos has emerged as a very important research topic recently, 
attracting the interest of many researchers across disciplines, and has been examined not only in quantum information physics but also in high energy physics. One of the main motivations for high-energy theorists to study quantum chaos is its connection to black holes. It has been conjectured that black holes are the fastest scramblers \cite{Sekino:2008he}, which means that for black holes, the scrambling time $t_{\rm sc}$, a time scale needed in quantum systems for a small perturbation to diffuse over all the degrees of freedom, scales as $t_{\rm sc} \sim \beta \log S$, where $S$ is the entropy of the system. This conjecture was motivated partly by 
the Hayden-Preskill model for black holes \cite{Hayden:2007cs}. 
Furthermore, progress in our understanding of out-of-time-order correlators (OTOCs) \cite{larkin1969quasiclassical, Kitaevtalk} has led to a better understanding of the connection between shock waves in black hole spacetime \cite{Dray:1984ha, tHooft:1990fkf, Kiem:1995iy, tHooft:1996rdg}, the butterfly effect on a two-sided correlation function in the thermofield double state 
\cite{Shenker:2013pqa, Shenker:2013yza, Shenker:2014cwa}, and the bound on the Lyapunov exponent \cite{Maldacena:2015waa}.
In these ways, examining quantum chaos gives us a better understanding of the quantum nature of black holes and quantum chaos itself.

Although OTOCs and the Lyapunov coefficient are very useful for the diagnosis of chaos, 
other diagnostics have been proposed as well. These include matrix elements of an operator between energy eigenstates for eigenstate thermalization hypothesis \cite{PhysRevA.43.2046, Srednicki:1994mfb}, 
the nearest neighbor level spacing, 
\cite{wigner1951class, Dyson:1962es, berry1977level, Bohigas:1983er}, and the spectral form factor (SFF) \cite{Brezin:1993lnq, Brezin:1996hze, PhysRevE.55.4067}. 
Especially, SFF has been studied from the perspective of black holes in \cite{Papadodimas:2015xma, Cotler:2016fpe}.  
The SFF is a partition function two-point function obtained by analytically continuing $\beta$ as 
$Z(\beta + i t) Z(\beta - i t)$. 
This is related to a thermal two-point function for local operators, although the matrix elements for local operators are removed in SFF.  
In chaotic systems, where the Hamiltonian eigenvalues are expected to obey the GUE ensemble, the late time behavior of the SFF exhibits ramp ($t$ linear growth) and plateau (constant behavior). 
Since the spectral form factor is one of the probes for quantum chaos, one might wonder if the thermal two-point function may also contain information about quantum chaos even though 
the thermal two-point function is less universal than the spectral form factor in the sense that the thermal two-point function depends on operators. 

The Lanczos coefficients and the Krylov complexity are proposed in \cite{Parker:2018yvk} as a diagnostic of chaos associated with the two-point function of an operator $\hat{\mathcal{O}}$. 
These measure the operator growth in the Krylov subspace for an operator $\hat{\mathcal{O}}$.  These measures are associated with the Krylov basis, which can be determined from the dynamics of a two-point function of $\hat{\mathcal{O}}$ only. 
In this way, the Lanczos coefficients depend on the operator $\hat{\mathcal{O}}$.

The key hypotheses regarding quantum chaos proposed in \cite{Parker:2018yvk} are as follows. 
$\vspace{-3mm}$
\begin{enumerate}
\item The Lanczos coefficients grow linearly with respect to $n$ up to log correction in chaotic quantum systems, where $n$ is a subscript to represent the Krylov basis. In such a case, the Krylov complexity grows exponentially in time. 
$\vspace{-3mm}$
\item The exponent of the Krylov complexity bounds the quantum Lyapunov exponent of OTOCs. $\vspace{-3mm}$
\end{enumerate}

The asymptotic growth behavior of the Lanczos coefficients can be determined by the high frequency behavior of a Fourier transformation of the two-point function, and thus the hypothesis 1.~above is regarded as a quantum version of the proposal \cite{Elsayed:2014chaos}. Hypothesis 2.~gives a tighter bound of the  
Lyapunov exponent than the bound obtained in 
\cite{Maldacena:2015waa}.

Motivated by these recent developments, we study the Lanczos coefficients and the Krylov complexity in the IP model \cite{Iizuka:2008hg}. As we mentioned, the IP model consists of a harmonic oscillator of the adjoint $X_{ij}$ with mass $m$ and a probe fundamental $\phi_i$ with mass $M$, and they interact through a trilinear coupling. We always consider the limit where the mass of the fundamental $M$ is much heavier than the mass of the fundamental, {\it i.e.,} $M \gg m$ limit. The effect of temperature $T  = 1/\beta$ is important in this model. We restrict to the situation $\beta M \gg 1$.

In the large $N$ limit, a two-point function of the fundamental $\phi_i$ can be computed by solving the Schwinger-Dyson equation in the planar limit.  From the resultant two-point function, one can see various behavior for the spectral density dependent on the parameters: 
$\vspace{-3mm}$

\begin{enumerate}
\item In massless limit for the adjoint, $m \to 0$, the spectral density for the fundamental follows the Wigner semi-circle law for any temperature $T$. 
$\vspace{-3mm}$
\item In the zero temperature limit $T=0$ with nonzero mass $m \neq 0$ for the adjoint, the spectral density for the fundamental shows a collection of the delta function, implying a discrete spectrum. 
$\vspace{-3mm}$
\item However as we increase the temperature $T$ with nonzero mass for adjoint $m \neq 0$, the spectrum changes from discrete to continuous. For small but nonzero $T$, the spectrum is a collection of the Wigner semi-circle type therefore it is continuous but gapped. As we increase the temperature furthermore, the gap is closed and at high enough temperature, it becomes continuous and gapless. The continuous spectrum without any gap results in a long-time exponential decay of the two-point function for the fundamental and this is a signature of thermalization and information loss \cite{Iizuka:2008hg}. $\vspace{-2mm}$
\end{enumerate}

As we will show, dependent on the parameters, the resultant behaviors of the Lanczos coefficients are quite different. From these, we study how the Krylov complexity and entropy grow as a function of time. Especially at sufficiently high temperature, the Lanczos coefficients grow
linearly in $n$ with log corrections, and the Krylov complexity grows exponentially in time as $e^{\order{(\sqrt{t})}}$. 
These studies reveal a rich structure of the model at various parameter limits.

Before we proceed, we emphasize the role of large $N$. In finite $N$, with $m \neq 0$, the IP model consists of a finite number of coupled harmonic oscillators. Only in the large $N$ limit, a planar approximation can be justified and one can solve the spectral density. Thus, the Lanczos coefficients we will evaluate are valid only in the large $N$ limit. 
We are interested in how the IP model, 
which is merely a collection of interacting harmonic oscillators at finite $N$, 
comes to exhibit the characteristics of chaos in the large $N$ limit. 
Thus by investigating the Lanczos coefficient and the Krylov complexity in the large $N$ limit, we will see if the $n$ linearity can appear\footnote{On the other hand, to evaluate the other chaos diagnostics such as nearest level spacing or SFF, we better solve the model at large but finite $N$ exactly to see the discrete nature of the spectrum, which is much harder analytically than evaluating Lanczos coefficients in the large $N$ limit.} and thus the Krylov complexity grows exponentially in time.

The organization of this work is as follows. We review how to calculate the Lanczos coefficients, the Krylov complexity, and the Krylov entropy in Section \ref{sec:reviewLanczos}. The IP model Hamiltonian and the spectral density of the fundamental are shown in Section \ref{sec:IP} for various values of temperature and mass. In Section \ref{sec:LanczosIP}, we investigate asymptotic behaviors of the Lanczos coefficients of the IP model by analyzing the behaviors of the spectral density at large frequency.  
The resultant Krylov complexity and entropy as a function of time are shown in Section \ref{sec:Kcomp}. We conclude this paper with discussions in Section \ref{sec:conc}.
The readers familiar with the Krylov complexity can skip Section \ref{sec:reviewLanczos}.



\section{Review of Lanczos coefficients and Krylov complexity}\label{sec:reviewLanczos} 
\subsection{Lanczos algorithm for pure states}
We briefly review how to compute the Lanczos coefficients and their connection to quantum chaos \cite{Parker:2018yvk, RecursionBook}. Let us consider a local operator $
\hat{\mathcal{O}}$. Its time evolution is given by the Baker-Campbell-Hausdorff formula
\begin{align}
\hat{\mathcal{O}}(t)&=e^{iHt}\hat{\mathcal{O}}e^{-iHt} = \hat{\mathcal{O}} +  it [H,  \hat{\mathcal{O}}] + \frac{(it)^2}{2!} [H, [ H, \hat{\mathcal{O}}]] + \cdots \nonumber \\
&= \sum_{n=0}^\infty (i t) ^n \mathcal{L}^n \hat{\mathcal{O}} =e^{i\mathcal{L}t}\hat{\mathcal{O}} \,, \quad \mbox{where}  \quad \mathcal{L}:=[H, \, \cdot \, \, ] \,.
\end{align} 
Here $H$ is a local Hamiltonian, which is Hermitian. 
Given an operator, one can construct the states by acting that operator on a pure state $\ket{\psi}$,  
\begin{align}
\vert \hat{\mathcal{O}}) := \hat{\mathcal{O}}  \ket{\psi}  \,, \quad  
\mathcal{L}^n \vert  \hat{\mathcal{O}}) : = [H \,,[ H \,, [H \cdots , [ H, \hat{\mathcal{O}} ]]]]  \ket{\psi}
\end{align}
where $n$ is a non-negative integer. In this way, the operator $\hat{\mathcal{O}}$ keeps spreading over the subspace of the Hilbert space and how quickly it spreads is our interest. 
Here the choice of pure state $\ket{\psi}$ depends on the choice of two-point function one considers. For example, for the zero temperature two-point function of $\hat{\mathcal{O}}$, one chooses $\ket{v}$, the free ground state for $\ket{\psi}$. 

A time-dependent state $\vert\hat{\mathcal{O}}(t)) : = \hat{\mathcal{O}}(t)  \ket{\psi} $ for $\hat{\mathcal{O}}(t)=e^{i\mathcal{L}t}\hat{\mathcal{O}}$ can be expanded by $ \mathcal{L}^n \vert\hat{\mathcal{O}})$. However, this itself is not an orthonormal basis. To make it orthonormal, one can use the Gram-Schmidt orthonormalization. Using that, we can construct the Krylov basis $\vert\hat{\mathcal{O}}_n)$, which is an orthonormal basis as $(\hat{\mathcal{O}}_m\vert\hat{\mathcal{O}}_n)=\delta_{mn}$. The Krylov subspace is a subspace spanned by $\{ \mathcal{L}^n \vert \hat{\mathcal{O}})\}$, and the Krylov basis follows
\begin{align}
& \vert\hat{\mathcal{O}}_0):=  \vert\hat{\mathcal{O}}) \,, \qquad \mathcal{L}\vert\hat{\mathcal{O}}_n)=\sum_{i=0}^{n+1}h_{i,n}\vert \hat{\mathcal{O}}_i) \;\;\;(n \ge0 ),\label{KrylovBasisArnoldi}\\
&(\hat{\mathcal{O}}_m\vert \mathcal{L}\vert \hat{\mathcal{O}}_n)=
\begin{pmatrix}
h_{0,0}&h_{0,1}&h_{0,2}&h_{0,3}&\cdots\\
h_{1,0}&h_{1,1}&h_{1,2}&h_{1,3}&\cdots\\
0&h_{2,1}&h_{2,2}&h_{2,3}&\cdots\\
0&0&h_{3,2}&h_{3,3}&\cdots\\
\vdots&\vdots&\vdots&\vdots&\ddots\\
\end{pmatrix}.\label{MatrixArnoldi}
\end{align}
We have also chosen an appropriate normalization such that $(\hat{\mathcal{O}}_0 \vert\hat{\mathcal{O}}_0)=1$.
This construction of the basis is called the Arnoldi iteration for general matrices \cite{WEARNOLDI}.

If $(\hat{\mathcal{O}}_m\vert \mathcal{L}\vert \hat{\mathcal{O}}_n)$ is a Hermitian matrix, then eq.~(\ref{KrylovBasisArnoldi}) and (\ref{MatrixArnoldi}) are simplified as 
\begin{align}
&\qquad \vert\hat{\mathcal{O}}_{-1}) := 0 \,, \quad \vert\hat{\mathcal{O}}_0):=  \vert\hat{\mathcal{O}}), \;\\
& \mathcal{L}\vert\hat{\mathcal{O}}_n)= a_n\vert\hat{\mathcal{O}}_n)+b_n\vert\hat{\mathcal{O}}_{n-1})+b_{n+1}\vert\hat{\mathcal{O}}_{n+1}) \;\;\;(n \ge0 ), \label{recursionOn}\\
&\qquad (\hat{\mathcal{O}}_m\vert \mathcal{L}\vert \hat{\mathcal{O}}_n)=
\begin{pmatrix}
a_0&b_1&0&0&\cdots\\
b_1&a_1&b_2&0&\cdots\\
0&b_2&a_2&b_3&\cdots\\
0&0&b_3&a_3&\cdots\\
\vdots&\vdots&\vdots&\vdots&\ddots\\
\end{pmatrix}.\label{MatrixLanczos}
\end{align}
This construction is called the Lanczos algorithm for Hermitian matrices \cite{CLanczos}. 
If $\vert \psi\rangle$ is an eigenstate of $H$ as $H\vert\psi\rangle=\lambda\vert \psi\rangle$, we obtain
\begin{align}
(\hat{\mathcal{O}}_m\vert \mathcal{L}\vert \hat{\mathcal{O}}_n)=\langle\psi\vert\hat{\mathcal{O}}_m^\dagger(H-\lambda)\hat{\mathcal{O}}_n\vert\psi\rangle,
\end{align}
which is Hermitian, therefore eq.~(\ref{recursionOn}) is valid. However, if $\vert \psi\rangle$ is not an eigenstate of $H$, then $(\hat{\mathcal{O}}_m\vert \mathcal{L}\vert \hat{\mathcal{O}}_n)$ need not be Hermitian, and in such case, one should use eq.~(\ref{KrylovBasisArnoldi}) instead of (\ref{recursionOn})\footnote{To measure the time evolution for the state $\vert \hat{\mathcal{O}})$, instead for an operator $ \hat{\mathcal{O}}$, the following definition of $\mathcal{L}$ for a state $\vert \hat{\mathcal{O}})$ is often used
\begin{align}
\mathcal{L}^n\vert \hat{\mathcal{O}}):=H^n\vert\hat{\mathcal{O}}) \,. \label{Lspread}
\end{align}
For example, for the Hamiltonian given by eq.~\eqref{linearH}, \cite{Balasubramanian:2022tpr} studied the Krylov complexity for states.}. We will see an example where one cannot use (\ref{recursionOn}) soon.

The coefficients $a_n$ and $b_n$ are called the Lanczos coefficients, which can be determined systematically as follows. From eq.~(\ref{recursionOn}) with $n=0$, $a_0$ is determined by
\begin{align}
a_0=(\hat{\mathcal{O}}_0\vert\mathcal{L}\vert\hat{\mathcal{O}}_0),
\end{align}
and $b_1$ is determined by normalization of $\vert \hat{A}_1)$
\begin{align}
\vert \hat{A}_1):=&\mathcal{L}\vert\hat{\mathcal{O}}_0)-a_0\vert\hat{\mathcal{O}}_0),\\
b_1=&\sqrt{(\hat{A}_1\vert \hat{A}_1)}, \;\;\; \vert\hat{\mathcal{O}}_{1})=b_1^{-1}\vert \hat{A}_1).
\end{align}
From (\ref{recursionOn}) with $n=1$, $a_1$ and $b_2$ are determined by
\begin{align}
a_1=&(\hat{\mathcal{O}}_1\vert\mathcal{L}\vert\hat{\mathcal{O}}_1), \;\;\; \vert \hat{A}_2):=\mathcal{L}\vert\hat{\mathcal{O}}_1)-a_1\vert\hat{\mathcal{O}}_1)-b_1\vert\hat{\mathcal{O}}_0),\\
b_2=&\sqrt{(\hat{A}_2\vert \hat{A}_2)}, \;\;\; \vert\hat{\mathcal{O}}_{2})=b_2^{-1}\vert \hat{A}_2).
\end{align}
Similarly, one can determine $a_n$ and $b_{n+1}$ for $n\ge2$. 

When an operator $\hat{\mathcal{O}}$, a Hamiltonian $H$, and an inner product between states are given, the Krylov basis $\vert\hat{\mathcal{O}}_n)$ and the Lanczos coefficients $a_n, b_n$ can be determined. Assuming (1) a Hermitian operator $\hat{\mathcal{O}}$, (2) a Hermitian Hamiltonian $H$, and (3) an appropriate inner product by trace and Hermite conjugation, $a_n$ becomes zero. However, in this paper, $a_n$ can be nonzero because we use creation operators of a vector field 
in the IP model.
Note that the Lanczos coefficients contain only information on the Krylov subspace for $\hat{\mathcal{O}}(t)$ and not on the total Hilbert space. We are interested in $b_n$, especially at large $n$, since it represents how much the operator spreads into orthogonal directions in the Hilbert space at a later time. On the other hand, $a_n$ is a Hamiltonian eigenvalue in the absence of $b_n$, which is not directly related to the spreads of the operators. \\
$\vspace{-4mm}$

To understand the Lanczos coefficients better,  let us work out several concrete examples. 
\begin{enumerate}
\item Let us first compute the Lanczos coefficients of a  free harmonic oscillator \cite{RecursionBook} whose Hamiltonian is
\begin{align}
\label{freeH}
H_0=\omega_0 \,\hat{a}^\dagger \hat{a} \,, \quad \;\;\; [\hat{a},\hat{a}^\dagger]=1.
\end{align}
Consider $\hat{\mathcal{O}}_0=\hat{a}^\dagger$ and choose $\ket{\psi} = \ket{v}$, where $\ket{v}$ is the ground state.  Then, 
\begin{align}
\vert \hat{\mathcal{O}}_0)=\hat{\mathcal{O}}_0\vert v\rangle=\hat{a}^\dagger\vert v\rangle, \;\;\; 
(\hat{\mathcal{O}}_0\vert\hat{\mathcal{O}}_0)=\langle v\vert \hat{a} \hat{a}^\dagger \vert v\rangle=1  \,.
\end{align}
One can obtain $a_0$ as 
\begin{align}
\mathcal{L}\vert\hat{\mathcal{O}}_0)= [H_0,\hat{a}^\dagger]\vert v\rangle= \omega_0 \hat{a}^\dagger \ket{v} =\omega_0\vert\hat{\mathcal{O}}_0), \;\;\; a_0 =  \omega_0.  
\end{align}
Thus, $\mathcal{L}$ does not create a new basis, {\it i.e.,} 
\begin{align}
\mathcal{L}^n\vert\hat{\mathcal{O}}_0)=[ H_0, [H_0,[ H_0, \cdots [H_0,\hat{a}^\dagger]  ]]] \vert v\rangle=\omega_0^n \vert\hat{\mathcal{O}}_0),
\end{align}
therefore for Krylov subspace, only $\vert \hat{\mathcal{O}}_0)$ exists, and all $\vert \hat{\mathcal{O}}_k)  = 0$ for $k \ge 1$. 
Thus we obtain 
\begin{align}
\label{anbn0}
a_n=b_n=0 \;\;\; (n\ge1) \,.
\end{align}

\item Suppose we modify the Hamiltonian by adding a linear term such as 
\begin{align}
\label{linearH}
\hspace{-5mm}
H =\omega_0 \, \hat{a}^\dagger \hat{a}   + g \left( \hat{a} + \hat{a}^\dagger \right) 
&= \omega_0 \left(\hat{a} + \frac{g}{\omega_0} \right)^\dagger \left(\hat{a} + \frac{g}{\omega_0} \right) + \mbox{const} 
\,, \quad 
[\hat{a},\hat{a}^\dagger]=1.
\end{align}
Neglecting the constant term, this Hamiltonian in \eqref{linearH} is essentially the same as free Hamiltonian \eqref{freeH} by simply re-defining the creation/annihilation operator as 
\begin{align}
\hat{A} := \hat{a} + \frac{g}{\omega_0} \,, \quad  [\hat{A},\hat{A}^\dagger]=1.
\end{align} 
Let us consider $\vert \psi\rangle=\vert v\rangle$ and $\vert\hat{\mathcal{O}}_0)=\hat{a}^\dagger \vert v\rangle$, where $\hat{a}\vert v\rangle=0$. Since the ground state is annihilated by $A$, not by $a$, $\vert v\rangle$ is not an eigenstate of $H$, and thus, the Hermitian nature used for eq.~\eqref{recursionOn} does not work. In this case, instead of eq.~\eqref{recursionOn}, we should use Arnoldi iteration eq.~(\ref{KrylovBasisArnoldi}), which gives   
\begin{align}
\mathcal{L}\vert\hat{\mathcal{O}}_0)=&[H,\hat{a}^\dagger]\vert v\rangle = \omega_0 \left(\hat{a} + \frac{g}{\omega_0} \right)^\dagger  \ket{v}=\omega_0\vert\hat{\mathcal{O}}_0)+g\vert\hat{\mathcal{O}}_1),\\
\vert\hat{\mathcal{O}}_1):=&\ket{v}\,, \qquad \mathcal{L}\vert\hat{\mathcal{O}}_1)=0 \,.
\end{align}
Thus, we obtain
\begin{align}
h_{0,0}=\omega_0 \,, \quad h_{1,0}=g \,, \quad h_{i,n}=0 \;\;\; (n\ge1).
\end{align}
In this example, due to the choice of 
$\vert \psi\rangle=\vert v\rangle$, which is not the eigenstate of the Hamiltonian, we could not use the Lanczos algorithm and instead, we use the Arnoldi iteration\footnote{In this paper, we always consider the case where Hamiltonian is Hermitian. For people who are interested in non-Hermitian cases, such as Hamiltonian for open systems, see \cite{Bhattacharya:2022gbz, Bhattacharjee:2022lzy} where one needs the Arnoldi iteration as well.}.

As another example, let us consider 
\begin{align}
\vert \psi\rangle=\vert v_A\rangle \,, \quad A  \vert v_A\rangle =  \left(\hat{a} +  \frac{g}{\omega_0} \right) \vert v_A\rangle=0 \,, \quad 
\vert\hat{\mathcal{O}}_0)=\frac{1}{\sqrt{1+\frac{g^2}{\omega_0^2}}}\hat{a}^\dagger \vert v_A\rangle 
\end{align} 
Since $\vert v_A\rangle$ is an eigenstate of $H$, in this case one can use the Lanczos algorithm eq.~(\ref{recursionOn}) and we obtain, 
\begin{align}
\langle v_A\vert \hat{a} \hat{a}^\dagger\vert v_A\rangle=&1+ \frac{g^2}{\omega_0^2} \,, \quad \langle v_A\vert \hat{a} \vert v_A\rangle=\langle v_A\vert \hat{a}^\dagger\vert v_A\rangle=- \frac{g}{\omega_0} \,,\\
\vert\hat{\mathcal{O}}_1):=&\left(\frac{g}{\omega_0}\frac{1}{\sqrt{1+\frac{g^2}{\omega_0^2}}}\hat{a}^\dagger+\sqrt{1+\frac{g^2}{\omega_0^2}}\right)\ket{v_A} , \\
  \quad (\hat{\mathcal{O}}_0\vert \hat{\mathcal{O}}_1)=&0 \,, \;\;\; (\hat{\mathcal{O}}_1\vert \hat{\mathcal{O}}_1)=1 \,, \\
\mathcal{L}\vert\hat{\mathcal{O}}_0)=&\frac{1}{\sqrt{1+\frac{g^2}{\omega_0^2}}}\left(\omega_0\hat{a}^\dagger + g \right)\ket{v_A}=\frac{\omega_0}{1+\frac{g^2}{\omega_0^2}}\vert\hat{\mathcal{O}}_0)+\frac{g}{1+\frac{g^2}{\omega_0^2}}\vert\hat{\mathcal{O}}_1),\\  
\mathcal{L}\vert\hat{\mathcal{O}}_1)=&\frac{g}{\omega_0}\mathcal{L}\vert\hat{\mathcal{O}}_0)=\frac{g}{1+\frac{g^2}{\omega_0^2}}\vert\hat{\mathcal{O}}_0)+\frac{\frac{g^2}{\omega_0}}{1+\frac{g^2}{\omega_0^2}}\vert\hat{\mathcal{O}}_1). 
\end{align}
Therefore, the Lanczos coefficients are given by
\begin{align}
a_0=\frac{\omega_0}{1+\frac{g^2}{\omega_0^2}}\,,\quad  a_1&=\frac{\frac{g^2}{\omega_0}}{1+\frac{g^2}{\omega_0^2}} \,,\quad b_1=\frac{g}{1+\frac{g^2}{\omega_0^2}}\,, \nonumber \\
 a_n&=b_n=0 \;\;\; (n\ge2).
 \label{zerotoy2Lanczos}
\end{align}

In these examples, the Krylov subspace is a two-dimensional space spanned by $\hat{a}^\dagger$ and the identity operator. $(\hat{a}^\dagger)^n$ for all $n \ge 2$ are not included in the Krylov subspace.

\item \label{ex3zerotemperature}
If we introduce cubic or higher couplings, the situation is different. Let us consider the following model of two coupled harmonic oscillators, 
\begin{align}
\label{doubleharmonics}
& \qquad H =\omega_0\, \hat{a}^\dagger \hat{a}   + M \, \hat{b}^\dagger \hat{b}  + g \left( \hat{a}^\dagger + \hat{a}\right) \hat{b}^\dagger \hat{b} \,, \\
& [\hat{a},\hat{a}^\dagger] = [\hat{b},\hat{b}^\dagger]=1\,, \quad 
\hat{\mathcal{O}} = \hat{b}^\dagger \,, \quad \ket{V} = \ket{v}_{\hat{a}} \otimes  \ket{v}_{\hat{b}} 
\end{align}
Here $ \ket{v}_{\hat{a}} $ is the state satisfying $\hat{a}  \ket{v}_{\hat{a}} =0$, similarly $\hat{b} \ket{v}_{\hat{b}} = 0$. As we will see, this model has a similar nature to the IP model \cite{Iizuka:2008hg}, where we consider $M \gg \omega_0$.  
In any case, by using $H\vert V\rangle=0$, one can use the Lanczos algorithm and we can explicitly construct
\begin{align}
\hat{\mathcal{O}}_n:=&\frac{1}{\sqrt{n!}}(\hat{a}^\dagger)^n\hat{b}^\dagger, \;\;\; 
\vert \hat{\mathcal{O}}_n)=\frac{1}{\sqrt{n!}}(\hat{a}^\dagger)^n\hat{b}^\dagger\vert V\rangle, \;\;\; 
(\hat{\mathcal{O}}_m\vert\hat{\mathcal{O}}_n)=\delta_{mn},\\
\mathcal{L}\vert \hat{\mathcal{O}}_n)=&H\vert \hat{\mathcal{O}}_n)=(\omega_0n+M)\vert \hat{\mathcal{O}}_n)+g\sqrt{n}\vert \hat{\mathcal{O}}_{n-1})+g\sqrt{n+1}\vert \hat{\mathcal{O}}_{n+1}).
\end{align}
Therefore, the Lanczos coefficients are given by
\begin{align}
a_n=\omega_0n+M, \;\;\; b_n=g\sqrt{n} \,.\label{LancozsToyzeroT}
\end{align}
The increase in $b_n$ with respect to $n$ means that the spread over the Krylov subspace accelerates as $n$ increases, which is related to cubic interactions in $H$. On the other hand, the increase of $a_n$ with respect to $n$ cannot be interpreted as the acceleration of the spread. This is because $a_n$ is simply the diagonal component $(\hat{\mathcal{O}}_n\vert \mathcal{L}\vert \hat{\mathcal{O}}_n)$, which represents the Hamiltonian eigenvalue of the state $\vert \hat{\mathcal{O}}_n)$ in the absence of $b_n$.
\end{enumerate}

\subsection{Lanczos algorithm for mixed states}

The above procedure can be generalized to the finite temperature case by introducing an inner product between operators at finite temperature. For canonical ensemble, one can define the following inner product between operators $\hat{A}$ and $\hat{B}$
\begin{align}\label{innerproductfiniteT}
(\hat{A}\vert \hat{B})_\beta:=\frac{1}{Z}\text{Tr}[e^{-\beta H}\hat{A}^\dagger \hat{B}], 
\;\;\;  Z:=\text{Tr}[e^{-\beta H}],
\end{align}
where $\beta$ is the inverse temperature and Tr is over all $H$ eigenstates. We can see 
\begin{align}
(\hat{A}\vert \mathcal{L}\hat{B})_\beta =  \frac{1}{Z}\text{Tr}[e^{-\beta H}\hat{A}^\dagger \left( H \hat{B} - \hat{B}H \right) ] 
= \frac{1}{Z}\text{Tr}[e^{-\beta H}\left( \hat{A}^\dagger H - H  \hat{A}^\dagger  \right) \hat{B} ]  =  (\mathcal{L}\hat{A}\vert \hat{B})_\beta
\end{align} 
and define
\begin{align}
(\hat{A}\vert\mathcal{L}^n\vert \hat{B})_\beta:=(\hat{A}\vert \mathcal{L}^n\hat{B})_\beta=(\mathcal{L}^n\hat{A}\vert \hat{B})_\beta.
\end{align}
Once the inner product is defined, we can construct the Krylov basis that follows 
\begin{align}
\label{orthonormalbasis}
(\hat{\mathcal{O}}_m\vert \hat{\mathcal{O}}_n)_\beta=\delta_{mn} \quad \mbox{(orthonormal basis)}
\end{align} in mixed states by using the same procedure as pure states. Moreover, 
\begin{align}
L_{mn}&:=(\hat{\mathcal{O}}_m\vert\mathcal{L}\vert\hat{\mathcal{O}}_n)_\beta=\frac{1}{Z}\text{Tr}[e^{-\beta H}((\hat{\mathcal{O}}_m)^\dagger H\hat{\mathcal{O}}_n-(\hat{\mathcal{O}}_m)^\dagger \hat{\mathcal{O}}_nH)],
\end{align}
which is Hermitian, and thus we can use the Lanczos algorithm  
with the inner product eq.~(\ref{innerproductfiniteT}). 

For mixed states,  
it is better to define the Lanczos coefficients by operator form as
\begin{align}
  \hat{\mathcal{O}}_{-1} &:= 0\,, \; \quad  \hat{\mathcal{O}}_0:=  \hat{\mathcal{O}}, \; \\
\mathcal{L}\hat{\mathcal{O}}_n 
&= a_n\hat{\mathcal{O}}_n+b_n\hat{\mathcal{O}}_{n-1}+b_{n+1}\hat{\mathcal{O}}_{n+1} \notag\\
&=\hat{\mathcal{O}}_nL_{n,n}+\hat{\mathcal{O}}_{n-1}L_{n-1,n}+\hat{\mathcal{O}}_{n+1}L_{n+1,n} \notag\\
&=\sum_{m=0}\hat{\mathcal{O}}_mL_{m,n} \;\;\;(n \ge0 ), \label{recursionOnfiniteT}
\end{align}
where $L_{m,n}$ is expressed by 
\begin{align}
L_{m,n}:= \, (\hat{\mathcal{O}}_m\vert \mathcal{L}\vert \hat{\mathcal{O}}_n)_\beta=&
\begin{pmatrix}
a_0&b_1&0&0&\cdots\\
b_1&a_1&b_2&0&\cdots\\
0&b_2&a_2&b_3&\cdots\\
0&0&b_3&a_3&\cdots\\
\vdots&\vdots&\vdots&\vdots&\ddots\\
\end{pmatrix}.\label{MatrixLanczosfiniteT}
\end{align}
By using (\ref{recursionOnfiniteT}), we also obtain
\begin{align}
(\hat{\mathcal{O}}_m\vert\mathcal{L}^k\vert\hat{\mathcal{O}}_n)_\beta=(L^k)_{mn}.\label{Lkmn}
\end{align}
Clearly in the zero temperature limit $\beta \to \infty$, these reduce to the Lanczos algorithm eq.~(\ref{recursionOn}) and eq.~(\ref{MatrixLanczos}) for the pure state case with the choice of $\ket{\psi}$ as a ground state. \\
$\vspace{-4mm}$

 Let us reconsider the previous examples at finite temperature.
 \begin{enumerate}
\item Hamiltonian of a free harmonic oscillator and its partition function are given by
\begin{align}
H_0=\omega_0 \,\hat{a}^\dagger \hat{a} \,,  \;\;\; [\hat{a},\hat{a}^\dagger]=1,  \;\;\; Z:=\text{Tr}[e^{-\beta H}]=\frac{1}{1-e^{-\beta\omega_0}}.
\end{align}
For $ \hat{\mathcal{O}}_0 \propto \hat{a}^\dagger$, the right normalization is 
\begin{align}
  \hat{\mathcal{O}}_0=Z^{-1/2}\hat{a}^\dagger \,, 
\end{align}
which satisfies 
\begin{align}
( \hat{\mathcal{O}}_0 \vert  \hat{\mathcal{O}}_0 )_\beta= \frac{\text{Tr}[e^{-\beta H}\hat{a}\hat{a}^\dagger]}{Z^2} = 1 \,.
\end{align}
In this free harmonic oscillator, the Lanczos coefficients do not depend on $\beta$ as
\begin{align}
\mathcal{L}\hat{\mathcal{O}}_0=[H,\hat{\mathcal{O}}_0]=\omega_0\hat{\mathcal{O}}_0, \;\;\; a_0=\omega_0, \;\;\; a_n=b_n=0 \;\;\; (n\ge1) \,.
\end{align}
\item Next, let us consider the modified Hamiltonian by a linear term
\begin{align}
H =&\omega_0 \, \hat{a}^\dagger \hat{a}   + g \left( \hat{a} + \hat{a}^\dagger \right) + \frac{g^2}{\omega_0}
= \omega_0 \left(\hat{a} + \frac{g}{\omega_0} \right)^\dagger \left(\hat{a} + \frac{g}{\omega_0} \right) = \omega_0\hat{A}^\dagger \hat{A} \,, \\
\hat{A}^\dagger:=&\hat{a}^\dagger + \frac{g}{\omega_0}, \;\;\; [\hat{a},\hat{a}^\dagger]=[\hat{A},\hat{A}^\dagger]=1, \;\;\; Z:=\text{Tr}[e^{-\beta H}]=\frac{1}{1-e^{-\beta\omega_0}}.
\end{align}
By using $\text{Tr}[e^{-\beta H}\hat{A}\hat{A}^\dagger]=Z^2$, we obtain
\begin{align}
\frac{\text{Tr}[e^{-\beta H}\hat{a}\hat{a}^\dagger]}{Z}=\frac{\text{Tr}[e^{-\beta H}(\hat{A}\hat{A}^\dagger+ \frac{g^2}{\omega_0^2})]}{Z}=Z+ \frac{g^2}{\omega_0^2}\,, 
\end{align}
thus, we choose a normalized operator 
\begin{align}
\hat{\mathcal{O}}_0=\frac{1}{\sqrt{Z+ \frac{g^2}{\omega_0^2}}}\hat{a}^\dagger=\frac{1}{\sqrt{Z+ \frac{g^2}{\omega_0^2}}} \left(\hat{A}^\dagger- \frac{g}{\omega_0} \right)
\end{align} 
From this operator, we can check that
\begin{align}
\hat{\mathcal{O}}_1:=&\frac{g}{\sqrt{Z}\omega_0\sqrt{Z+ \frac{g^2}{\omega_0^2}}}(\hat{A}^\dagger+\frac{\omega_0Z}{g}), \;\;\; (\hat{\mathcal{O}}_0\vert \hat{\mathcal{O}}_1)_\beta=0, \;\;\; (\hat{\mathcal{O}}_1\vert \hat{\mathcal{O}}_1)_\beta=1,\\
\mathcal{L}\hat{\mathcal{O}}_0=&[H,\hat{\mathcal{O}}_0]=\frac{\omega_0\hat{A}^\dagger}{\sqrt{Z+ \frac{g^2}{\omega_0^2}}}\notag\\
=&\frac{\omega_0Z}{\left(Z+ \frac{g^2}{\omega_0^2} \right)^{3/2}}(\hat{A}^\dagger- \frac{g}{\omega_0})+\frac{ \frac{g^2}{\omega_0}}{\left(Z+ \frac{g^2}{\omega_0^2} \right)^{3/2}}(\hat{A}^\dagger+ \frac{\omega_0Z}{g})\notag\\
=&\frac{\omega_0Z}{Z+ \frac{g^2}{\omega_0^2}}\hat{\mathcal{O}}_0+\frac{g\sqrt{Z}}{Z+ \frac{g^2}{\omega_0^2}}\hat{\mathcal{O}}_1 \,, \\
\mathcal{L}\hat{\mathcal{O}}_1=&\frac{g}{\sqrt{Z}\omega_0}\mathcal{L}\hat{\mathcal{O}}_0=\frac{g\sqrt{Z}}{Z+ \frac{g^2}{\omega_0^2}}\hat{\mathcal{O}}_0+\frac{ \frac{g^2}{\omega_0}}{Z+ \frac{g^2}{\omega_0^2}}\hat{\mathcal{O}}_1  \,.
\end{align}
Therefore, the Lanczos coefficients are given by
\begin{align}
a_0=\frac{\omega_0Z}{Z+\frac{g^2}{\omega_0^2}} \,, \quad a_1 &=\frac{\frac{g^2}{\omega_0}}{Z+ \frac{g^2}{\omega_0^2}} \,, \quad b_1=\frac{g\sqrt{Z}}{Z+ \frac{g^2}{\omega_0^2}} \,, \\
  \qquad a_n &=b_n=0 \;\;\; (n\ge2) \,.
\end{align}
In the zero temperature limit $\beta \to \infty$, $Z \to 1$ and this reduces to eq.~\eqref{zerotoy2Lanczos}. 
\item
In general, it is difficult to solve eigenvalues of the Hamiltonian which contains cubic or higher interactions, such as\footnote{We can apparently simplify this Hamiltonian as 
\begin{align}
H 
&=\omega_0\hat{B}^\dagger \hat{B} -g^2(\hat{b}^\dagger \hat{b})^2/\omega_0+M\hat{b}^\dagger \hat{b}\,,\\
\hat{B}^\dagger &:=\hat{a}^\dagger + \frac{g}{\omega_0}\hat{b}^\dagger \hat{b}, \;\;\; [\hat{B},\hat{B}^\dagger]=[\hat{a},\hat{a}^\dagger]=1, \;\;\; [\hat{B},\hat{b}^\dagger]=\frac{g}{\omega_0}b^\dagger,
\end{align}
However since $\hat{B}$ does not commute with $\hat{b}^\dagger$, we cannot diagonalize it.}   
\begin{align}\label{doubleharmonicsfiniteT}
H=\omega_0\, \hat{a}^\dagger \hat{a}   + M \, \hat{b}^\dagger \hat{b}  + g \left( \hat{a}^\dagger + \hat{a}\right) \hat{b}^\dagger \hat{b} \,, \;\; M \gg \omega_0 \,, 
\;\;  [\hat{a},\hat{a}^\dagger] = [\hat{b},\hat{b}^\dagger]=1\,. 
\end{align}
In such a case, analytical calculation of the inner product is difficult unless some approximation is used, and thus we cannot determine the normalization of $\hat{\mathcal{O}}_0$ as well as the orthonormal condition $(\hat{\mathcal{O}}_m\vert\hat{\mathcal{O}}_n)_\beta=\delta_{mn}$.

As we will see, the IP model \cite{Iizuka:2008hg} has this type of interaction. Although they are matrix-valued and vector-valued operators, {\it i.e.,} we enhance $\hat{a}$ into a matrix and $\hat{b}$ into a vector. 
The Hamiltonian eq.~(\ref{doubleharmonicsfiniteT}) is not solvable. However the IP model, the matrix-valued version of eq.~(\ref{doubleharmonicsfiniteT}) is solvable at the large $N$ limit. 
\end{enumerate}

Before we proceed we also review how to obtain the Lanczos coefficients given a two-point function.

\subsection{Moment method}
We review how the Lanczos coefficients can be calculated from a two-point function $C(t;\beta):=(\hat{\mathcal{O}}\vert\hat{\mathcal{O}}(-t))_\beta$ for a given operator $\hat{\mathcal{O}}$.  Here $\hat{\mathcal{O}}$ is a normalized operator such that $C(t;\beta)=1$. Now, we consider the finite temperature case, and the zero temperature limit is given by $\beta\to\infty$. We can compute $a_n$ and $b_n$ for $\hat{\mathcal{O}}_0=\hat{\mathcal{O}}$ by the following moment method. Let us define moments $M_{n}$ by using the Taylor expansion coefficients of $C(t;\beta)$ at $t=0$:
\begin{align}
C(t;\beta)=\sum_{n=0}M_n\frac{(-it)^n}{n!}, \;\;\; M_n:=\frac{1}{(-i)^{n}}\frac{d^{n}C(t;\beta)}{dt^{n}}\Big\vert_{t=0}=(\hat{\mathcal{O}}_0\vert\mathcal{L}^n\vert\hat{\mathcal{O}}_0)_\beta.
\end{align}
One can also compute moments $M_{n}$ by using a Fourier transformation of $C(t;\beta)$:
\begin{align}
M_{n}=&\int_{-\infty}^{\infty}\frac{d\omega}{2\pi}\,\omega^{n}f(\omega),\label{mufw}\\
 f(\omega):=&\int_{-\infty}^{\infty}dt\,e^{i\omega t}C(t;\beta).
\end{align}
Then, by using eq.~(\ref{MatrixLanczosfiniteT}) and (\ref{Lkmn}), one can obtain relations between the Moments and the Lanczos coefficients. For example,
\begin{align}
M_1=(\hat{\mathcal{O}}_0\vert\mathcal{L}\vert\hat{\mathcal{O}}_0)_\beta=&a_0,\\
M_2=(\hat{\mathcal{O}}_0\vert\mathcal{L}^2\vert\hat{\mathcal{O}}_0)_\beta=&a_0^2+b_1^2,\\
M_3=(\hat{\mathcal{O}}_0\vert\mathcal{L}^3\vert\hat{\mathcal{O}}_0)_\beta=&a_0^3+2a_0b_1^2+a_1b_1^2 \,, \\
M_4 =(\hat{\mathcal{O}}_0\vert\mathcal{L}^4\vert\hat{\mathcal{O}}_0)_\beta=&(a_0+a_1)^2b_1^2+(a_0^2+b_1^2)^2+b_1^2b_2^2.
\end{align}
Note that $M_{2n+1}$ is zero if $f(\omega)$ is an even function, which implies that $a_n=0$.

Conveniently, there exists an algorithm to compute the Lanczos coefficients from the moments \cite{Gordon1968ErrorBI, RecursionBook}. Suppose that a set of the moments $M_0, M_1, \dots, M_{2K+1}$ is given. Then, a set of the Lanczos coefficients $a_0, \dots, a_K$ and $b_1, \dots, b_K$ is obtained by the following recursion relation 
\begin{align}
M_k^{(0)}:=&(-1)^kM_k \;\;\;\;\;\;\;\;\;\;\;\;\;\;\;\;\;\;\;\;\;\;\;\; (k \le 2 K  + 1 )\\
L_k^{(0)}:=&(-1)^{k+1}M_{k+1} \;\;\;\;\;\;\;\;\;\;\;\;\;\;\;\;\;\; (k \le 2 K) \\
M_k^{(n)}:=&L_{k}^{(n-1)}-L_{n-1}^{(n-1)}\frac{M_k^{(n-1)}}{M_{n-1}^{(n-1)}}  \quad \,(k \le 2 K - n + 1)\\
L_k^{(n)}:=&\frac{M_{k+1}^{(n)}}{M_n^{(n)}}-\frac{M_k^{(n-1)}}{M_{n-1}^{(n-1)}} \,\,\,\, \,\quad \qquad (k \le 2 K - n) 
\end{align}
Then the Lanczos coefficient can be read-off as 
\begin{align}
a_n=&-L_n^{(n)}, \;\;\; b_n=\sqrt{M_n^{(n)}}\,, \;\;\; (n=0, \dots, K).
\end{align}
In summary, when a two-point function $C(t;\beta)$ or $f(\omega)$ is given, we can compute the moments $M_n$ and then use the above algorithm to compute the Lanczos coefficients $a_n$ and $b_n$.

Through eq.~(\ref{mufw}), the high-frequency behavior of $f(\omega)$ and the asymptotic behavior of $b_n$ at large $n$ are correlated \cite{Lubinsky:1988}. Historically in classical systems such as quantum spin systems in the limit of infinite spin, the exponential tail of $f(\omega)$ has been proposed as a probe of chaos \cite{Elsayed:2014chaos}. Thus a relationship between chaos and the behavior of $b_n$ is expected in quantum systems with large degrees of freedom. Then it is conjectured in \cite{Parker:2018yvk} that, in chaotic quantum many-body systems, $b_n$ grows as fast as possible with respect to $n$. In fact, it can be shown that the maximal growth of $b_n$ is at most linear in $n$ given $r$-local lattice Hamiltonians \cite{Abanin2015}.

\subsection{Krylov complexity}

Let us first decompose the operator $\hat{\mathcal{O}}(t)$ into orthonormal bases; 
\begin{align}
\hat{\mathcal{O}}(t): =\sum_{n=0}i^n\varphi_n(t)\hat{\mathcal{O}}_n \,,
\end{align}
$i^n \varphi_n(t)$ is defined as a coefficient of the orthonormal basis. From the orthonormality eq.~\eqref{orthonormalbasis}, we obtain 
\begin{align}
\varphi_n(t)=i^{-n}(\hat{\mathcal{O}}_n\vert\hat{\mathcal{O}}(t))_\beta.
\end{align}
The time evolution of the operator follows 
\begin{align}
\frac{d \hat{\mathcal{O}}(t)}{dt} &= \sum_{n=0}i^n \frac{d \varphi_n(t)}{dt} \hat{\mathcal{O}}_n \nonumber \\
&= i [H \,, \hat{\mathcal{O}}(t)] = i \mathcal{L}  \hat{\mathcal{O}}(t) = \sum_{n=0}i^{n+1} \varphi_n(t)  \mathcal{L}\hat{\mathcal{O}}_n 
\end{align}
using eq.~(\ref{recursionOnfiniteT}) and its orthonormality eq.~\eqref{orthonormalbasis},  
$\varphi_n(t)$ satisfies
\begin{align}
\frac{d\varphi_{n}(t)}{dt}=ia_n \varphi_{n}(t)-b_{n+1}\varphi_{n+1}(t) +b_{n}\varphi_{n-1}(t) \,,\label{recursionwf}\\
\varphi_{-1}(t):=0, \;\;\; \varphi_{0}(t)=C(-t;\beta) \,.  \label{recursionwf2}
\end{align}
The Krylov complexity $K(t)$ was introduced in \cite{Parker:2018yvk}, as 
\begin{align}
\label{Krylovdef}
K(t):=\sum_{n=1}^\infty n\vert\varphi_n(t)\vert^2 \,,
\end{align}
and it is conjectured that this $K(t)$ is a good diagnostic for operator growth in the Krylov basis.

Intuitively, $K(t)$ measure how much averaged ``distances'' for an operator $\hat{\mathcal{O}}(t)$ to moves in an orthonormal basis; $n$ corresponds to the distance, and larger $n$ corresponds to a larger distance. The implicit assumption here is that in each step of the action of $\mathcal{L}$, the distance increases.  Initially, an operator is at the origin, and $\varphi_n(t)$ is the probability amplitude at the distance $n$.   The larger $b_n$ growth is, the faster the scrambling is, therefore the system is more chaotic\footnote{There is an example for a system which is not chaotic but shows the exponential growth of the Krylov complexity, see for example, \cite{Xu:2019lhc,Bhattacharjee:2022vlt}. These are associated with the instability of the system.}.

To understand the Krylov complexity better, let us consider the SYK models \cite{Sachdev:1992fk, Kitaevtalk} as an example. 
$K(t)$ has some similar characteristic to the mean size $A(t)$ studied in \cite{Roberts:2018mnp}, however they are slightly different.  
Let us consider SYK $q$ model where the Hamiltonian is  
\begin{align}
	H_{{\text{SYK}_q}} = i^{q/2} \sum_{1 \le i_1< i_2< \cdots i_q \le N} j_{{i_1 i_2 \cdots i_q}} \psi_{i_1} \psi_{i_2}  \cdots \psi_{i_q}  \,, \quad \{ \psi_i \,, \psi_j \} = \delta_{ij} \,,
\end{align}
with $q$ is even, and $\psi_i, i=1, \cdots , N$ are Majorana fermions. 
A complete basis in the space of operators can be made up by 
\begin{align}
\hat{O}_{s=1,i} = 2^{1/2} \psi_i \,, \quad  \hat{O}_{s=2, ij} = 2 \psi_i \psi_j \,, \quad  \hat{O}_{s=3, ijk} = 2^{3/2} \psi_i \psi_j \psi_k \,,  \quad \cdots 
\end{align}
which satisfy eq.~\eqref{orthonormalbasis} with $\beta= 0$. 
Then starting with a single local fermion operator $\hat{\mathcal{O}} = 2^{1/2} \psi_i$, 
one can easily understand that it evolves as
\begin{align}
 \hat{\mathcal{O}}(t) = 2^{1/2} e^{i H t} \psi_i e^{-i H t} =\sum_{s,a_{1}<\cdots<a_{s}} c^{\{i\}}_{a_{1}\cdots a_{s}}(t)  \left[ 2^{\frac{s}{2}}\psi_{a_{1}}\psi_{a_{2}}\cdots \psi_{a_{s}} \right] \,.
\end{align}
Clearly the probability distribution of having operators of size $s$ is given by 
\begin{align}
P_{s}(t)=\sum_{a_{1}<\cdots<a_{s}}|c^{\{i\}}_{a_{1}\cdots a_{s}}(t)|^{2}  \,,
\end{align}
and the mean size $A(t)$  of an operator $ \hat{\mathcal{O}}(t) $ is given by \cite{Roberts:2018mnp}
\begin{align}
A(t) = \frac{1}{N} \sum_{s,a_{1}<\cdots<a_{s}} s |c^{\{i\}}_{a_{1}\cdots a_{s}}(t)|^{2}  
= \frac{1}{N} \sum_{s,a_{1}<\cdots<a_{s}} s P_{s}(t)
= \frac{\braket{s}}{N} \,.
\end{align}
The expression for this $A(t)$ has some analogy to Krylov complexity $K(t)$ define in eq.~\eqref{Krylovdef} but they are different as follows; in the mean size $A(t)$, the weight for operator size $s$, $\hat{O}_{s, ij \cdots}$ is just $s$. However for the Krylov complexity, we give different weights for all the independent operators. For example, there are ${}_NC_s$ number of different operators for the size $s$, $\hat{O}_{s, ij\cdots}$ but all of them are weighted as $s$ in $A(t)$. 
In general, as we act $\mathcal{L} \psi_i\sim [\psi^q,   \psi ]$, the operator size $s$ keep growing as $s \to s+ q-2$ at early times. So initially one can expect that they behave similarly. However, the mean size $A(t)$ can saturate at $n = \order(N) = \order(S)$ steps, and this ends up $A(t) = \order(1)$. Here $S$ is the entropy. This is because the size is bounded above as $s \le N$. On the other hand, $K(t)$ can keep growing at a much later time, since $n \lesssim e^S$, although the growth of $K(t)$ changes drastically around $n = \order(S)$ steps. See \cite{Barbon:2019wsy, Jian:2020qpp, Rabinovici:2020ryf, Rabinovici:2022beu}.  Thus, the magnitude of $K(t)$ can grow up to $K(t) = \order(e^S)$. Thus both $A(t)$ and $K(t)$ grows at an early time, and they yield significant difference at later time\footnote{In the large $q$ limit, one can obtain the analytic expression for the Krylov complexity and one can see the direct correspondence between $A(t)$ and $K(t)$ in the leading order in $1/q$ expansion \cite{Roberts:2018mnp, Parker:2018yvk}. See \cite{Bhattacharjee:2022lzy} as well.}.  

The late time behavior of $K(t)$ can be determined by the smooth asymptotic behavior of $b_n$ at large $n$. See the Appendix \ref{app1}.
Especially, if the system is chaotic, it is expected that $b_n$ shows a linear growth with respect to $n$, which leads to an exponential growth of the Krylov complexity $K(t)$,   
\begin{align}
b_n \sim \alpha n \,, \quad K(t) \sim e^{2 \alpha t} \,.  
\end{align}
For example in SYK model, the Lyapunov exponent $\lambda$ relates to $\alpha$ as $\lambda = 2 \alpha$.  
In \cite{Parker:2018yvk}, it has been conjectured that the exponential growth rate of $K(t)$ bounds the quantum Lyapunov exponent of four-point OTOCs. 
If the two-point function in chaotic systems has no singularity in the entire complex time plane, then $b_n$ has log correction at large $n$:
\begin{align}
b_n \sim \frac{ \alpha n }{\log n} \,, \quad K(t) \sim e^{\sqrt{ 4 \alpha t}} \,. 
\end{align}
It is known that in the $d=1$ spatial dimensional lattice system, this is the fastest growth of the Lanczos coefficients \cite{Parker:2018yvk, Avdoshkin:2019trj}. A famous example of this is the 1d Ising model with both transverse and parallel magnetic fields, the `chaotic Ising chain' model \cite{Banulsetal, Cao:2020zls}. 
On the other hand, for integrable systems, typically \cite{Barbon:2019wsy, Fan:2022xaa}
\begin{align}\label{bnintegrable}
b_n \sim \alpha   n^p \,,  \quad K(t) \sim  t^\gamma \,, \quad \gamma = \frac{1}{1 - p} \quad \mbox{(integrable systems)}\quad.
\end{align}

If $a_n$ is nonzero, the situation is different. To illustrate how $K(t)$ behaves with nonzero $a_n$, we need to calculate $\varphi_n$ in the presence of $a_n$ and $b_n$ by solving (\ref{recursionwf}) and (\ref{recursionwf2}). 

\begin{figure}
\centering
     \begin{subfigure}[b]{0.4\textwidth}
         \centering
         \includegraphics[width=\textwidth]{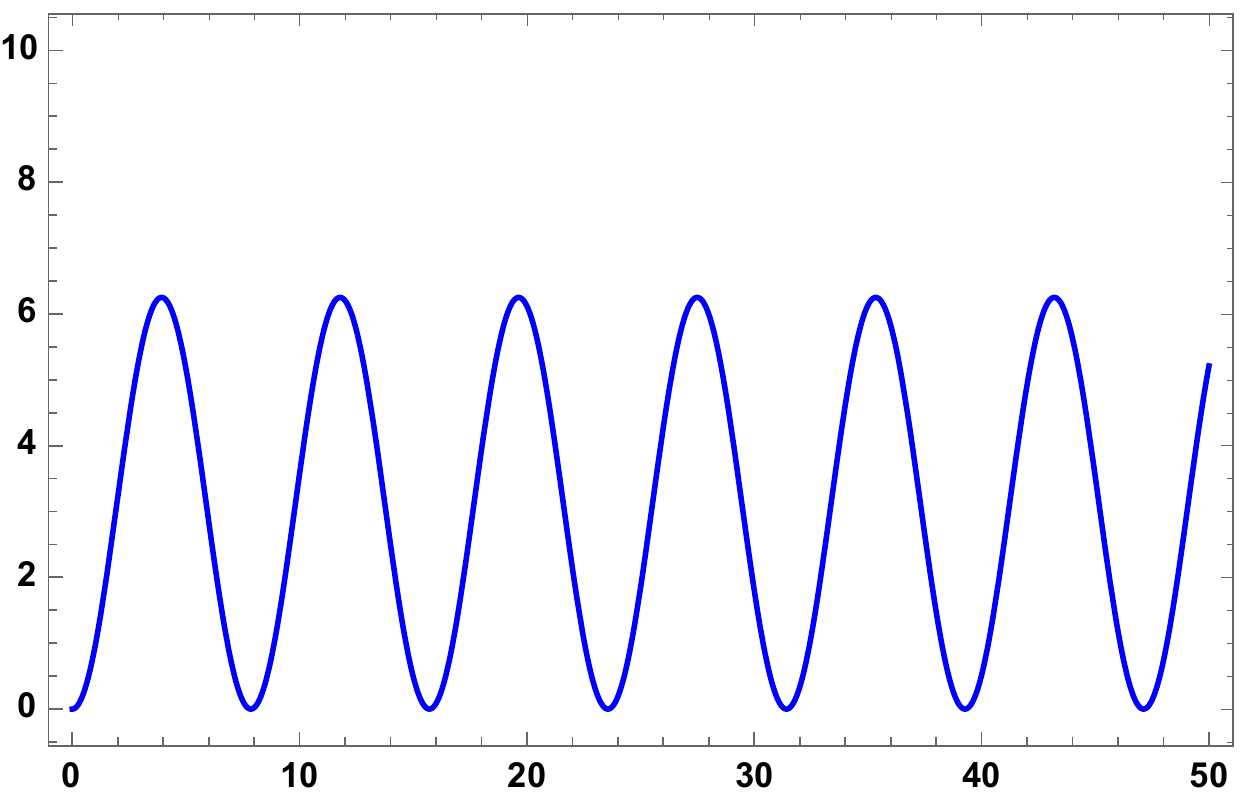}
       \put(5,0){$t$}
    \put(-175,115){$K(t)$}
       \caption{$\omega_0=0.8, \; M=0, \;g=1$}\label{fig:KrylovComplexityEx3(a)}
     \end{subfigure}
      \quad\quad 
           \begin{subfigure}[b]{0.4\textwidth}
         \centering
         \includegraphics[width=\textwidth]{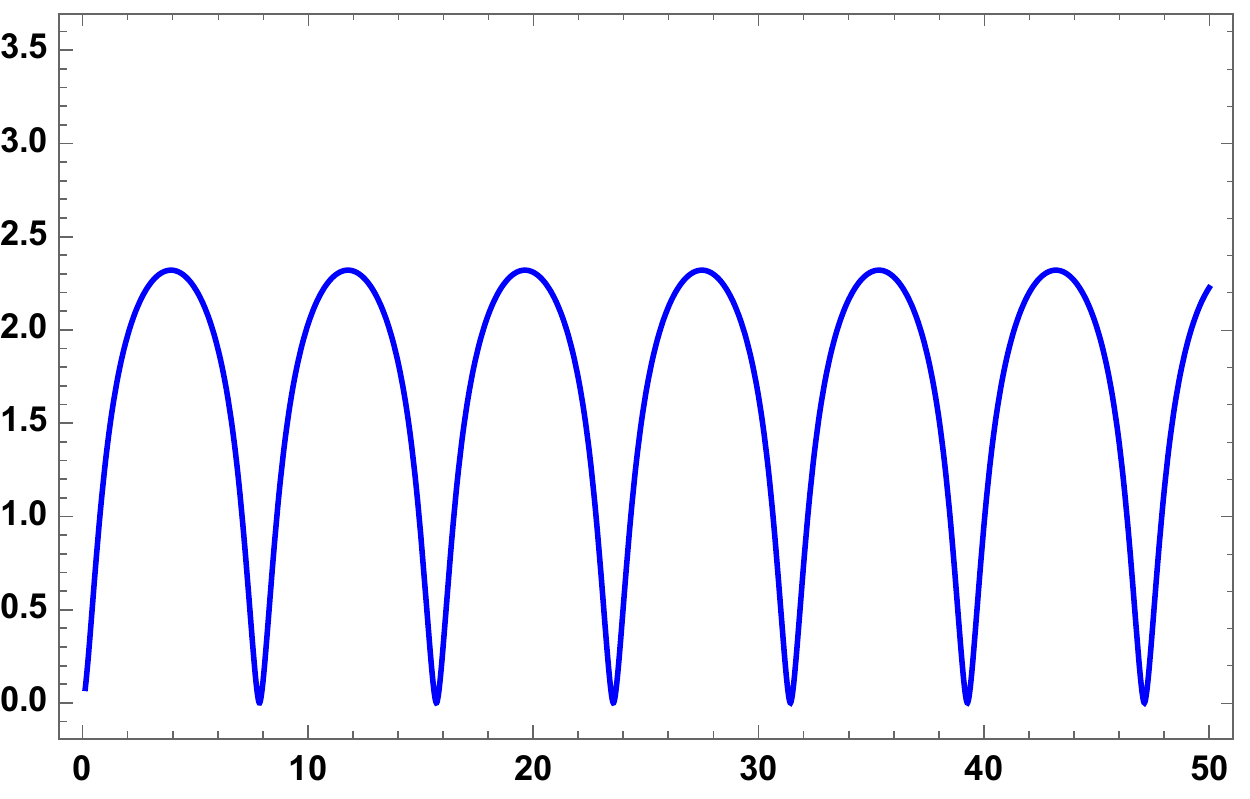}
       \put(5,0){$t$}
    \put(-175,115){$S(t)$}
       \caption{$\omega_0=0.8, \; M=0, \;g=1$}\label{fig:KrylovEntropyEx3(a)}
     \end{subfigure}
             \caption{The Krylov complexity $K(t)$ (left) and 
             the Krylov entropy $S(t)$ (right) for the Lanczos coefficients given by eq.~(\ref{LancozsToyzeroT}).
We fix parameters as $\omega_0 = 0.8$, $M=0$, $g=1$, although there is no $M$-dependence.}
        \label{fig:KrylovComplexityEx3}
\end{figure}

As an example, for the case where the Lanczos coefficients $a_n$ and $b_n$ are given by (\ref{LancozsToyzeroT})\footnote{This corresponds to the case where $b_n$ grows as eq.~(\ref{bnintegrable}) with $p=1/2$.}, the equation for  $\varphi_n$ becomes 
\begin{align}
\frac{d\varphi_{n}(t)}{dt}=i \left( \omega_0 n + M \right) \varphi_{n}(t)- g \sqrt{n+1} \, \varphi_{n+1}(t) + g \sqrt{n} \, \varphi_{n-1}(t) \,,
\end{align}
and its solution is given as \cite{Balasubramanian:2022tpr} 
\begin{align}
&\qquad \qquad \qquad  \varphi_{n}(t) = i^n e^{\alpha(t)} \frac{\left(\beta(t)\right)^n}{\sqrt{n!}} \,, \qquad \mbox{where}  \\
&\alpha(t) := i M t + \frac{g^2}{\omega_0^2} \left( - i \omega_0 t - 1 + e^{i \omega_0 t}\right) \,, \quad \beta(t):= \frac{g}{\omega_0} \left( 1 - e^{i \omega_0 t} \right) \,,
\end{align}
and the Krylov complexity $K(t)$ is 
\begin{align}
K(t) 
=   \sum_{n=1}^\infty n \frac{|\beta(t)|^{2n}}{n!}  \exp\left({2 \mbox{Re}[\alpha(t)]}\right)
=  |\beta(t)|^2  =  \frac{4 g^2}{\omega_0^2} \sin^2 \left( \frac{\omega_0 t}{2} \right)\,.
\end{align}
We plot in Figure \ref{fig:KrylovComplexityEx3}(a) the time evolution of $K(t)$.  
Due to nonzero $a_n$, $K(t)$ in Figure \ref{fig:KrylovComplexityEx3}(a) oscillates and does not grow with respect to $t$.  $K(t)$ does not depend on the value of $M$. 

\subsection{Krylov entropy}
Krylov entropy $S(t)$ was introduced by \cite{Barbon:2019wsy} as a diagnostic of randomization of $\hat{\mathcal{O}}(t)$ in the Krylov subspace, which is defined by
\begin{align}
S(t):=-\sum_{n=0}^\infty\vert\varphi_{n}(t)\vert^2\log\vert\varphi_{n}(t)\vert^2.
\end{align}
As usual Shannon entropy, $S(t)$ measures randomness of $\varphi_{n}(t)$, which are coefficients of the expansion by the Krylov basis, and is maximized when $\varphi_{n}(t)$ follows a uniform distribution. In a similar way to determine the late time behavior of $K(t)$ from the smooth asymptotic behavior of $b_n$, typical behaviors of $S(t)$ with $a_n=0$ at late times are obtained by \cite{Barbon:2019wsy, Fan:2022xaa}
\begin{align}
b_n \sim \alpha n \,, \quad  \quad S(t) &\sim 2 \alpha t  \,,\\
b_n \sim \frac{\alpha n }{\log n} \,, \quad S(t) &\sim \sqrt{ 4 \alpha t}  \,,\\
b_n \sim \alpha   n^p \,, \,\,\, \quad S(t) &\sim  \log t \,,
\end{align}
where generically $S(t) \sim \log K(t)$.

Similar to $K(t)$, the Krylov entropy $S(t)$ also changes the behavior when $a_n$ is nonzero. 
For the case where the Lanczos coefficients given by (\ref{LancozsToyzeroT}), $S(t)$ is 
\begin{align}
S(t) = - \sum_{n=0}^\infty  \frac{|\beta(t)|^{2n}}{n!}  \exp\left({2 \mbox{Re}[\alpha(t)]}\right) \log \left(  \frac{|\beta(t)|^{2n}}{n!}  \exp\left({2 \mbox{Re}[\alpha(t)]}\right) \right) \nonumber \\
= - 2 \mbox{Re}[\alpha(t)]   -  \exp\left({2 \mbox{Re}[\alpha(t)]}\right) \sum_{n=0}^\infty  \frac{|\beta(t)|^{2n}}{n!}  \log \left(  \frac{|\beta(t)|^{2n}}{n!} \right)\,.
\end{align}
Figure \ref{fig:KrylovComplexityEx3}(b)  shows the time evolution of $S(t)$, which is independent of  $M$. As in Figure \ref{fig:KrylovComplexityEx3}(a) for $K(t)$, $S(t)$ oscillates and does not grow with respect to $t$.



\section{The IP model}\label{sec:IP}
\subsection{The model}
\begin{figure}
\center \includegraphics[width=20pc]{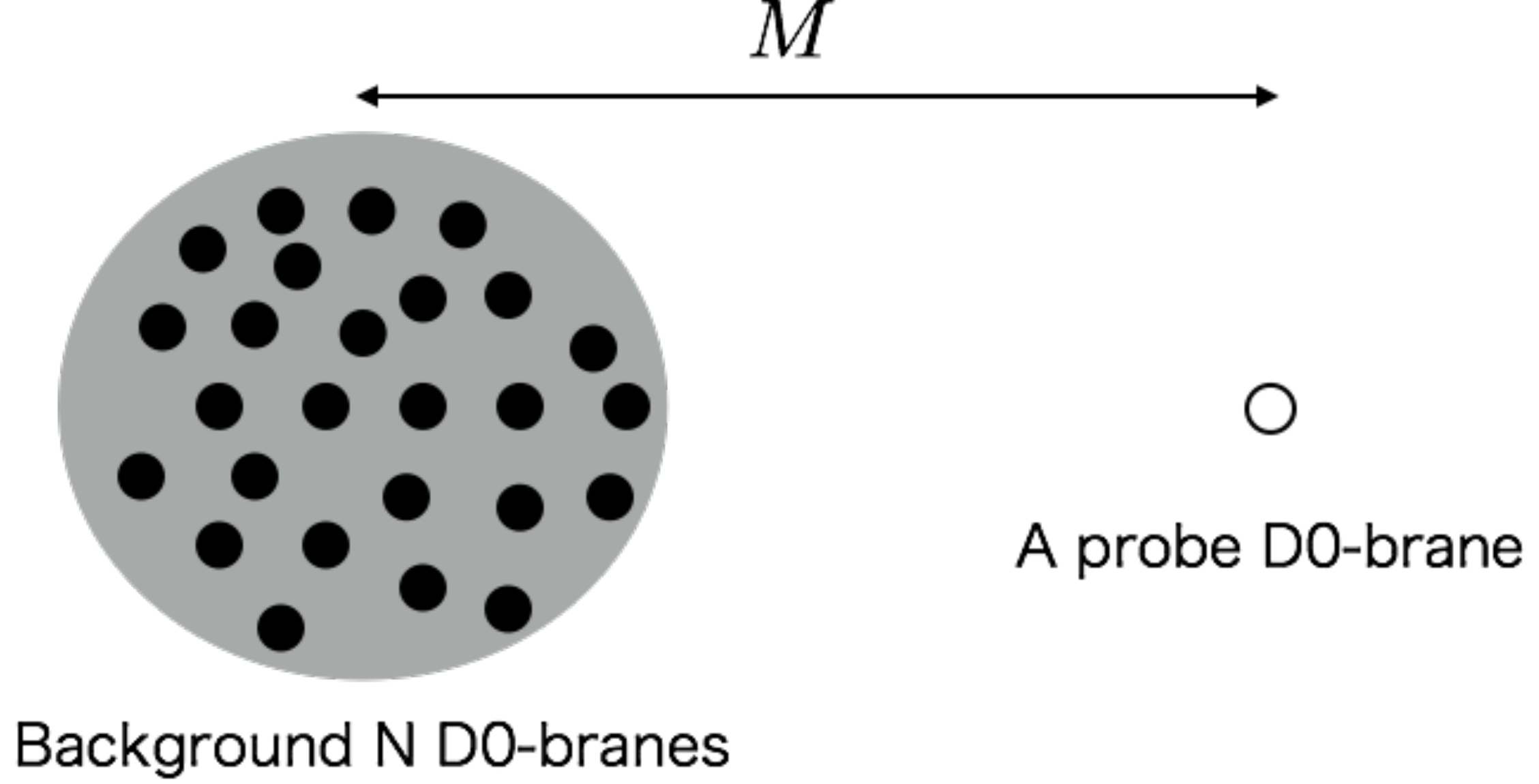}
\caption[]{The model can be regarded as a reduced version of a large $N$ D0-brane background black hole with a D0 probe \cite{Iizuka:2001cw}. The mass scale for D0-D0 fields is set to $m$, while the mass scale for D0-D0$_{\rm probe}$ fields is set to $M$. We consider the limit $M \gg m$ and $M \gg T$ at various $m/T$.} 
\label{fig:Dbranes} 
\end{figure}

The IP model contains 
a Hermitian matrix field $X_{ij}(t)$ and a complex vector field $\phi_i(t)$. $X_{ij}(t)$ and $\phi_i(t)$ are harmonic oscillators in the $U(N)$ adjoint and fundamental representations respectively. 
They obey the conventional quantization condition with their conjugate momenta, as 
\begin{equation}
[ X_{ij}, \Pi_{kl} ] = i \delta_{il} \delta_{jk}\ ,\quad [\phi_i, \pi_j] = i \delta_{ij}\ .
\end{equation}
The IP model Hamiltonian is 
\begin{equation}
H = \frac{1}{2} {\rm Tr}(\Pi^2) + \frac{m^2}{2} {\rm Tr}(X^2) +  \pi^\dagger (1+gX/M) \pi
+ M^2 \phi^\dagger(1+gX/M) \phi\ .
\end{equation}
This Hamiltonian can be re-written in terms of the creation and annihilation operators for the fundamental and antifundamental as 
\begin{equation}
H = \frac{1}{2} {\rm Tr}(\Pi^2) + \frac{m^2}{2} {\rm Tr}(X^2) + M(a^\dagger a + \bar a^\dagger \bar a) + g (a^\dagger X a + \bar a^\dagger X^T \bar a)\ . \label{ham}
\end{equation}
where we have 
\begin{equation}
a_i = \frac{ \pi_i^\dagger  - i M\phi_i}{\sqrt{2M}}\ ,\quad
\bar a_i = \frac{ \pi_i  - i M\phi_i^\dagger}{\sqrt{2M}}\ ,
\end{equation}
$X$ can be interpreted as a background $N$ D0-branes and $\phi$ can be interpreted as a probe to it. 
See Figure \ref{fig:Dbranes} as well as appendix \ref{sec:AppIP} for the motivation for this model related to D0-brane pictures and also the related models.

\begin{figure}
\center \includegraphics[width=28pc]{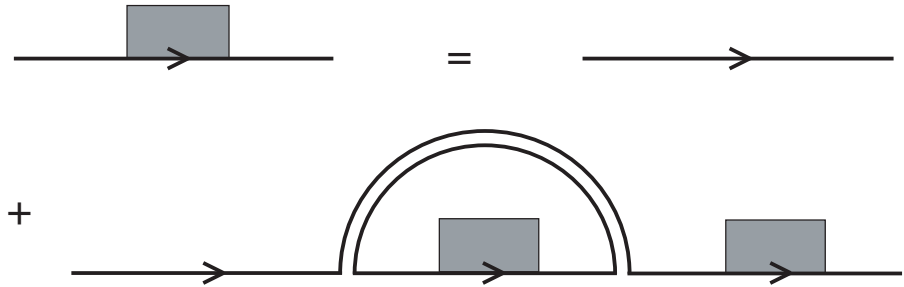}
\caption[]{Schwinger-Dyson equation for planar contributions to $\tilde G(\omega)$ (propagator with shaded rectangle) in terms of $\tilde G_0(\omega)$ and $\tilde K(\omega)$.}
\label{SDeq} 
\end{figure}

Consider the following observable, 
\be\label{tpa}
e^{i M (t-t')} \braket{ \mbox{T} \, a_i(t)\, a_j^\dagger(t')}_T := \delta_{ij} G(T, t-t')
\ee
where $a^\dagger_i$ and $a_i$ are creation/annihilation operator for a vector field $\phi_i$. 
With 't Hooft coupling $\lambda := g^2 N$, the Schwinger-Dyson equations for this vector field in the large $N$ limit becomes 
\be
\label{IPSDeq}
\tilde{G}(\omega) = \tilde{G}_0(\omega) - \lambda \tilde{G}_0(\omega) \tilde{G}(\omega) \int_{-\infty}^{\infty}  \frac{d \omega'}{2 \pi} \tilde{G}(\omega') \tilde{K}(T, \omega- \omega')
\ee
where $\tilde{G}$ is dressed propagator and $\tilde{G}_0$ is bare propagator, 
\be\label{G0}
\tilde{G}_0  = \frac{i}{\omega + i \epsilon} \,.
\ee
$\tilde{K}$ is propagator for the $N \times N$ matrix $X$, which is free. 
See Figure \ref{SDeq}. 
\\

A few comments are in order;
\begin{enumerate}
\item We choose $X_{ij}$ free. 
Note that in the large $N$ limit, the backreaction of vector $\phi_i$ on $X_{ij}$  is suppressed in $1/N$. 
\item We choose $\beta M$ very large, which corresponds to that a probe is far away from the background $N$ D0-branes, see Figure \ref{fig:Dbranes}. Therefore, a bare propagator for vector $\phi_i$ has no temperature dependence. However, $\beta m$ is not necessarily large. Thus, even though $X_{ij}$ is free, its correlator is given by the thermal one
\begin{align} 
\tilde K(T,\omega) = \frac{i}{1 - e^{- m/T}} \left( \frac{1}{\omega^2 - m^2 + i\epsilon}
- \frac{e^{- m/T}}{\omega^2 - m^2 - i\epsilon} \right) \ . \label{ktherm}
\end{align}
Note that even in the case of nonzero temperature, the propagator for $X_{ij}$ eq.~\eqref{ktherm} has a discrete pole at $\omega = \pm m$.   
\item 
The real-time thermal correlator $\tilde{G}(\omega)$ for $\phi_i$ can be obtained by solving the Schwinger-Dyson equation \eqref{IPSDeq}. Note that since $X_{ij}$ is free and the bare propagator for $\phi_i$ has no temperature dependence in our setting, the Schwinger-Keldysh formalism~\cite{Schwinger:1960qe,Keldysh:1964ud,Niemi:1983nf} simplifies. 
\item  Only through this $\tilde{K}$ in eq.~\eqref{ktherm},  the Schwinger-Dyson equation depends on the temperature. As is studied in detail in \cite{Iizuka:2008hg}, the spectrum of $\tilde{G}(\omega)$ behaves very differently depending on the parameters. $m=0$ is different significantly. Even for $m\neq0$, dependent on the temperature, the spectral density behaves very differently. In the infinite temperature limit, the spectral density shows a gapless spectrum and this results in that the correlator $G(t)$ decays exponentially in time. 
\end{enumerate}

\subsection{The spectral density}
Here we review the key nature of the IP model. We comment on zero temperature and nonzero temperature separately. 

\subsubsection{Zero temperature $T=0$}
In the zero temperature case,  solving the Schwinger-Dyson equation \eqref{IPSDeq} is easy. 
In zero temperature, \eqref{ktherm} reduces to 
\be
\label{kzero0}
\tilde{K}(T=0, \omega) = \frac{i}{\omega^2 - m^2 + i \epsilon} \,.
\ee

By conducting the $\omega$ integration in the upper half-plane, the Schwinger-Dyson equation 
eq.~\eqref{IPSDeq} becomes 
\be
\label{zeroTdifeqfortildeG}
\tilde{G}(\omega) = \frac{i}{\omega} \left( 1 - \frac{\lambda}{2 m} \tilde{G}(\omega)  \tilde{G}(\omega - m) \right)  \,,
\ee
One can solve this equation analytically by mapping this equation to the Bessel recursion relation based on the canonical calculation, See \S 2.3 in \cite{Iizuka:2008hg}. The result is 
\begin{align}\label{GzeroT}
\tilde G(\omega) = \frac{2i}{\nu}\frac{ J_{-\omega/m}(-\nu/m)}{J_{-1-\omega/m}(-\nu/m)}= -\frac{2i}{\nu}\frac{ J_{-\omega/m}(\nu/m)}{J_{-1-\omega/m}(\nu/m)}\ .
\end{align}
where $J$ is a Bessel function. 
The spectrum is determined by the pole of $\tilde G(\omega) $ which is discrete for nonzero $m \neq 0$. There are infinite poles, which are determined by the zeros of the denominator as 
\begin{align}
\label{zeroTpoles}
J_{-1-\omega/m}(-\nu/m) = 0  \,.
\end{align}
The numerators in $\tilde G(\omega)$ can not cancel these zeros due to Bourget's hypothesis.

Let us take the following massless double scaling limit 
\be
\label{GUEmodellimit}
m \to 0 \,,  \quad \lambda \to 0 \,, \quad \nu^2 := \frac{2 \lambda}{m}  = \mbox{fixed} 
\ee
Then we can obtain the dressed propagator from eq.~\eqref{zeroTdifeqfortildeG} as 
\be
\tilde{G}(\omega) = \frac{2 i}{\nu^2} \left({ \omega - \sqrt{\omega^2 - \nu^2} }\right) \,, 
\ee
and from this, we obtain the spectrum 
\be
\label{wignermassless}
\rho(\omega) = - \frac{1}{\pi} \mbox{Im} \left(-i  \tilde{G}(\omega) \right) = \frac{2}{ \nu^2 \pi}  \sqrt{\nu^2 - \omega^2} \,.
\ee
This shows that the pole at $\omega = 0$ has been broadened into a branch cut of width $\order(\nu)$. 
However note that this is simply due to the non-compact spreading of $X$ at $m \to 0$, which is not related to the large $N$ phase transition. To see this,  
at finite $m$, the zero temperature wave function of $X$ is Gaussian and therefore follows Wigner's semi-circle eigenvalue distribution of width $\order(\sqrt{N/m})$.  
Since the effective mass is given by $\sim \sqrt{g} \braket{X}$, 
at the limit given by eq.~\eqref{GUEmodellimit}, an effective mass spread whose width is given by  $\sqrt{g} \braket{X} = \order(\nu)$. 
\\

Note that 
\begin{enumerate}
\item Even though the spectrum is continuous in the massless limit of $X$, {\it i.e.,} $m\to 0$, the spectrum of $\phi$ is Wigner's semi-circle type. This gives only a power-law decay for $G(t)$ and it does not give an exponential decay. 
\item We are interested in how the discrete and mass-gapped system at zero coupling shows a continuum spectrum in the large $N$ limit at strong coupling. For that purpose, we study finite temperature effects with $m\neq 0$. 
\end{enumerate}

\subsubsection{Nonzero temperature $T \neq 0$}
We now consider nonzero temperature case $T \neq 0$ with mass gapped $m \neq 0$ where $X$ correlator is given by \eqref{ktherm}. Again using the fact that 
the time-ordered correlator $G(T,t)$ vanishes at $t < 0$ (because $ M \gg T$),  by closing the contour in the upper half-plane, we have 
\begin{align}
\tilde G(T,\omega - m)  - \frac{4}{\nu_T^2}\frac{1}{ \tilde G(T,\omega)} + e^{- m/T} \tilde G(T,\omega + m) 
&= \frac{4 i \omega}{\nu_T^2}  \label{sde4} \\
\mbox{where} \qquad \nu_T^2 = \nu^2/(1- e^{- m/T}) \,. \qquad \qquad  & 
\end{align}
This is key recursion relation to solve to obtain the spectrum.

Note that in the limit $m \to 0$, with $\nu_T^2$ fixed, the recursion relation becomes 
\begin{equation}
2{\nu_T^2} \tilde G^2(T,\omega)  - 4 i \omega \tilde G(T,\omega)  - 4 = 0\ . \label{alg2}
\end{equation}
and this gives the solution
\begin{equation}
\tilde G(T ,\omega) = \frac{i}{\nu_T^2} \left(
\omega - \sqrt{\omega^2 - 2 \nu_T^2} \right) \ . \label{m0t}
\end{equation}
and the spectrum 
\be
\label{wignermassless2}
\rho(\omega) = - \frac{1}{\pi} \mbox{Im} \left(-i  \tilde{G}(\omega) \right) = \frac{1}{ \nu_T^2 \pi}  \sqrt{2 \nu_T^2 - \omega^2} \,.
\ee
which is the same type of Wigner's semi-circle distribution at zero temperature. However, this is not our main interest.  
Therefore we focus on finite mass case $m\neq 0$. 

Since the real part of $\tilde{G}$ is spectral density, defining 
\begin{align}
{\rm Re}\,\tilde G(T,\omega) =  \pi \pho(\omega) 
\end{align}
eq.~\eqref{sde4} reduces to 
\begin{equation}
\pho(\omega - m)  - \frac{4}{\nu_T^2 | \tilde G(T,\omega) |^2} {\pho(\omega)} + e^{- m/T} \pho(\omega + m) = 0 \ .
\label{real} 
\end{equation}
This recursion relation shows the behavior we want; in large $\omega$, it decays exponentially as $\pho(\omega) \sim |\omega|^{-|\omega|}$. 
In fact, asymptotic solution can be obtained as  
\begin{align}
\pho(\omega) \sim& \exp\left[-\frac{2\vert\omega\vert}{m}\log\left(\frac{2\vert\omega\vert}{e^{- m/2T}\nu_T}\right)\right] \;\;\; (\omega\to-\infty)\label{sed-}\\
\sim&\exp\left[-\frac{2\vert\omega\vert}{m}\log\left(\frac{2\vert\omega\vert}{\nu_T}\right)\right] \;\;\; (\omega\to+\infty).\label{sed+}
\end{align}
There is a difference in the asymptotic behaviors due to $e^{- m/T}\ne1$, but this is a sub-leading contribution at sufficiently high temperature and negligible in large $|\omega|$. This is because  
\begin{align}
\log\left(\frac{2\vert\omega\vert}{e^{- m/2T}\nu_T}\right) = \log\left(\frac{2\vert\omega\vert}{\nu_T}\right) + \frac{m}{2 T} \,, 
\end{align}
thus it is negligible at sufficiently high frequency where 
\begin{align}
\frac{2\vert\omega\vert}{\nu_T} \gg e^\frac{m}{2 T} \sim 1 + \frac{m}{2T} \quad \mbox{(at $T \gg m$)}\,. 
\end{align}

\begin{figure}[p]
\vspace{-8mm}
\centering
     \begin{subfigure}[b]{0.35\textwidth}
         \centering
         \includegraphics[width=\textwidth]{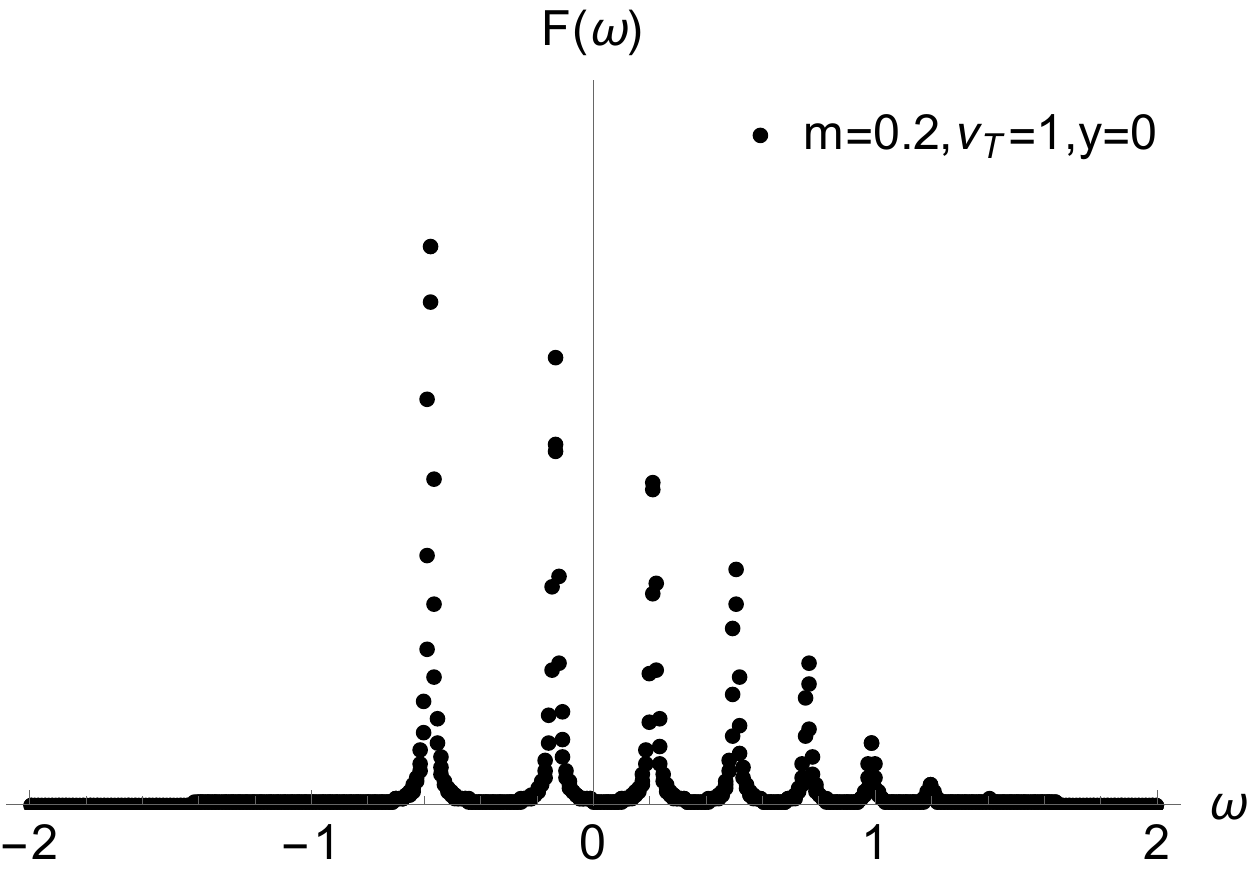}
     \end{subfigure}
      \quad\quad\quad
          \begin{subfigure}[b]{0.35\textwidth}
         \centering
         \includegraphics[width=\textwidth]{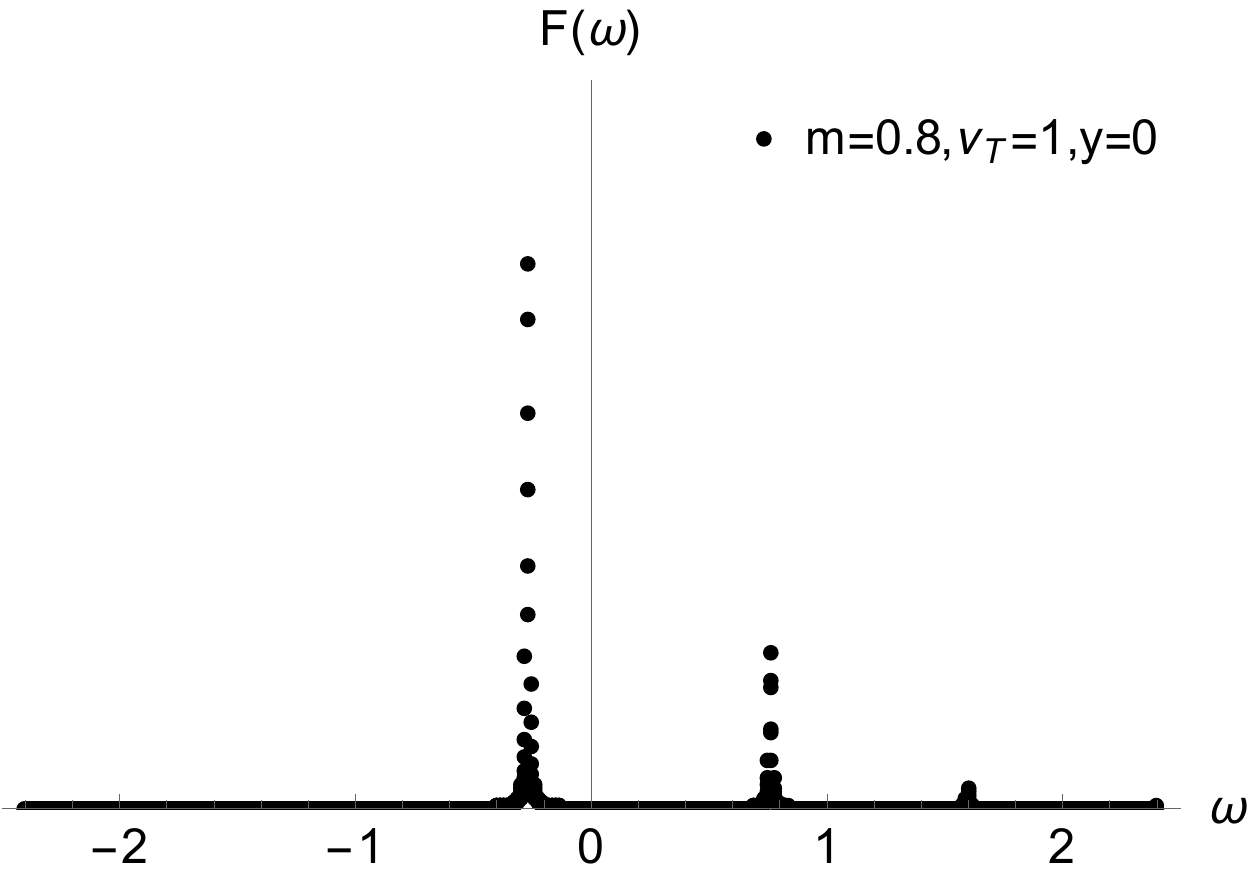}
     \end{subfigure}
     \begin{subfigure}[b]{0.35\textwidth}
         \centering
         \includegraphics[width=\textwidth]{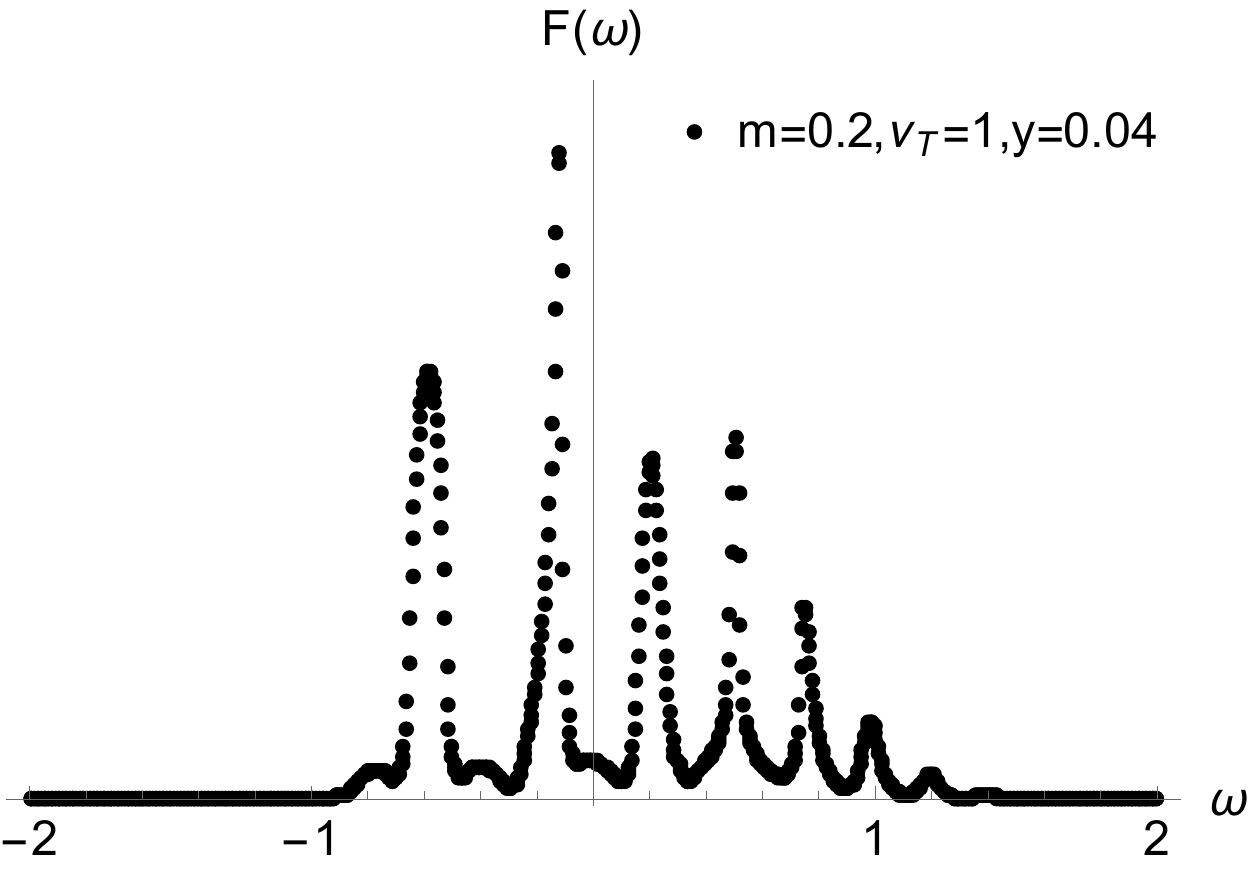}
     \end{subfigure}
      \quad\quad\quad
        \begin{subfigure}[b]{0.35\textwidth}
         \centering
         \includegraphics[width=\textwidth]{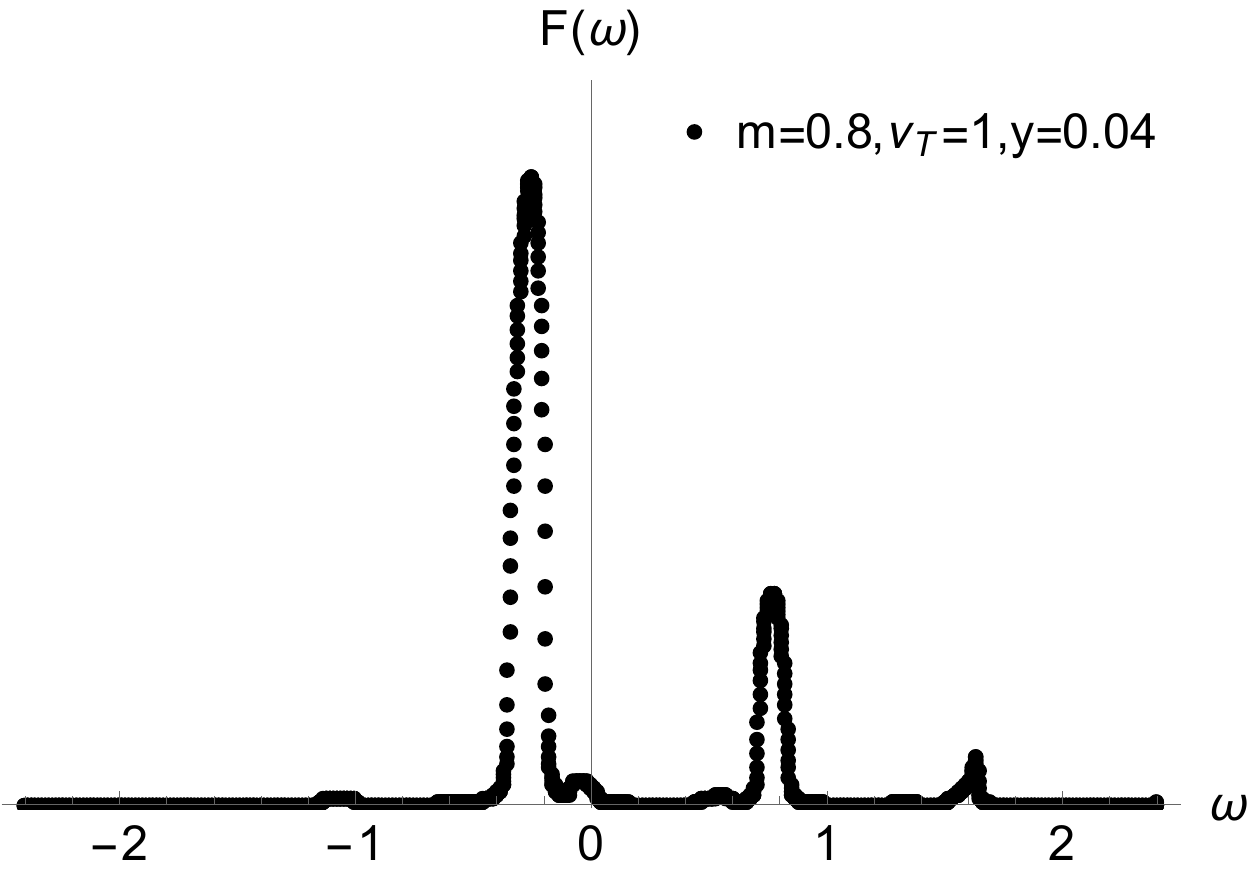}
     \end{subfigure}
         \begin{subfigure}[b]{0.35\textwidth}
         \centering
         \includegraphics[width=\textwidth]{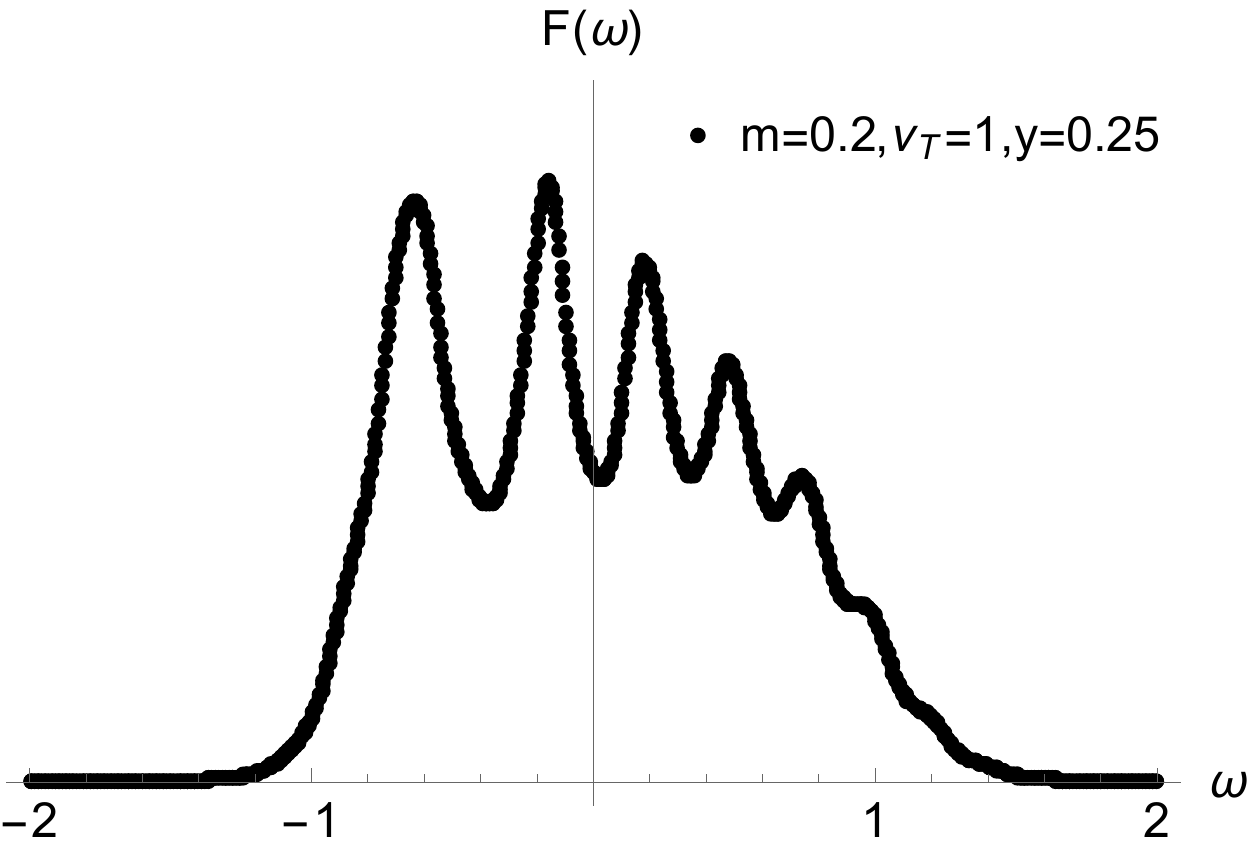}
     \end{subfigure}
      \quad\quad\quad
       \begin{subfigure}[b]{0.35\textwidth}
         \centering
         \includegraphics[width=\textwidth]{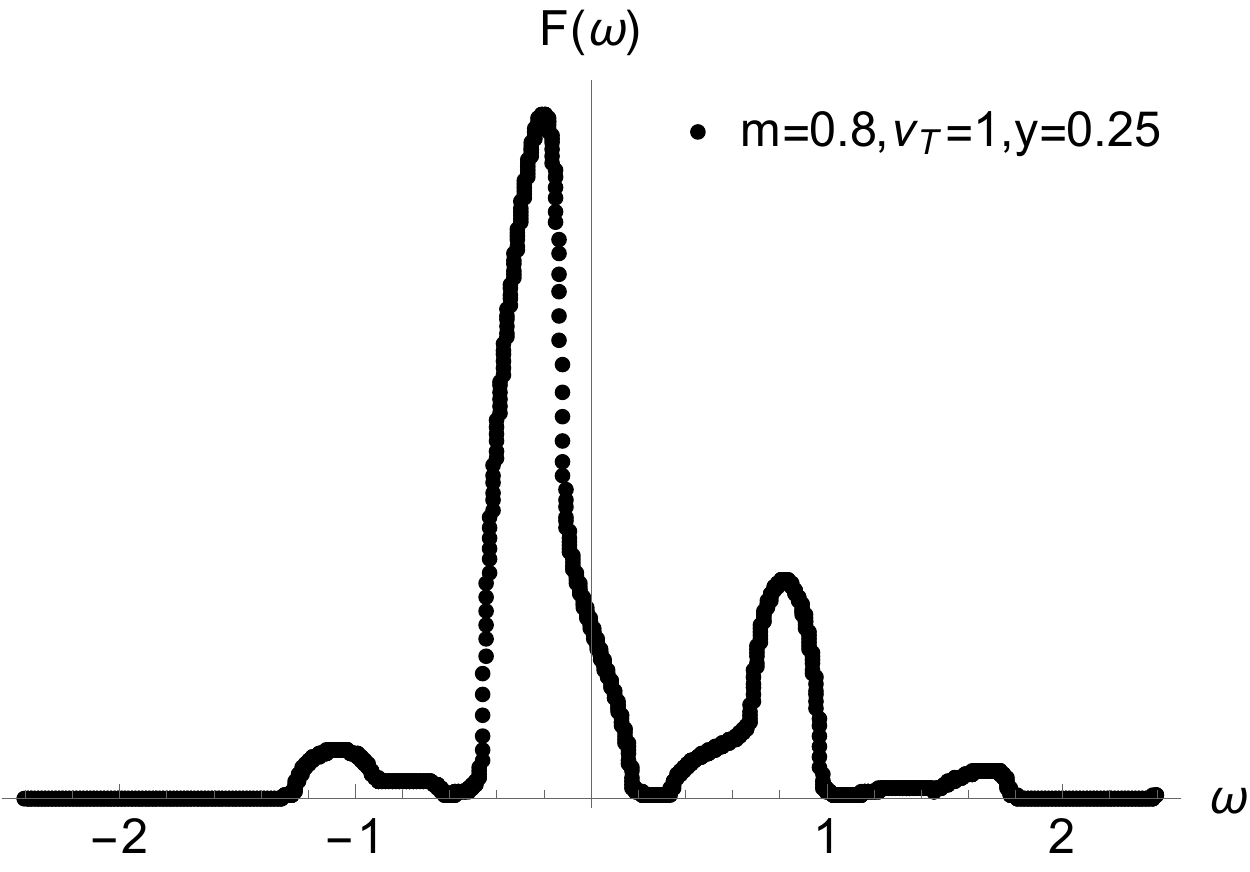}
     \end{subfigure}
            \begin{subfigure}[b]{0.35\textwidth}
         \centering
         \includegraphics[width=\textwidth]{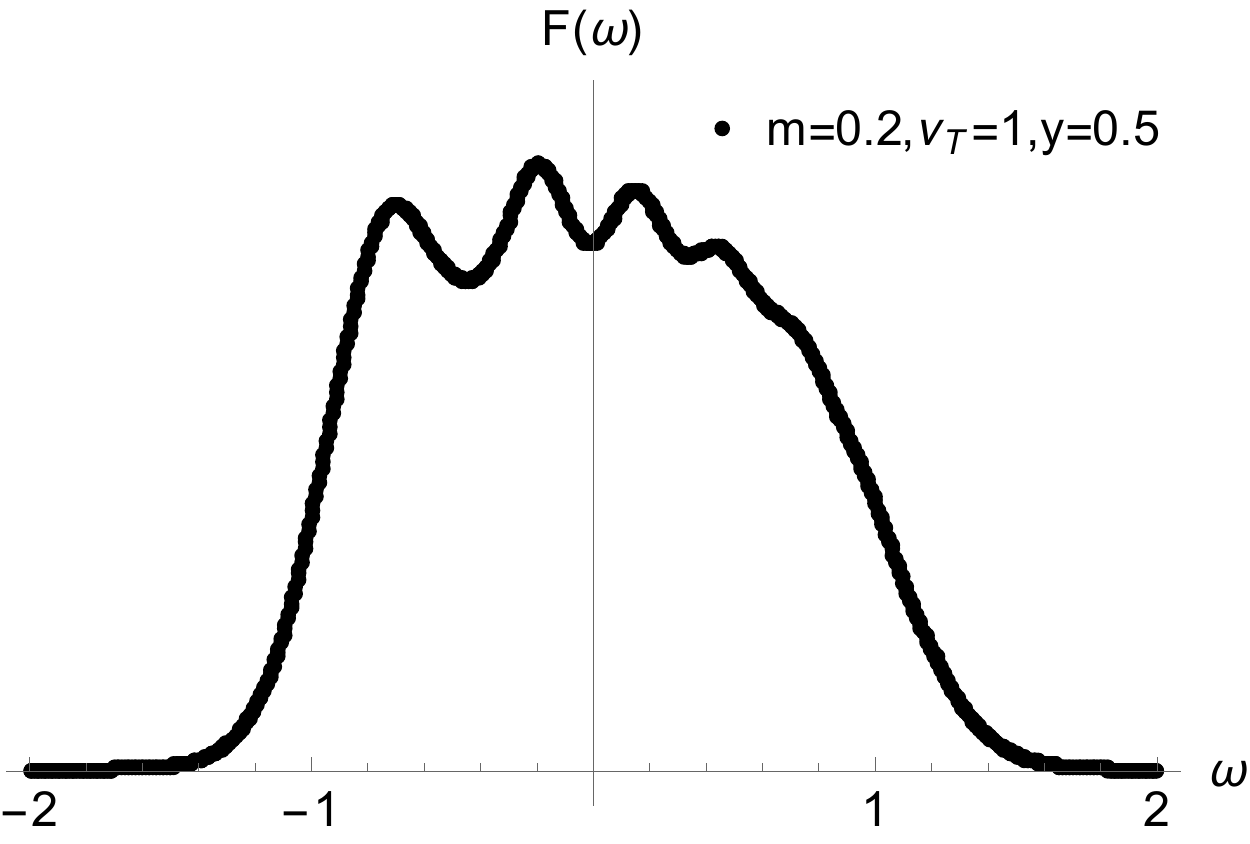}
     \end{subfigure}
      \quad\quad\quad
       \begin{subfigure}[b]{0.35\textwidth}
         \centering
         \includegraphics[width=\textwidth]{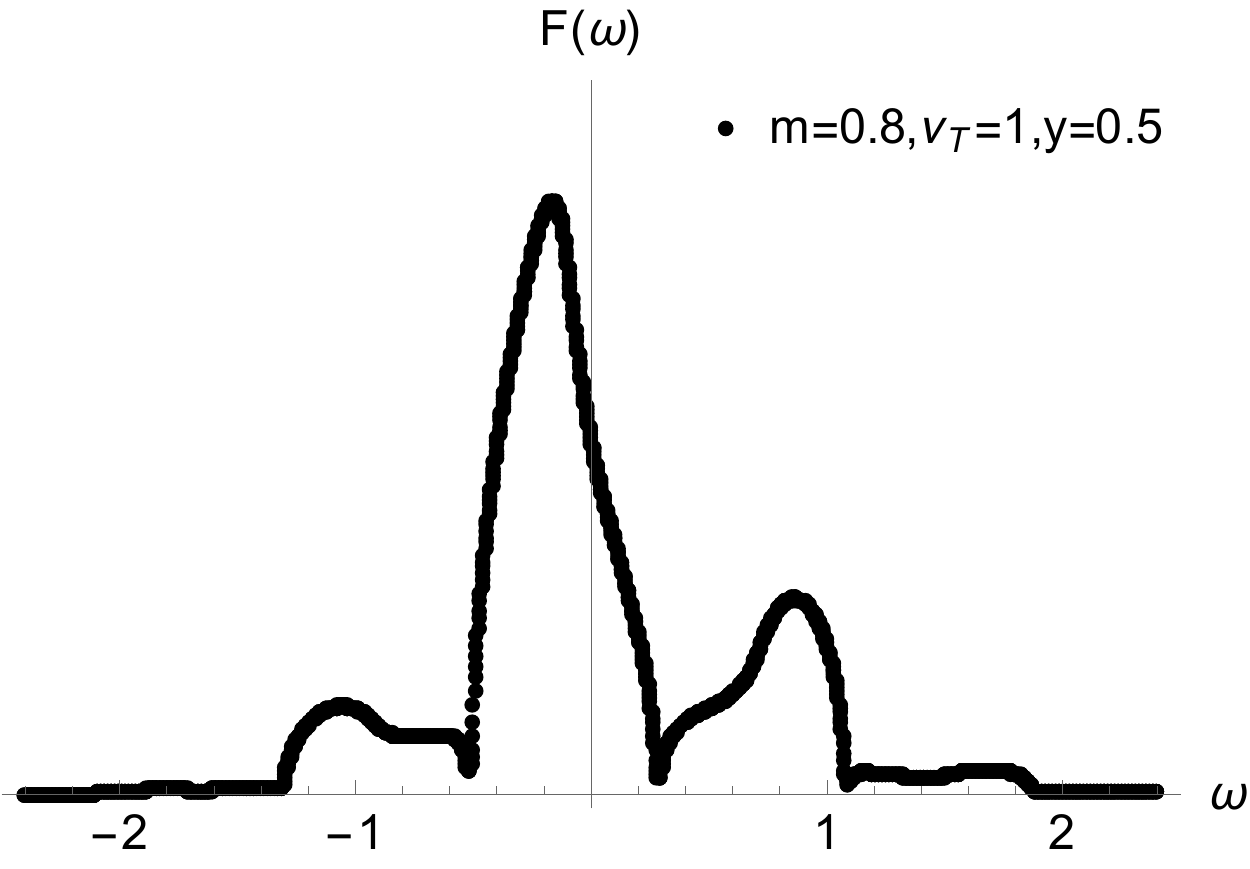}
     \end{subfigure}
       \begin{subfigure}[b]{0.35\textwidth}
         \centering
         \includegraphics[width=\textwidth]{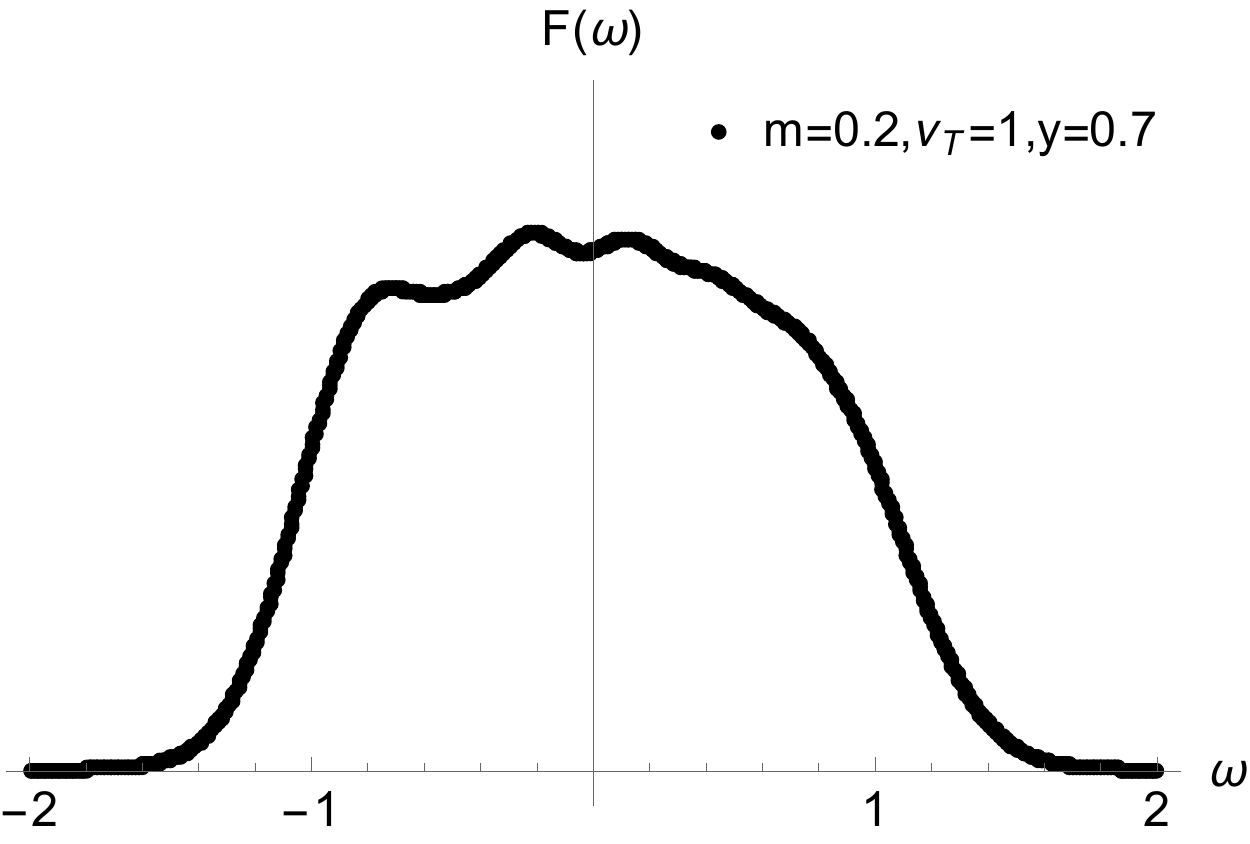}
     \end{subfigure}
      \quad\quad\quad
       \begin{subfigure}[b]{0.35\textwidth}
         \centering
         \includegraphics[width=\textwidth]{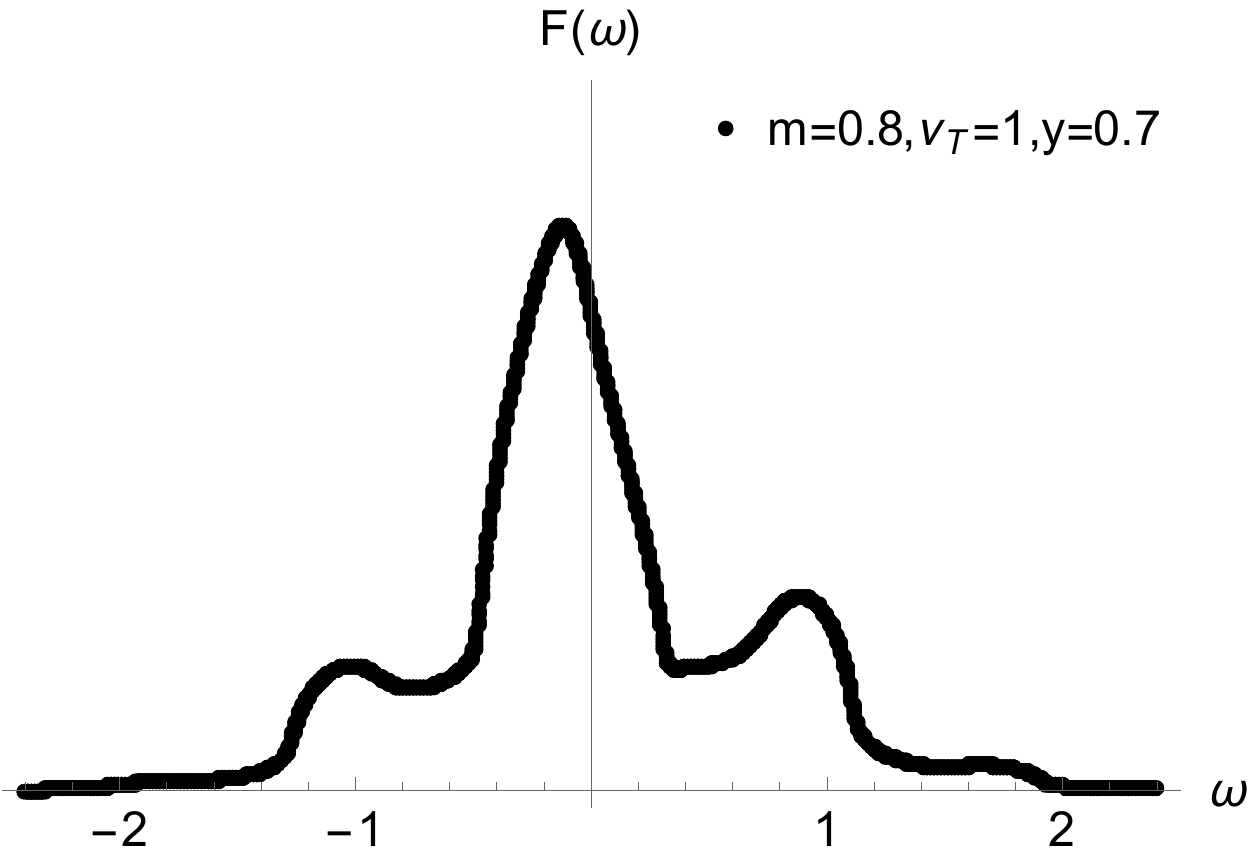}
     \end{subfigure}
              \begin{subfigure}[b]{0.35\textwidth}
         \centering
         \includegraphics[width=\textwidth]{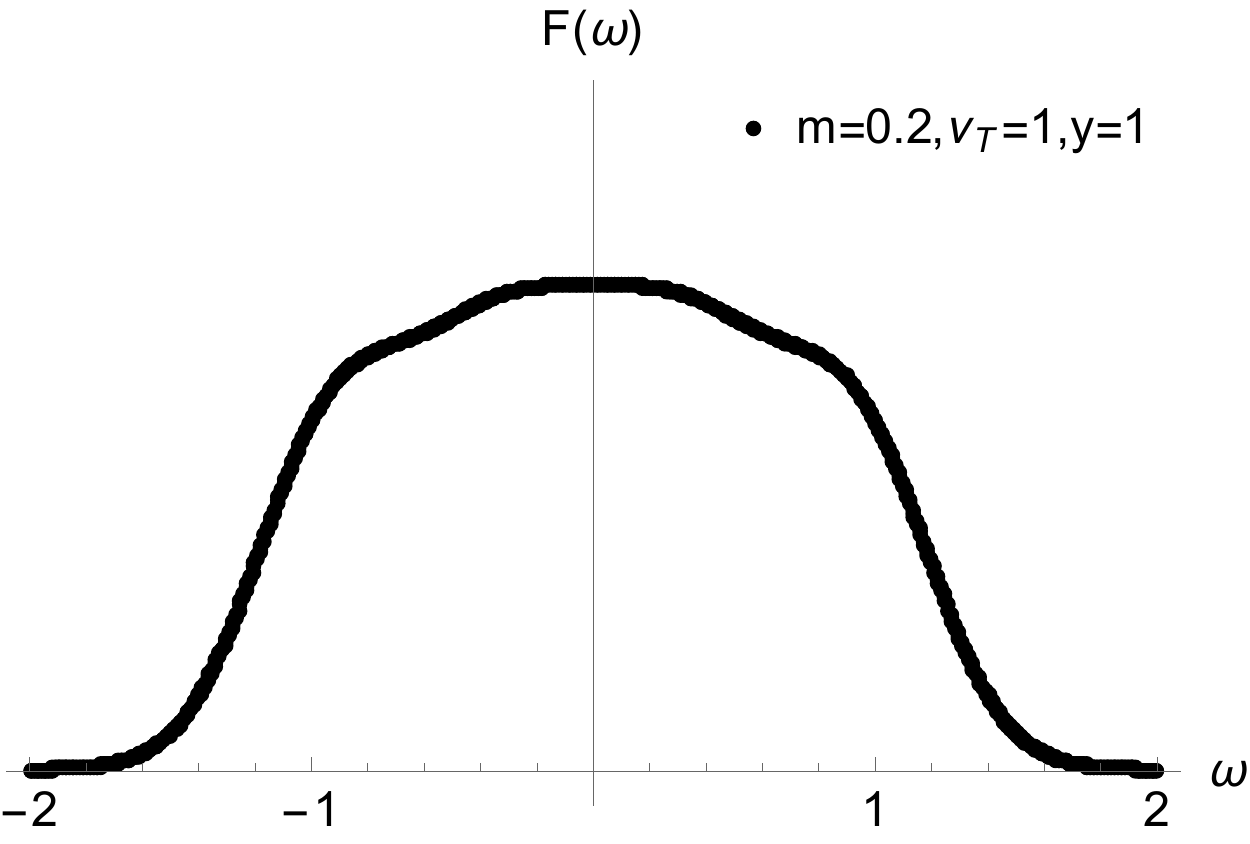}
              \end{subfigure}
         \quad\quad\quad
          \begin{subfigure}[b]{0.35\textwidth}
         \centering
         \includegraphics[width=\textwidth]{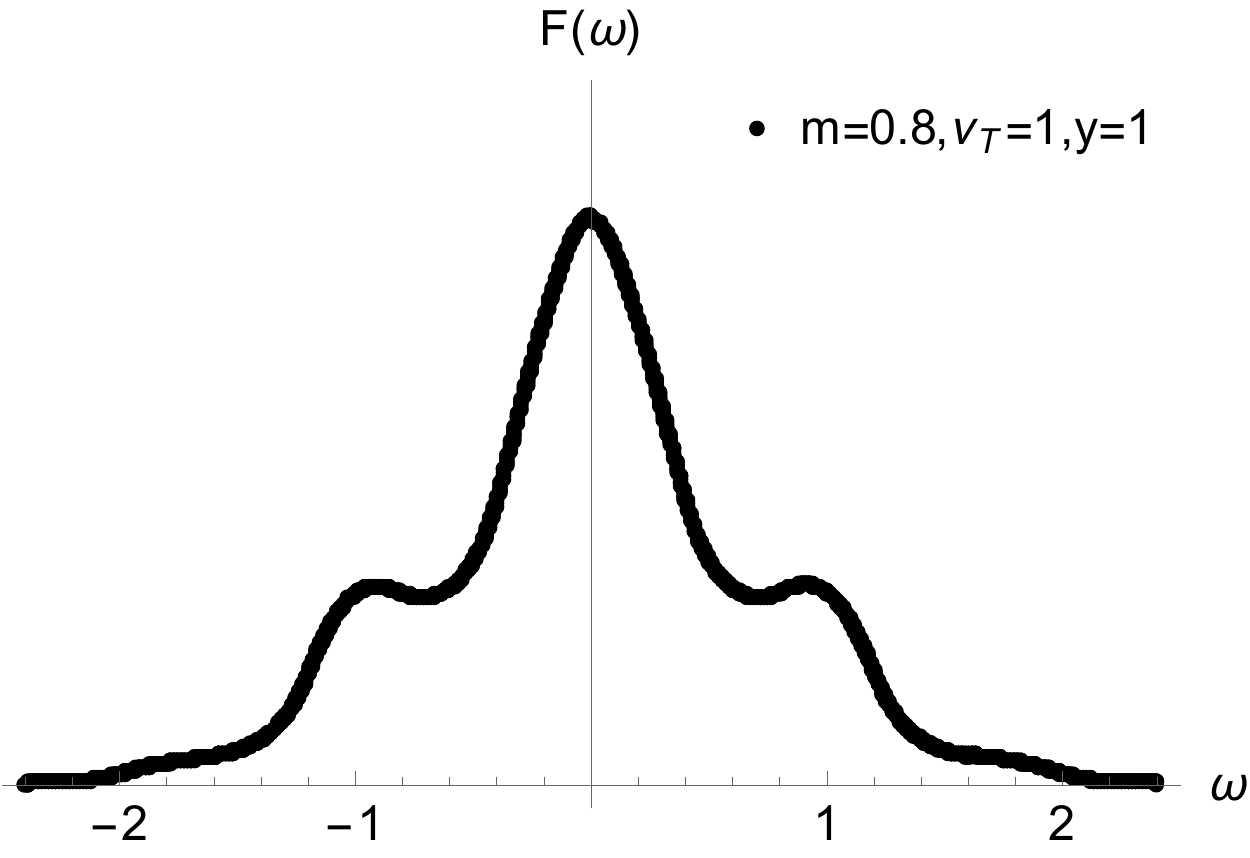}
     \end{subfigure}
             \caption{Numerical plots of $F(\omega)$ for $\nu_T = 1$, $m=0.2$ and $m=0.8$ at various temperatures from $T=0$ to $T=\infty$ where $y=e^{-m/T}$. For visibility, the scale of the vertical axis is changed as appropriate for each figure. To visualize poles at zero temperature, we set the imaginary part of $\omega$ to a small positive nonzero value.}
        \label{fig:F(w)}
\end{figure}

In Figure \ref{fig:F(w)}, we plot $F(\omega)$ with various values of $y:=e^{- m/T}$ by solving the recursion relation (\ref{real}) numerically. The left figures shows the $y$-dependence of $F(\omega)$ with $m=0.2, \nu_T=1$, and the right figures show the $y$-dependence of $F(\omega)$ with $m=0.8, \nu_T=1$. For numerical computations, we use a boundary condition (\ref{G0}) of $\tilde{G}(\omega)$ at large $\vert\omega\vert$ with small $\epsilon>0$. At the zero temperature $y=0$, the numerical plot of $F(\omega)$ has peaks that correspond to poles. As $y$ increases, the spread of peaks becomes larger, which means that the poles become branch cuts. Furthermore, as $y$ increases, the brunch cuts merge, and $F(\omega)$ becomes gapless. At the infinite temperature $y=1$, $F(\omega)$ becomes a positive, smooth, and even function.

\clearpage

\section{Lanczos coefficients in the IP model}\label{sec:LanczosIP} 
In this section, we evaluate the Lanczos coefficients of the IP model in the large $N$ limit associated with $\hat{\mathcal{O}}=\hat{a}^\dagger_j$ and its two-point function 
\begin{align}
C(t;\beta):=e^{i M t}(\hat{a}_j^\dagger(t)\vert \hat{a}_j^\dagger(0))_\beta \,.
\end{align}
where $C(t;\beta)$ is {\it not} a time-ordered correlator, and thus different from $G(t)$ which is a time-ordered one as eq.~\eqref{tpa}.

In the previous section, we reviewed the recursion relation (\ref{real}) to compute the real part ${\rm Re}\,\tilde G(T,\omega) =  \pi \pho(\omega)$ of the Fourier transformation of $G(T,t)$ (\ref{tpa}). We show that the Fourier transformation $f(\omega)$ of $C(t;\beta)$ can be expressed by $\pho(\omega)$. First, one can show that
\begin{align}\label{timereversal}
C^*(t;\beta)=e^{-i M t}(\hat{a}_j^\dagger(0)\vert \hat{a}_j^\dagger(t))_\beta=e^{-i M t}(\hat{a}_j^\dagger(-t)\vert \hat{a}_j^\dagger(0))_\beta=C(-t;\beta).
\end{align}
Next, by using (\ref{timereversal}), we show that
\begin{align} 
f(\omega) &= \int_{0}^{\infty}dt\,e^{i\omega t}C(t;\beta)+\int_{-\infty}^{0}dt\,e^{i\omega t}C(t;\beta)\notag\\
&=\int_{0}^{\infty}dt\,e^{i\omega t}C(t;\beta)+\int_{0}^{\infty}dt\,e^{-i\omega t}C(-t;\beta)\notag\\
&=\int_{0}^{\infty}dt\,e^{i\omega t}C(t;\beta)+\left(\int_{0}^{\infty}dt\,e^{i\omega t}C(t;\beta)\right)^*\notag\\
&=\tilde G(T,\omega)+\tilde G^*(T,\omega)=2\pi \pho(\omega),
\end{align}
where we use the fact that $G(T,t)$ vanishes at $t<0$ and $G(T,t)=C(t;\beta)$ at $t>0$. Therefore, we can compute $f(\omega)=2\pi \pho(\omega)$ by solving the recursion relation (\ref{real}) at least numerically. The Lanczos coefficients can then be calculated from the obtained $f(\omega)$ by using the moment method.

\subsection{Lanczos coefficients in the massless limit}
In the massless limit for an adjoint, $m \to 0$, with $\nu_T^2$ fixed, the spectral density $F(\omega)$ is given by
\begin{align}
F(\omega) = \frac{1}{\pi\nu_T^2} \left(
\sqrt{ 2 \nu_T^2 -\omega^2} \right) \;\;\; \left(\omega^2\le2 \nu_T^2\right) .
\end{align}
With the above spectrum, the Lanczos coefficients are given by an $n$-independent constant \cite{RecursionBook}
\begin{align}\label{bndsl}
b_n= \frac{\nu_T}{\sqrt{2}} \,.
\end{align}
We note that $a_n=0$ because $F(\omega)$ is an even function.
Therefore, in the massless limit, the Lanczos coefficients do not grow due to the continuous spectrum with bounded support. 

\subsection{Lanczos coefficients at zero temperature for nonzero mass}
At zero temperature, let $\vert v\rangle$ be the free ground state. Then, consider excited states $\vert j, n\rangle$ with a single excitation by $\hat{a}^\dagger_i$ and $n$-excitations by $\hat{A}^\dagger_{ij}$ such as
\begin{align}\label{IPbasiszerotemperature}
\vert j, n\rangle:=i^{-n}N^{-n/2}\hat{a}_i^\dagger (\hat{A}^{\dagger n})_{ij}\vert v\rangle.
\end{align}
In the large $N$ limit, the states $\vert j, n\rangle$ span an orthonormal basis for the two-point function. For example, at $n=2$, one can obtain
\begin{align}
\langle j', 2\vert j,2\rangle=&\frac{1}{N^2}\langle v\vert  \hat{A}_{j'k'} \hat{A}_{k'i'} \hat{a}_{i'} \hat{a}_i^\dagger  \hat{A}^{\dagger}_{ik} \hat{A}^{\dagger}_{kj}\vert v\rangle\notag\\
=&\frac{\delta_{ii'}}{N^2}\left([ \hat{A}_{k'i'}, \hat{A}^{\dagger}_{ik}][ \hat{A}_{j'k'},  \hat{A}^{\dagger}_{kj}]+[ \hat{A}_{k'i'}, \hat{A}^{\dagger}_{kj}][ \hat{A}_{j'k'},  \hat{A}^{\dagger}_{ik}]\right)\notag\\
=&\frac{N^2+1}{N^2}\delta_{jj'},
\end{align}
which is normalized to one in the large $N$ limit.
Moreover, in the large $N$ limit, we obtain \cite{Iizuka:2008hg}
\begin{align}
(H-M)\vert j,n\rangle=mn \vert j,n\rangle+\frac{\nu}{2}\vert j,n-1\rangle+\frac{\nu}{2}\vert j,n+1\rangle,
\end{align}
which has the same structure as (\ref{recursionOn}) with $\mathcal{L}=H-M$. 
Therefore, the Krylov basis for $\hat{\mathcal{O}}=\hat{a}^\dagger_j$ is $\vert \hat{\mathcal{O}}_n)=\vert j, n\rangle$ given by eq.~\eqref{IPbasiszerotemperature}. Then the Lanczos coefficients are given by
\begin{align}\label{LczeroT}
a_n=mn, \;\;\; b_n=\frac{\nu}{2}.
\end{align}
This zero temperature case is very similar to Example \ref{ex3zerotemperature} at zero temperature in Section \ref{sec:reviewLanczos}, but $b_n$ in the IP model does not depend on $n$ because we determine the normalization of (\ref{IPbasiszerotemperature}) in the large $N$ limit. 

Note that the Krylov subspace is infinite-dimensional, given by eq.~\eqref{IPbasiszerotemperature} with $n\ge 0$. This corresponds to the fact that there are infinite poles in the Green function as we have seen in eq.~\eqref{zeroTpoles}.

\subsection{Lanczos coefficients in the infinite temperature limit for nonzero mass}
We evaluate the Lanczos coefficients of the IP model in the infinite temperature limit by analyzing the asymptotic behavior of the spectral density.
In the infinite temperature limit $e^{-m/T}\to1$, as reviewed in Section \ref{sec:IP}, the spectral density $F(\omega)$ is a positive and smooth function everywhere \cite{Iizuka:2008hg}. Moreover, due to the symmetry of the recursion relation, $F(\omega)$ is an even function with respect to $\omega$, which leads to $a_n=0$. The asymptotic behavior in the infinite temperature limit is 
\begin{align}
F(\omega)\sim \exp\left[-\frac{2\vert\omega\vert}{m}\log\left(\frac{2\vert\omega\vert}{\nu_T}\right)\right] \;\;\; (\vert\omega\vert\to\infty).\label{abIP}
\end{align} 
This exponential suppression is known as the slowest decay for the situation where $C(t)$ is analytic in the entire complex time plane \cite{Parker:2018yvk, Avdoshkin:2019trj}. The corresponding maximal growth of the Lanczos coefficients $b_n$ is \begin{align}
b_n &\sim A \frac{n}{W(n)}\sim A \frac{n}{\log n} \;\;\; (n\to\infty), 
\end{align}
Here $W(n)$ is defined by 
\begin{align}
z=W(z e^z),
\end{align}
which is called the Lambert W function or the product-log function. In the infinite temperature limit, the IP model depends on only one dimensionless parameter $\frac{m}{\nu_T}$. In fact, (\ref{abIP}) can be expressed by $\frac{m}{\nu_T}$ and dimensionless frequency $\frac{\omega}{\nu_T}$.

Let us determine the constant $A$ by using a mathematical prescription \cite{Lubinsky:1988}. See the appendix \S \ref{appftobn}. Suppose that the spectral density is given by $F(\omega)=\exp\left[-2Q(\omega)\right]$, where $Q(\omega)$ satisfies some restrictions for a mathematical proof. 
For the asymptotic behavior (\ref{abIP}) of the IP model, we choose $2Q(\omega)=\frac{2\vert\omega\vert}{m}\log\left[2\vert\omega\vert/\nu_T\right]$, and a solution of (\ref{rootBn}) is
\begin{align}
B_n=\frac{m\pi n}{2W(2m\pi n/\nu_T)}.
\end{align}
Therefore, we conclude that the asymptotic behavior of $b_n$ of the IP model in the infinite temperature limit is
\begin{align}
b_n\sim \frac{m\pi n}{4W(2m\pi n/\nu_T)}\sim \frac{m\pi n}{4\log n}  \;\;\; (n\to\infty).\label{abbnIP}
\end{align}
From eq.~(\ref{recursionOnfiniteT}), $b_n$ has the same mass dimension as $H$, and the dimensionless Lanczos coefficient $\frac{b_n}{\nu_T}$ depends on $n$ and $\frac{m}{\nu_T}$. In the following numerical computations, we fix a scale of $\nu_T$ as $\nu_T=1$.

\begin{figure}
     \begin{subfigure}[b]{0.45\textwidth}
         \centering
         \includegraphics[width=\textwidth]{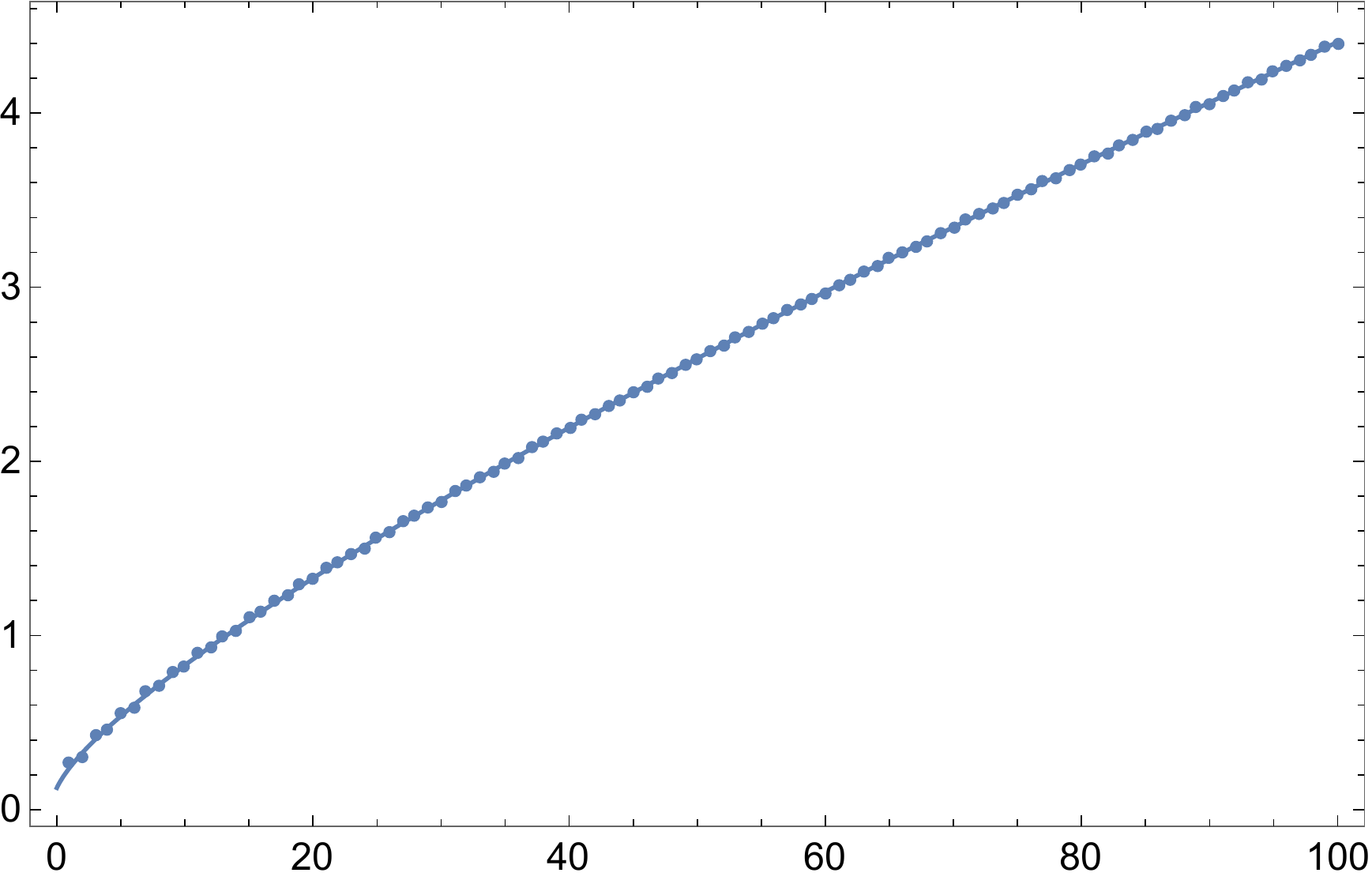}
       \put(5,0){$n$}
    \put(-195,130){$b_n$}
       \caption{$m=0.2, \;\nu_T=1$}\label{fig:LanczosToymodel(a)}
     \end{subfigure}
      \hfill
     \begin{subfigure}[b]{0.45\textwidth}
         \centering
         \includegraphics[width=\textwidth]{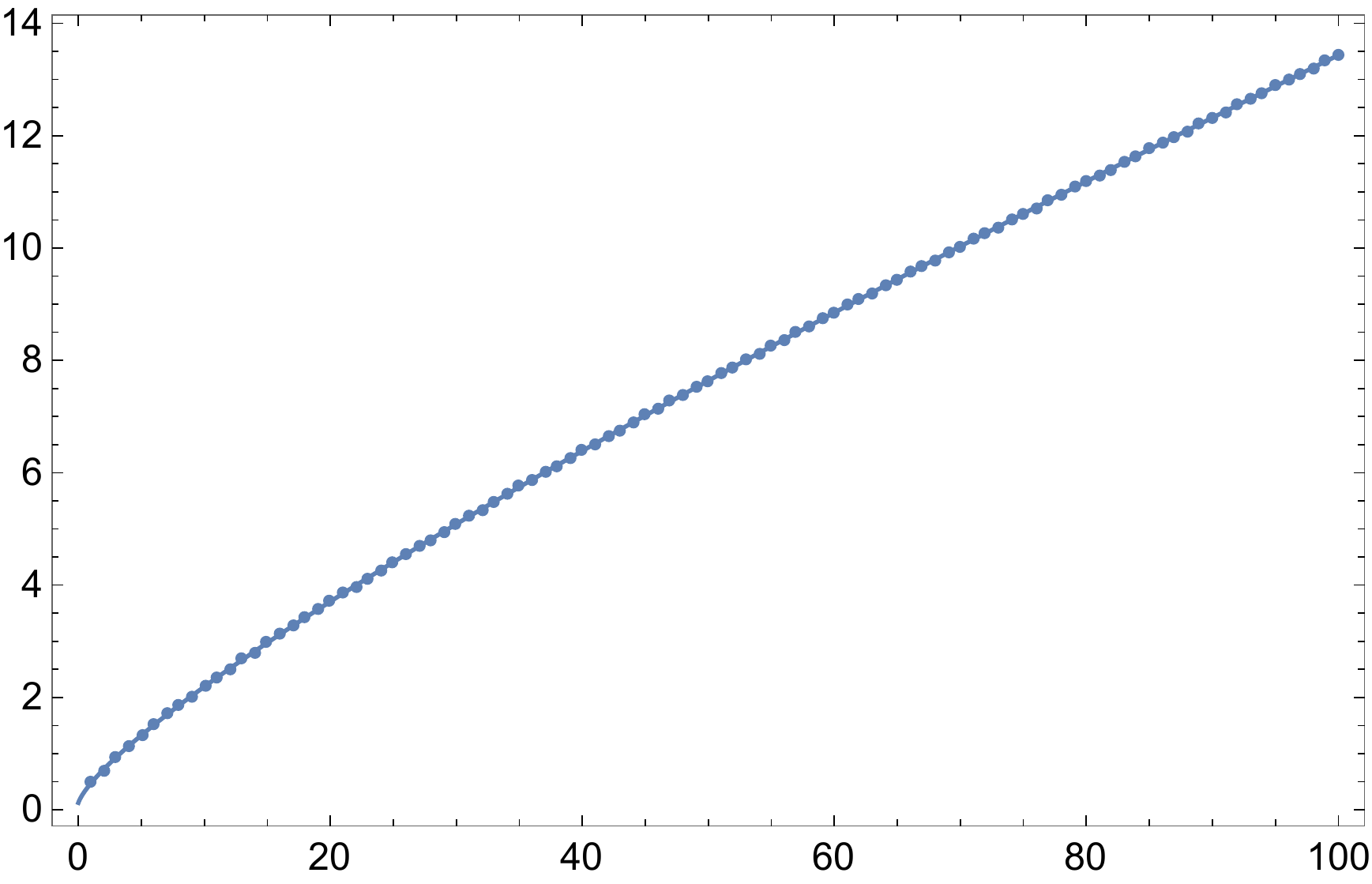}
          \put(5,0){$n$}
    \put(-195,130){$b_n$}
         \caption{$m=0.8,\; \nu_T=1$}\label{fig:LanczosToymodel(b)}
     \end{subfigure}
             \caption{Lanczos coefficients of the toy function (\ref{ToymodelF}). Each dot represents a numerical value of $b_n$. We also plot a curve $b_n=\frac{m\pi n}{4W(2m\pi n/\nu_T)}$, which is consistent with the numerical plots.}
        \label{fig:LanczosToymodel}
\end{figure}

In order to confirm our statement, we numerically compute the Lanczos coefficients of a toy function
\begin{align}
f(\omega)=C \exp\left[-\frac{2\vert\omega\vert}{m}\log\left(\frac{2\vert\omega\vert}{\nu_T}\right)\right],\label{ToymodelF}
\end{align}
where its asymptotic behavior is the same one of the IP model (\ref{abIP}), and $C$ is a normalization factor such that $\int_{-\infty}^{\infty}\frac{d\omega}{2\pi}\,f(\omega)=1$. Figure \ref{fig:LanczosToymodel} shows numerical plots of the Lanczos coefficients for the toy function (\ref{ToymodelF}). Each dot represents a numerical value of $b_n$, which is well fitted by the asymptotic behavior (\ref{abbnIP}).

\begin{figure}
     \begin{subfigure}[b]{0.45\textwidth}
         \centering
         \includegraphics[width=\textwidth]{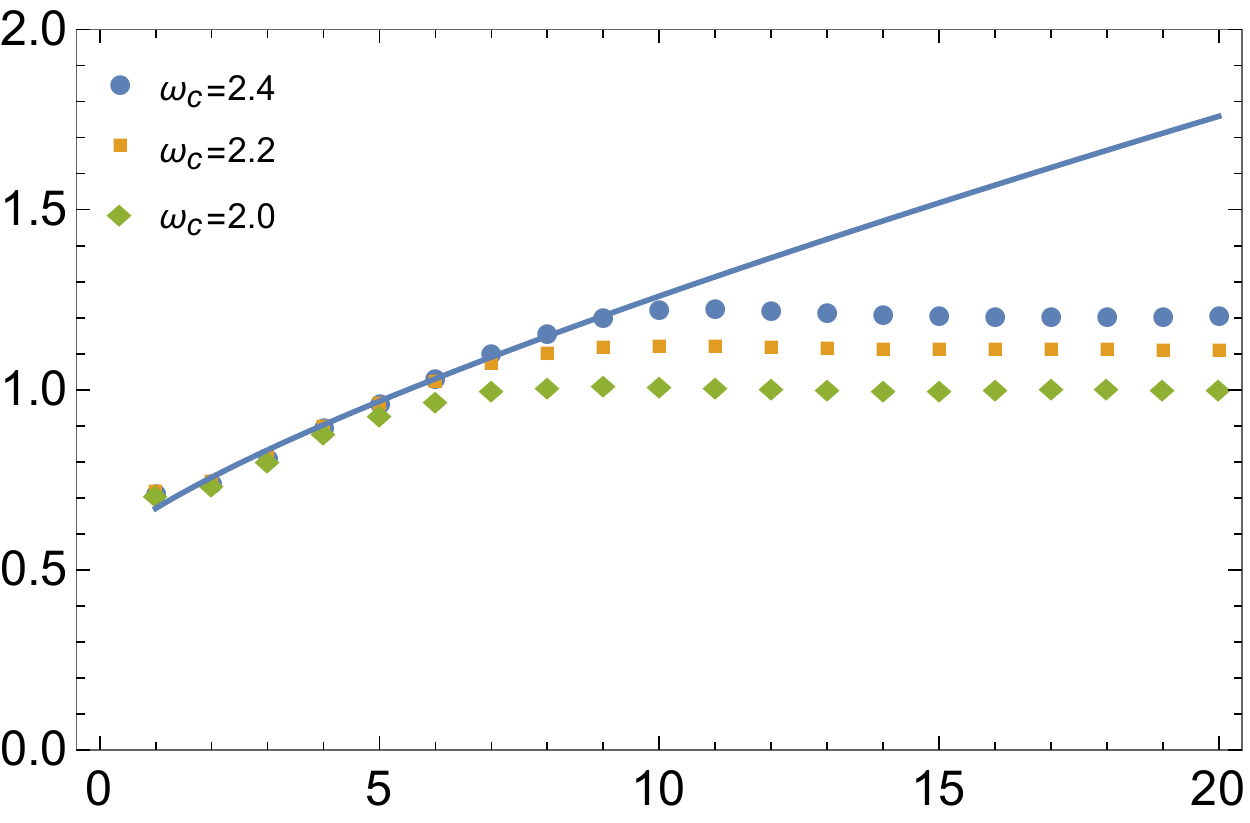}
       \put(5,0){$n$}
    \put(-195,130){$b_n$}
       \caption{$m=0.2, \;\nu_T=1$}\label{fig:LanczosIPmodel(a)}
     \end{subfigure}
      \hfill
     \begin{subfigure}[b]{0.45\textwidth}
         \centering
         \includegraphics[width=\textwidth]{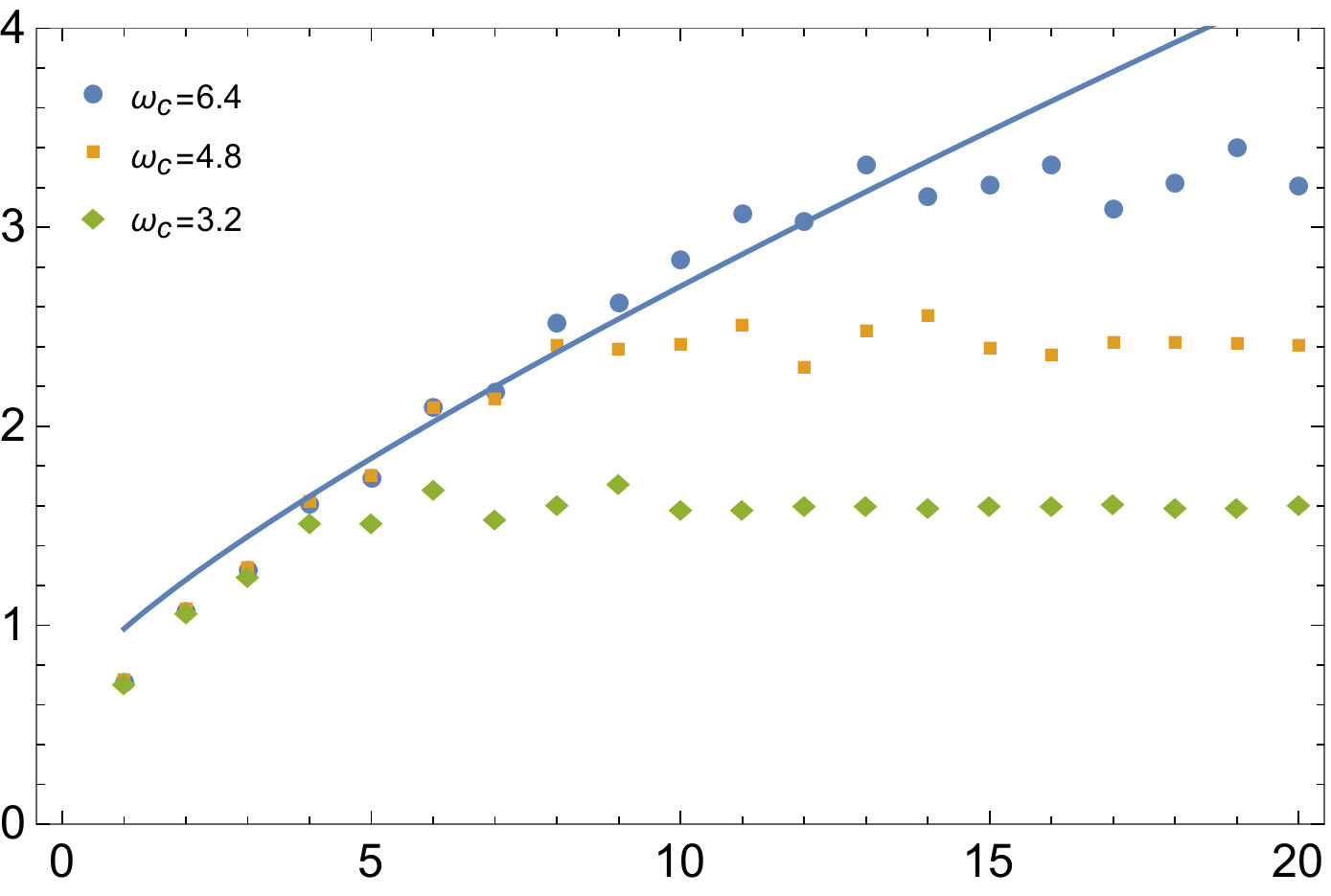}
          \put(5,0){$n$}
    \put(-195,130){$b_n$}
         \caption{$m=0.8,\; \nu_T=1$}\label{fig:LanczosIPmodel(b)}
     \end{subfigure}
             \caption{Lanczos coefficients of the IP model whose asymptotic behavior of the spectral density is (\ref{abIP}). Each dot represents a numerical value of $b_n$, where $\omega_c$ is a cutoff such that $F(\omega)=0$ in $\vert\omega\vert>\omega_c$. With the finite cutoff $\omega_c$, the Lanczos coefficients $b_n$ saturate as $b_n\sim \omega_c/2$. We also plot a curve $b_n=b_0+\frac{m\pi n}{4W(2m\pi n/\nu_T)}$, where $b_0=0.43$ for (a) and $b_0=0.51$ for (b).}
        \label{fig:LanczosIPmodel}
\end{figure}

It is challenging to perform numerical calculations with high precision for $F(\omega)$ of the IP model due to the exponential decay of the asymptotic behavior (\ref{abIP}). To avoid this difficulty, we introduce a cutoff scale $\omega_c$ and calculate $b_n$ of the IP model from $F(\omega)$ in $\vert\omega\vert\le\omega_c$, where we set $F(\omega)=0$ in $\vert\omega\vert>\omega_c$. Figure \ref{fig:LanczosIPmodel} shows numerical plots of the Lanczos coefficients of the IP model with different values of the cutoff $\omega_c$. The Lanczos coefficients $b_n$ for small $n$ do not depend on the cutoff $\omega_c$, but as $n$ becomes larger, $b_n$ begins to saturate like $b_n\sim\omega_c/2$. Thus, in the limit that the cutoff goes to infinite $\omega_c\to\infty$, the Lanczos coefficient $b_n$ keeps growing forever as $b_n \propto n/\log n$.

In Figure \ref{fig:LanczosIPmodel}, we also plot a curve $b_n=b_0+\frac{m\pi n}{4W(2m\pi n/\nu_T)}$ to confirm the asymptotic behavior (\ref{abbnIP}) for the IP model, where $b_0$ is a constant that can be ignored at large $n$. In Figure \ref{fig:LanczosIPmodel(a)} with $m=0.2$ and $\nu_T=1$, the numerical plots of $b_n$ for the IP model agree well with $b_n=b_0+\frac{m\pi n}{4W(2m\pi n/\nu_T)}$ despite the small value of $n$. In the case of Figure \ref{fig:LanczosIPmodel(b)} with $m=0.8$ and $\nu_T=1$, the numerical plots of $b_n$ are not well fitted by $b_n=b_0+\frac{m\pi n}{4W(2m\pi n/\nu_T)}$. The reason for the slight discrepancy is that (\ref{abbnIP}) is the asymptotic behavior at large $n$, and we expect the discrepancy to be smaller if we consider the fitting of $b_n$ at large $n$.

\section{Krylov complexity and entropy in the IP model}\label{sec:Kcomp}
In this section, we analyze Krylov complexity $K(t)=\sum_{n=1}^\infty n\vert\varphi_{n}(t)\vert^2$ and Krylov entropy $S(t)=-\sum_{n=0}^\infty\vert\varphi_{n}(t)\vert^2\log\vert\varphi_{n}(t)\vert^2$ of the IP model in three limits: massless limit, zero temperature limit for nonzero mass, and infinite temperature limit. 
\subsection{Krylov complexity in the massless limit}
\label{subsec:masslessKcomp}
In the massless limit, as we have seen in eq.~(\ref{bndsl}), the Lanczos coefficients do not depend on $n$ as $a_n=0$ and $b_n=\nu_T/\sqrt{2}$. In this case, the recursion relation of $\varphi_n$ eq.~(\ref{recursionwf}) is given by
\begin{align}
\frac{d\varphi_{n}(t)}{dt}=\frac{\nu_T}{\sqrt{2}}\left(-\varphi_{n+1}(t) +\varphi_{n-1}(t)\right).
\end{align}
Setting $\sqrt{2} \nu_T t := \tau$, this is the Bessel recursion relation. Therefore the solution is
\begin{align}
\varphi_{n}(t) =  J_n(\sqrt{2} \nu_T t) +  J_{n+2}(\sqrt{2} \nu_T t) = \frac{\sqrt{2} (n+1) }{\nu_T t} J_{n+1}(\sqrt{2} \nu_T t) 
\end{align} 
The reason why we add two Bessel functions is to satisfy the boundary condition \cite{Barbon:2019wsy} 
\begin{align}
\varphi_{-1}(t) = 0 \,,
\end{align}
and this boundary condition is achieved since  $J_{-1}(z) = - J_1(z)$. 
Since for $n > 0$, 
\begin{align}
J_n(t) &= \sum_{k=0}^\infty \frac{(-1)^k}{\Gamma[n + 1 + k]} \left(\frac{t}{2} \right)^{n+2 k} 
=  \frac{1}{\Gamma[n+1]}\left(\frac{t}{2} \right)^{n} + O(t^{n+2})\quad  (\mbox{at } t \to 0)
\end{align}
one can also confirm this solution satisfies the boundary condition at $t=0$,  
\begin{align}
\varphi_n(t=0) = \delta_{n,0} \, \quad \mbox{(for $n \ge -1$)}
\end{align}

\begin{figure}
              \centering
         \includegraphics[width=0.5\textwidth]{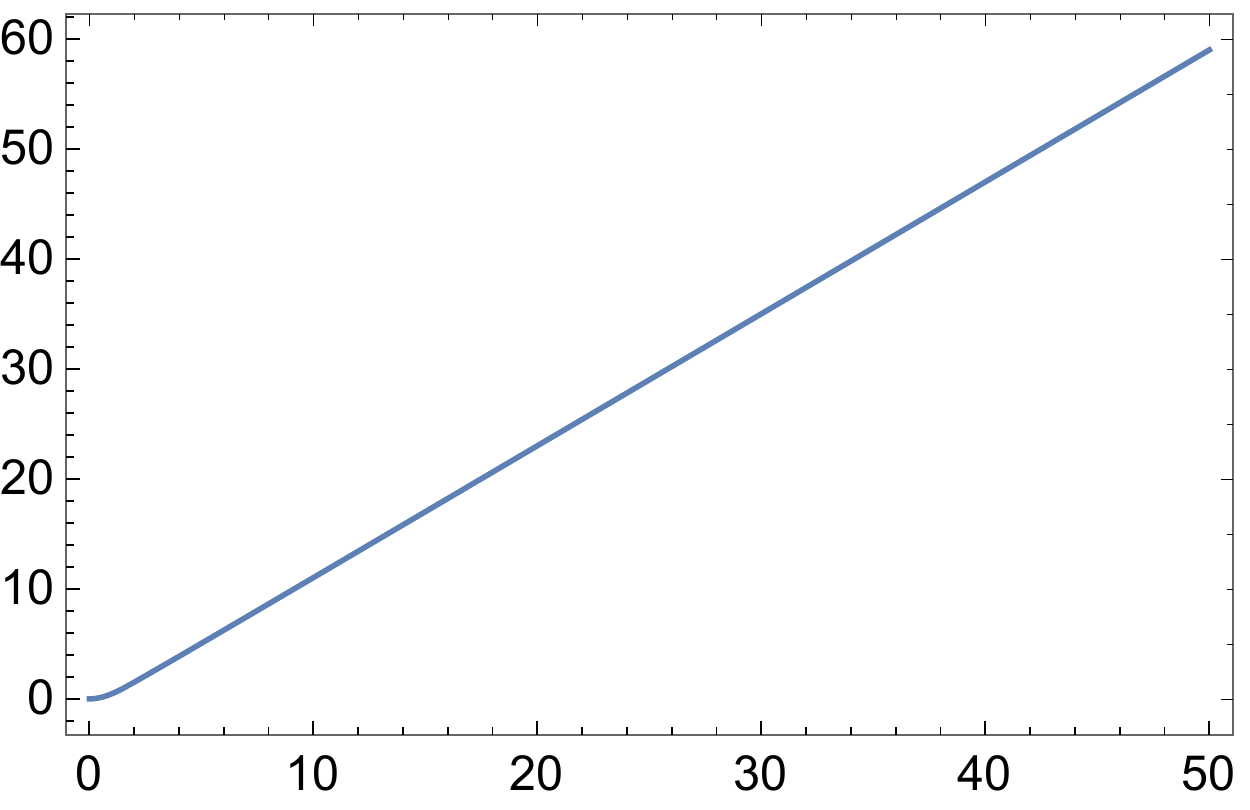}
       \put(5,10){$t$}
    \put(-220,145){$K(t)$}
       \caption{Krylov complexity in the massless limit (\ref{KrylovMassless}) with $\nu_T=1$.}\label{fig:KrylovMassless}
\end{figure}

The Krylov complexity $K(t)$ is
\begin{align}
K(t)&=\sum_{n=1}^\infty n\vert\varphi_{n}(t)\vert^2=\sum_{n=1}^\infty \frac{2n(n+1)^2}{\nu_T^2 t^2} \left(J_{n+1}(\sqrt{2}\nu_T t) \right)^2, \label{KrylovMassless}
\end{align}
which is a function of $\nu_T t$. Since $b_n$ does not depend on $n$, the late time behavior of $K(t)$ is $\delta\to0$ limit of (\ref{Ktintegrable})
\begin{align}
K(t)\propto \nu_T t.
\label{casezeroKrylovtime}
\end{align}
We plot $K(t)$ (\ref{KrylovMassless}) with $\nu_T=1$ in Figure \ref{fig:KrylovMassless}, and one can confirm the linear growth of $K(t)$ at late times.

\begin{figure}
              \centering
         \includegraphics[width=0.5\textwidth]{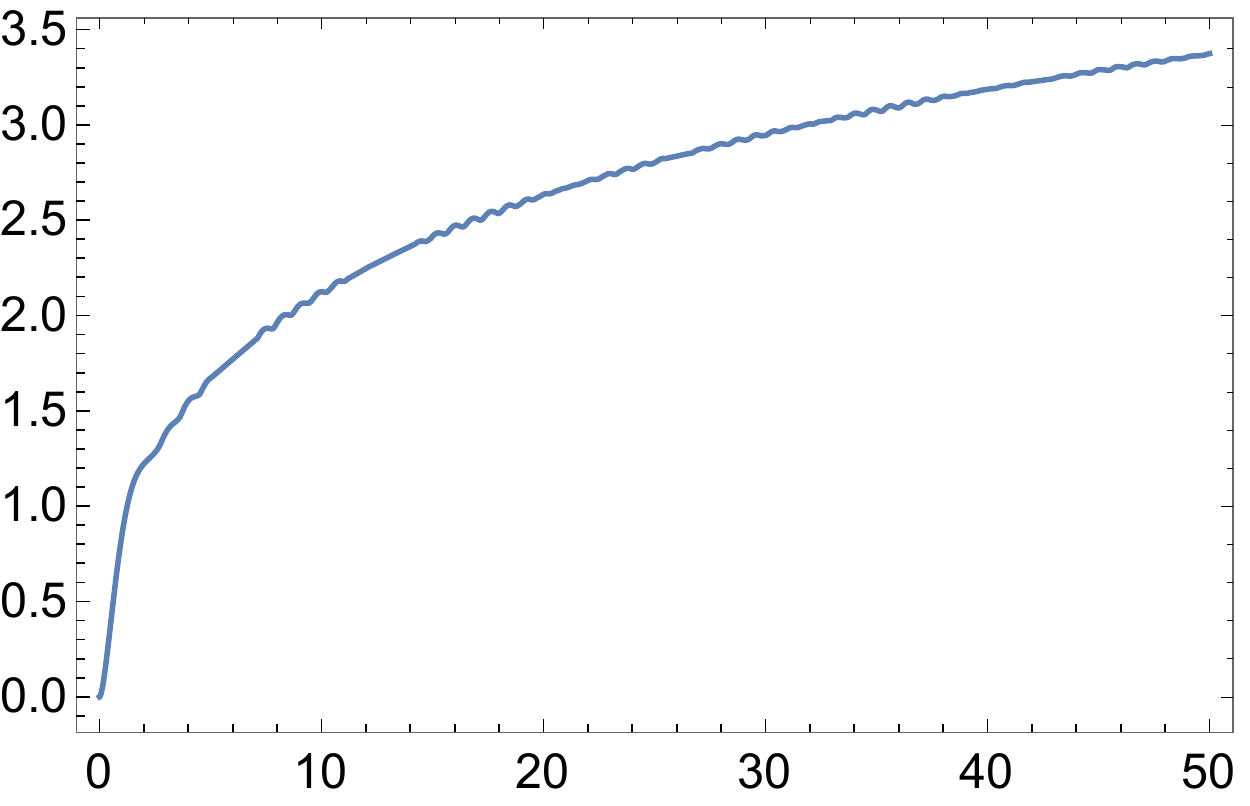}
       \put(5,10){$t$}
    \put(-220,145){$S(t)$}
       \caption{Krylov entropy in the massless limit (\ref{KrylovEntropyMassless}) with $\nu_T=1$.}\label{fig:KrylovEntropyMassless}
\end{figure}

The Krylov entropy $S(t)$ is
\begin{align}
S(t)=&-\sum_{n=0}^\infty\vert\varphi_{n}(t)\vert^2\log\vert\varphi_{n}(t)\vert^2\notag\\
=&-\sum_{n=0}^\infty\frac{2(n+1)^2}{\nu_T^2 t^2} \left(J_{n+1}(\sqrt{2}\nu_T t) \right)^2\log\left(\frac{2(n+1)^2}{\nu_T^2 t^2} \left(J_{n+1}(\sqrt{2}\nu_T t) \right)^2\right).\label{KrylovEntropyMassless}
\end{align} 
Figure \ref{fig:KrylovEntropyMassless} shows the time evolution of the Krylov entropy $S(t)$ in the massless limit with $\nu_T=1$. We note that the Krylov entropy $S(t)$ with $a_n=0, b_n=\text{const}$ grow logarithmic way at late times with respect to $t$ \cite{Barbon:2019wsy}. 

\subsection{Krylov complexity at zero temperature for nonzero mass}

To determine the Krylov complexity, let us first determine $\varphi_{n}(t)$ which obeys eq.~\eqref{recursionwf}, \eqref{recursionwf2}, 
\begin{align}
\label{zeroTphi1}
\frac{d\varphi_{n}(t)}{dt}=i m n  \varphi_{n}(t) +\frac{\nu}{2} \left( \varphi_{n-1}(t) - \varphi_{n+1}(t) \right)  \,,\\
\varphi_{-1}(t):=0, \;\;\; \varphi_{0}(t)=C(-t;\beta = \infty).\label{bczeroTphi}
\end{align}
Here we use the $a_n$ and $b_n$ given by eq.~\eqref{LczeroT}. 

Let us set its Fourier transformation as 
\begin{align}
\varphi_{n}(t) : = \int \frac{d \omega}{2 \pi} \varphi_{n}(\omega) e^{-i \omega t} 
\end{align}
Then eq.~\eqref{zeroTphi1} reduces to 
\begin{align}\label{eqzeroTphi2}
\left( \omega  + m n \right) \varphi_{n}(\omega)=\frac{i \nu}{2} \left( \varphi_{n-1}(\omega) - \varphi_{n+1}(\omega) \right) \,,
\end{align}
and this is essentially the Bessel recursion relation. 
Thus, we find a solution that is  
proportional to the Bessel function as 
\begin{align}
\varphi_{n}^{(1)}(\omega) = \frac{2i}{i^{n} \nu} \frac{J_{n + \omega/m}(-\frac{\nu}{m})}{  J_{-1 + \omega/m}(-\frac{\nu}{m}) }  = \frac{2 i^{n-1}}{ \nu} \frac{J_{n + \omega/m}(\frac{\nu}{m})}{  J_{-1 + \omega/m}(\frac{\nu}{m}) } \,,
\end{align} 
which is consistent with  eq.~(19) in \cite{Iizuka:2008hg}\footnote{The sign difference in front of $\omega$ is due to the fact that $\varphi_0(t) = G(-t)$, instead of $G(t)$. $i^n$ in the denominator is due to the difference of the phase factor in the wave function definition. There are minor sign errors in eqs.~(15) - (19) in \cite{Iizuka:2008hg} since $X_{ij} = \frac{i}{\sqrt{2m}} ( A_{ij} -  A^\dagger_{ij})$.  This can be corrected by $\nu \to -\nu$ there.}. 
By taking the complex conjugate of (\ref{eqzeroTphi2}), we construct another solution $\varphi^{(2)}_{n}(\omega)$, 
\begin{align}
\varphi^{(2)}_{n}(\omega):=(-1)^n\varphi^{(1)*}_{n}(\omega)=-\frac{2i}{i^{n} \nu} \left(\frac{J_{n + \omega/m}(-\frac{\nu}{m})}{  J_{-1 + \omega/m}(-\frac{\nu}{m}) }\right)^*  = -\frac{2 i^{n-1}}{ \nu} \left(\frac{J_{n + \omega/m}(\frac{\nu}{m})}{  J_{-1 + \omega/m}(\frac{\nu}{m}) }\right)^*,
\end{align}
where one can check that
\begin{align}
\varphi_{-1}^{(1)}(\omega)+\varphi^{(2)}_{-1}(\omega)=0, \;\;\; \varphi^{(1)}_{0}(\omega)=\tilde{G}(-\omega), \;\;\; \varphi^{(2)}_{0}(\omega)=\tilde{G}^*(-\omega).
\end{align}
Thus, the solution of (\ref{zeroTphi1}) with the boundary condition (\ref{bczeroTphi}) is
\begin{align}
\varphi_{n}(t) &= \int \frac{d\omega}{2\pi} (\varphi_{n}^{(1)}(\omega)+\varphi^{(2)}_{n}(\omega)) e^{-i \omega t}\notag\\
&=\frac{2i}{i^{n} \nu} \int \frac{d\omega}{2\pi} \left(\frac{J_{n + \omega/m}(-\frac{\nu}{m})}{  J_{-1 + \omega/m}(-\frac{\nu}{m}) }- \left(\frac{J_{n + \omega/m}(-\frac{\nu}{m})}{  J_{-1 + \omega/m}(-\frac{\nu}{m}) }\right)^* \right)e^{-i \omega t}   \,.
\end{align}

\begin{figure}
              \centering
         \includegraphics[width=0.5\textwidth]{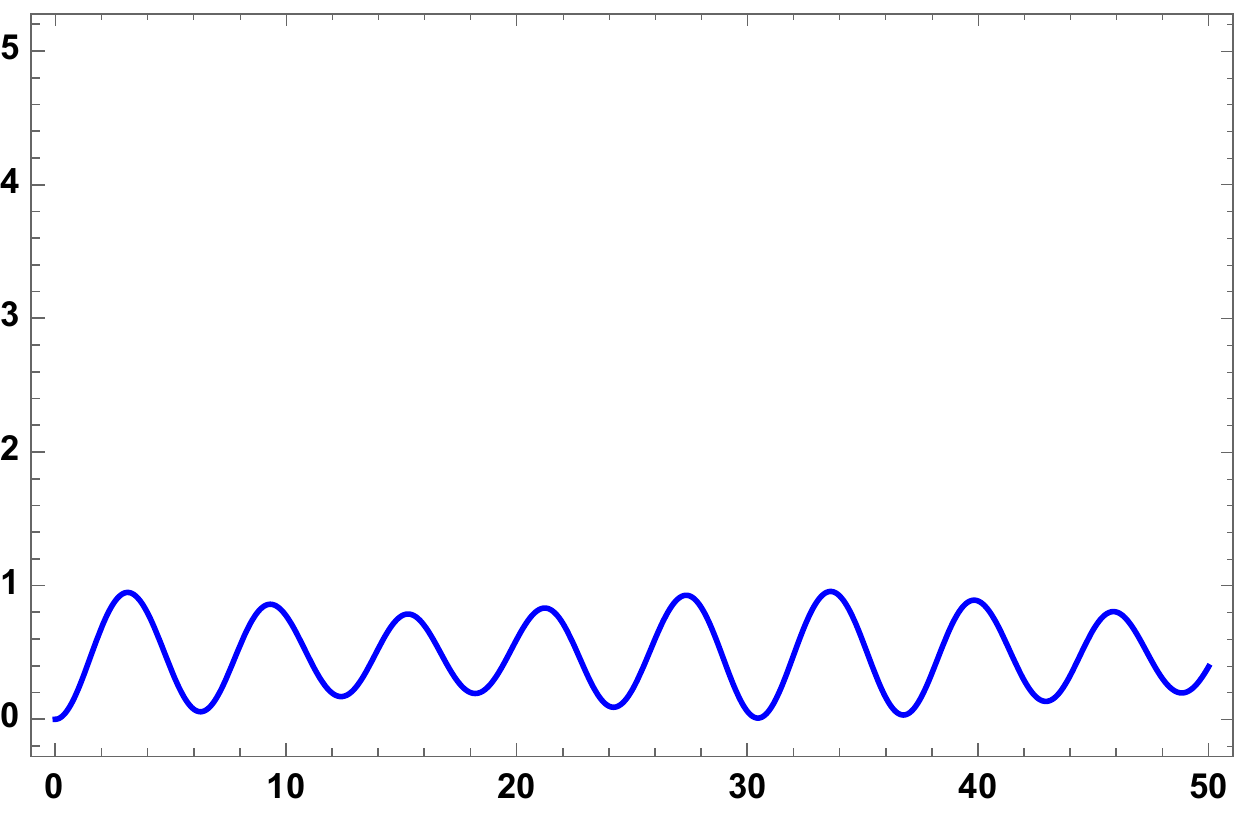}
       \put(5,5){$t$}
    \put(-220,145){$K(t)$}
       \caption{Krylov complexity in the zero temperature limit with $m=0.8, \nu=1$.}\label{fig:KrylovZeroT}
\end{figure}

Although we obtain analytic results for $\varphi_n(t)$, in order to see the time evolution of Krylov complexity $K(t)$, these expressions are not as useful as one might expect. Instead of using these expressions, we perform numerical calculations of $K(t)$ by using a method in \cite{Bhattacharya:2023zqt}. Specifically, we numerically solve (\ref{zeroTphi1}) with the following initial condition
\begin{align}
\varphi_{0}(0)=1, \;\;\; \varphi_{n}(0)=0 \;\;\; (n>0),
\end{align}
and then compute $K(t)$. Figure \ref{fig:KrylovZeroT} shows the time evolution of $K(t)$ in the zero temperature limit with $m=0.8, \nu=1$. The Krylov complexity $K(t)$ oscillates due to nonzero $a_n$ and does not grow at late times.

\begin{figure}
              \centering
         \includegraphics[width=0.5\textwidth]{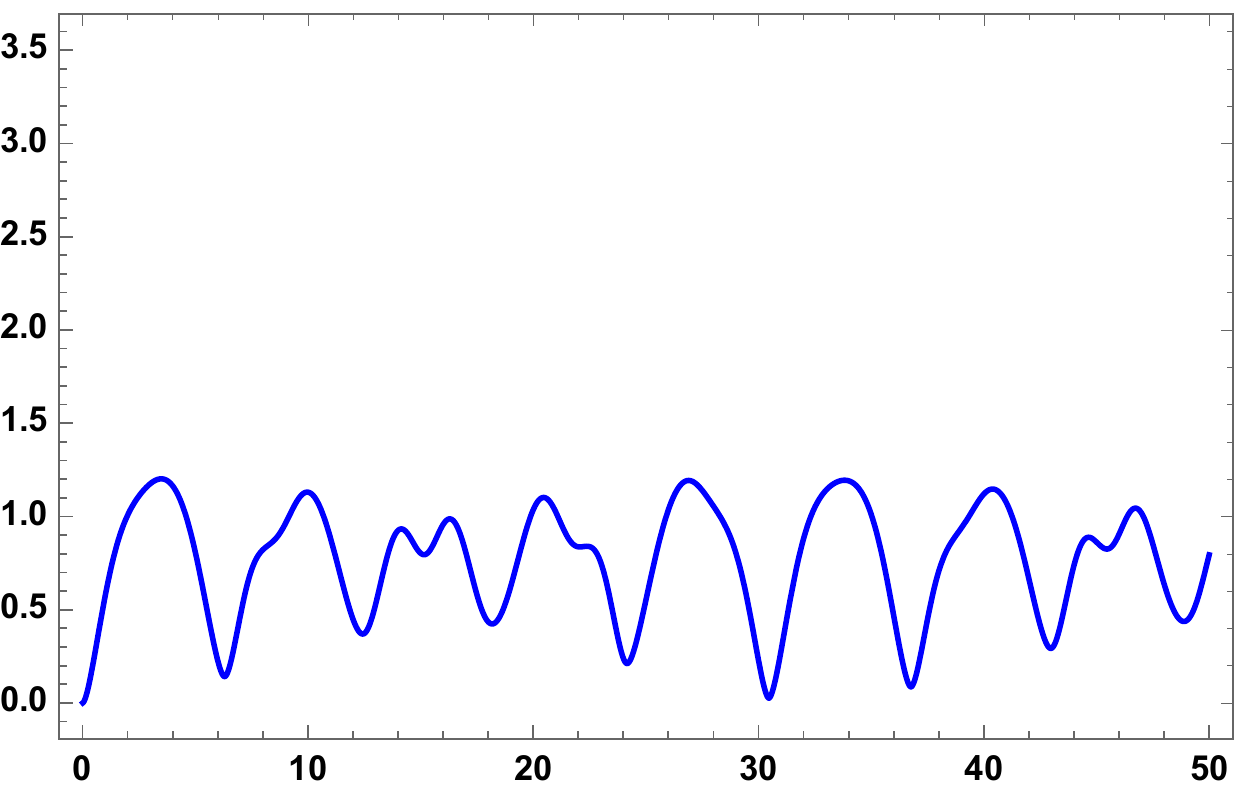}
       \put(5,10){$t$}
    \put(-220,145){$S(t)$}
       \caption{Krylov entropy in the zero temperature limit with $m=0.8, \nu=1$.}\label{fig:KrylovEntropyZeroT}
\end{figure}

We also plot the Krylov entropy $S(t)=-\sum_{n=0}^\infty\vert\varphi_{n}(t)\vert^2\log\vert\varphi_{n}(t)\vert^2$ in the zero temperature limit with $m=0.8, \nu=1$ in Figure \ref{fig:KrylovEntropyZeroT}. Since $K(t)$ in the zero temperature limit is bounded as seen in Figure \ref{fig:KrylovZeroT}, the growth of the operator does not propagate well to $\varphi_n(t)$ where $n$ is large. Therefore, $S(t)$ is also bounded because the entropy becomes large when $\varphi_n(t)$ follows a uniform distribution.

\subsection{Krylov complexity in the infinite temperature limit for nonzero mass}
In the infinite temperature limit, the asymptotic behavior of $b_n$ at large $n$ is given by 
\begin{align}
b_n\sim \frac{m\pi n}{4W(2m\pi n/\nu_T)}\sim \frac{m\pi n}{4\log n}\label{AsymptoticbnInfiniteT}
\end{align} 
with $a_n=0$ as eq.~(\ref{abbnIP}). 
Then $\varphi_{n}(t)$ at large $n$ obeys eq.~\eqref{recursionwf}, \eqref{recursionwf2}, 
\begin{align}
\label{zeroTphi2}
\frac{d\varphi_{n}(t)}{dt}\sim  -  \frac{m\pi (n+1)}{4\log (n+1)} \varphi_{n+1}(t)+  \frac{m\pi n}{4\log n}  \varphi_{n-1}(t)  \,,\\
\varphi_{-1}(t):=0, \;\;\; \varphi_{0}(t)=C(-t;\beta).
\end{align}
Thus, from (\ref{Kt1dchaos}), the late time behavior of $K(t)$ is 
\begin{align}
K(t)\sim e^{\sqrt{m\pi t}} = \sum_n \frac{\left( m \pi t\right)^{\frac{n}{2}}}{n!}\,.\label{Ktroott}
\end{align}
This growth behavior is slower than the exponential growth $e^{2\alpha t}$, but it is faster than any power low growth behavior (\ref{Ktintegrable}) for integral systems. With the growth of $b_n$ as $b_n\sim \frac{m\pi n}{4\log n}$, our analysis strongly suggests that the IP model in the infinite temperature limit is chaotic.

\begin{figure}
              \centering
         \includegraphics[width=0.5\textwidth]{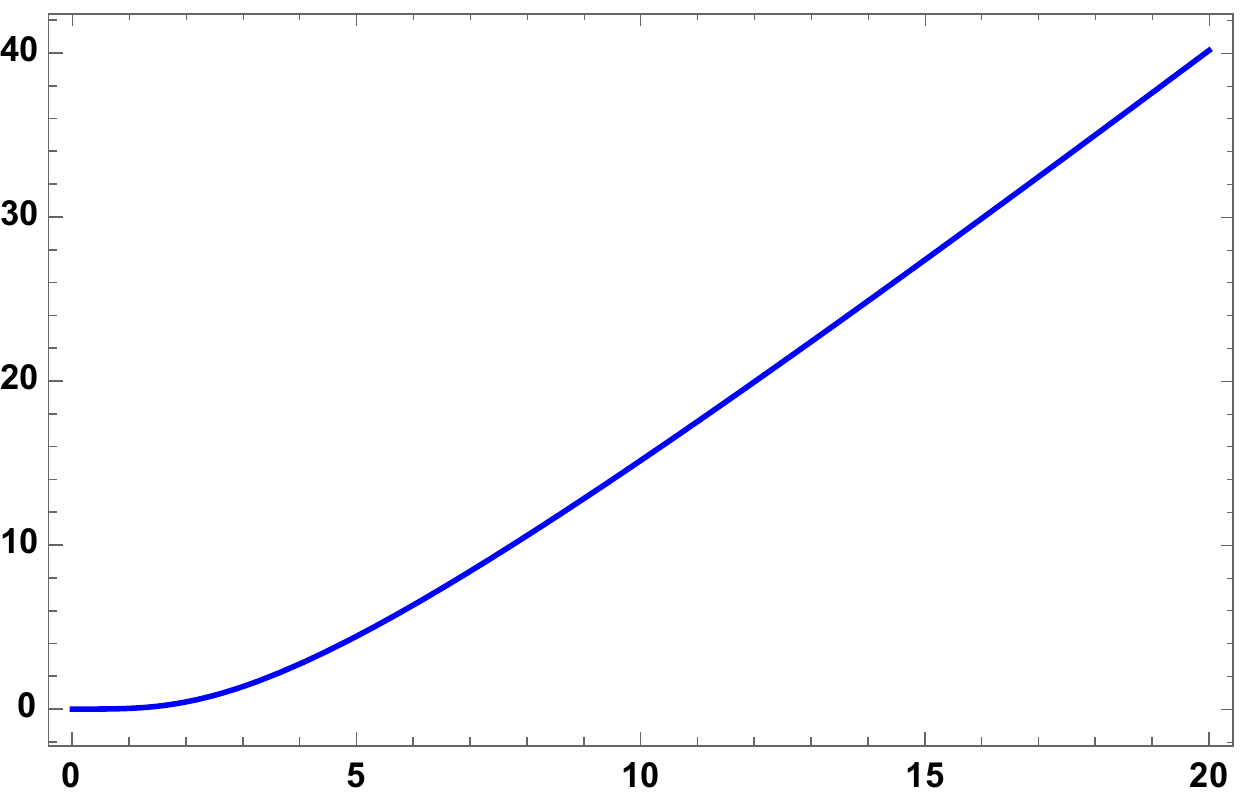}
       \put(5,5){$t$}
    \put(-230,145){$\left(\log[1+K(t)]\right)^2$}
       \caption{Krylov complexity of the toy model (\ref{ToyInfiniteT}) for the infinite temperature limit with $m=0.8, \nu_T=1$. The linear growth of $\left(\log[1+K(t)]\right)^2$at late times means that $K(T)$ grows exponentially with respect to $\sqrt{t}$ at late times. }\label{fig:KrylovInfiniteT}
\end{figure}

To confirm the growth behavior of $K(t)$, let us do a similar numerical computation as in the previous subsection with a simple toy model of $b_n$. We numerically solve (\ref{recursionwf}) and compute $K(t)$ with the following Lanczos coefficients
\begin{align}
a_n=0, \;\;\; b_n=\frac{m\pi n}{4W(2m\pi n/\nu_T)} \,,\label{ToyInfiniteT}
\end{align}
where the asymptotic behavior of $b_n$ is the same as (\ref{AsymptoticbnInfiniteT}). Figure \ref{fig:KrylovInfiniteT} shows the growth behavior of $\left(\log[1+K(t)]\right)^2$ of the toy model (\ref{ToyInfiniteT}) for the infinite temperature limit with $m=0.8, \nu_T=1$. One can see the linear growth at late times in the figure, which shows the exponential growth of $K(t)$ with respect to $\sqrt{t}$. 

\begin{figure}
              \centering
         \includegraphics[width=0.5\textwidth]{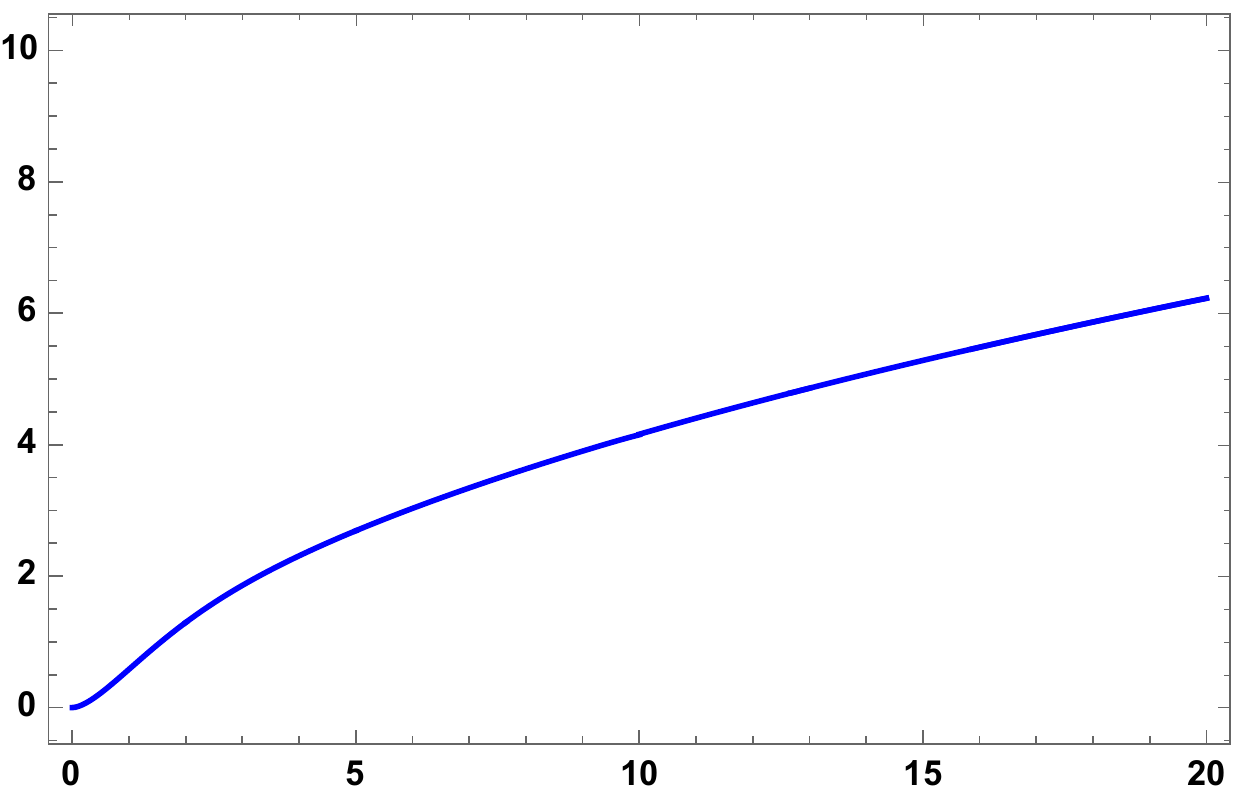}
       \put(5,10){$t$}
    \put(-220,145){$S(t)$}
       \caption{Krylov entropy of the toy model (\ref{ToyInfiniteT}) for the infinite temperature limit with $m=0.8, \nu_T=1$.}\label{fig:KrylovEntropyInfiniteT}
\end{figure}

Figure \ref{fig:KrylovEntropyInfiniteT} shows the Krylov entropy $S(t)$ of the toy model (\ref{ToyInfiniteT}) with $m=0.8, \nu_T=1$. The growth behavior of $S(t)$ at late times associated with (\ref{AsymptoticbnInfiniteT}) and (\ref{Ktroott}) is $S(t)\propto \sqrt{m \pi t}$ \cite{Barbon:2019wsy, Fan:2022xaa}.

\section{Conclusion and discussion}\label{sec:conc}

In this paper, we study the Krylov subspace and the Lanczos coefficients in the IP matrix model, which is a toy model of the gauge theory dual of an AdS black hole. This model consists of a mass $m$ matrix $X_{ij}$ which is adjoint, plus a mass $M$ probe complex vector $\phi_i$ which is fundamental.  
Our focus is in the parameter range where the fundamental is much heavier than the adjoint and temperature $M \gg m$ and $M \gg T$. In this limit, the Krylov subspace for a creation operator $a^\dagger_i$ for the fundamental is restricted to the sector where we have only one fundamental, $a^\dagger_i a_i =1 $. 
In the large $N$ limit, one can solve the Schwinger-Dyson equation for
the correlator for the fundamental field, and obtain the spectral density for the fundamental field.  
The resultant spectral density changes are drastically dependent on the parameters, and this results in a rich behavior for the Lanczos coefficients and Krylov complexity and entropy. Our results can be summarized as follows;
$\vspace{-3mm}$

\begin{enumerate}
\item In massless limit for the adjoint, $m \to 0$, the spectral density for the fundamental follows the Wigner semi-circle law for any temperature $T$. As a result, the Lanczos coefficients $a_n=0$ and $b_n$ is a nonzero constant given by eq.~\eqref{bndsl}. Thus, the Krylov complexity grows linearly in time, eq.~\eqref{casezeroKrylovtime}. In this case, the Krylov entropy $S(t)$ grows logarithmic way in time $t$ at late times.  
$\vspace{-3mm}$
\item In the zero temperature limit $T=0$ with nonzero mass $m \neq 0$ for the adjoint, the spectral density for the fundamental shows a collection of the delta-function, due to poles in the correlator, thus in this case it has a discrete spectrum.  In this case, we can solve the system canonically, and 
the Krylov basis for $\hat{\mathcal{O}}=\hat{a}^\dagger_j$ can be obtained as $\vert \hat{\mathcal{O}}_n)=\vert j, n\rangle$, given by eq.~\eqref{IPbasiszerotemperature}. 
The Lanczos coefficients are $a_n = m n$ and $b_n$ is a nonzero constant given by eq.~\eqref{LczeroT}. 
In this case, both the Krylov complexity and entropy do not grow much in time, they just oscillate in order one. 
$\vspace{-3mm}$
\item Finally as we increase the temperature $T$ with nonzero mass for adjoint $m \neq 0$, the spectrum changes from discrete to continuous. At sufficiently high temperature limit $T \to \infty$, the spectral density becomes continuous and gapless. This continuous spectrum results in an exponential decay of the correlator and this decay of the correlator is a key signature of thermalization and information loss \cite{Maldacena:2001kr}. In the high temperature limit, the spectral function becomes an even function in $\omega$ and the Lanczos coefficients are $a_n=0$, and $b_n$ is linear in $n$ with log correction, given by eq.~\eqref{AsymptoticbnInfiniteT}. This comes from the fact that the spectral density decays as $e^{- \order(|\omega| \log |\omega|)}$. Thus, the Krylov complexity grows exponentially in $\sqrt{t}$, as $K(t) \propto e^{\order({\sqrt{t}})}$, given by  eq.~\eqref{Ktroott}. In this case, the Krylov entropy $S(t)$ also grows linearly in $\sqrt{t}$ at late times.  
$\vspace{-2mm}$
\end{enumerate}

Before we discuss the implication of Krylov complexity in high temperature limit, 
let us discuss the Krylov basis. In the zero temperature limit, we can explicitly construct the Krylov basis, given by eq.~\eqref{IPbasiszerotemperature}.  The structure is clear. It consists of one fundamental and many adjoints. Since the number of adjoints can be any nonzero integers, it is infinite-dimensional due to the bosonic nature of the model.  
Intuitively at sufficiently high temperature, there are so many operators we have to think of in mixed states, which behaves like a thermal bath, 
and thus, the operator grows exponentially. Since the Lanczos coefficients $b_n \sim n/\log n$ is the maximal growth in the absence of singularity in the correlator \cite{Parker:2018yvk}, our results suggest that the IP model at sufficiently high temperature is chaotic. There are two crucial ingredients: large $N$ and sufficiently high temperature. Only in the large $N$ limit, the spectral density can be continuous. We need a high temperature to make many microstates interact with each other so that the operator growth shows rapid scrambling.  

Note that there are examples of the maximal growth of $b_n$ {\it without chaos}. However, we believe that the IP model does not belong to such non-chaotic examples for the following reasons:

The first example of maximal growth of $b_n$ without chaos occurs due to the instability; unstable saddle point dominance as seen in \cite{Xu:2019lhc,Bhattacharjee:2022vlt} for example. The IP model, especially the Krylov subspace which we consider, does not suffer any such instability. For example in the IP model Hamiltonian \eqref{ham}, the highest term is cubic. This leads to the lack of a sensible ground state in this model. However one can cure this by simply adding a quartic stabilizing term as \cite{Iizuka:2008hg} 
\begin{align}
\Delta H_{\rm stab} = c  \left( N_{\phi} +N_{\bar{\phi}} \right)  ( N_{\phi} + N_{\bar{\phi}} - 1) \,, \quad N_{\phi} = \sum_i  a_i^\dagger a_i \,, \quad N_{\bar{\phi}} = \sum_i  \bar{a}_i^\dagger \bar{a}_i \,.
\end{align}  
Since the Hamiltonian \eqref{ham} commute with the number operator $N_{{\phi}}$, $N_{\bar{\phi}}$, in each sector labeled by $(N_{{\phi}}, N_{\bar{\phi}})$ there is a ground state. 
Especially for sufficiently large $c > 0$ and $M$, there is a sensible ground state 
at $N_{{\phi}}  = N_{\bar{\phi}} =0$.  
Thus, this quartic term stabilizes the ground state and furthermore, it vanishes in our Krylov subspace \eqref{IPbasiszerotemperature} where $N_\phi = 1$. Thus this stabilizing term does not change our results and we have a sensible ground state in the Krylov subspace.  

Furthermore, as we have shown, at zero temperature, one can construct the Krylov basis by using the free ground state, and $b_n$  is constant there.  The maximal growth of $b_n$ appears only at nonzero temperature, especially at the high temperature limit. An unstable ground state can be cured usually by introducing nonzero temperature since thermal mass can make the unstable ground state stable. Due to these reasonings, we believe that the maximal growth of $b_n$ appearing only in the high temperature limit is not associated with instability.

The second example of maximal growth without chaos is seen in free QFTs \cite{Dymarsky:2021bjq, Avdoshkin:2022xuw, Camargo:2022rnt}. In QFTs, due to the divergence of operators at the same point, a two-point correlator can have a pole and that can cause the $b_n$ to grow as linear to $n$. Since this occurs even in free theory, clearly this is not associated with chaos.  On the other hand, the IP model is a zero spatial dimensional quantum mechanical model, and therefore there is no such divergence associated with operators at the same time. Furthermore,     
the growth of $b_n$ is not universal. The maximal growth of $b_n$ appears only in the high temperature limit, it is the limit where the spectrum becomes continuous and two-point function decay exponentially as a function of time \cite{Iizuka:2008hg}. Thus the natural interpretation of the maximal growth of $b_n$ appearing at sufficiently high temperature is the indication of chaos.

Note that our results are consistent with the analysis done in \cite{Michel:2016kwn}. In \cite{Michel:2016kwn}, the OTOCs are evaluated for the case where adjoints become massless in the IP model, and also for the generic parameter case in the IOP model. In the massless limit for the adjoint in the IP model, $m=0$, the spectral density is Wigner semi-circle type, and the fundamental correlator decay only by the power as $\sim t^{-3/2}$. In this massless limit, it is shown in \cite{Michel:2016kwn} that the OTOCs do not give rise to exponential growth. As we have shown in \S \ref{subsec:masslessKcomp}, in the massless limit $m=  0$, the Krylov complexity grows only linearly in time and does not show exponential growth either.  As we have seen in this paper, the chaotic signal can appear only for $m\neq0$ and at sufficiently high temperature limit in the IP model. There is no signal of chaos in the IP model in the massless limit $m=0$.  
The IOP model \cite{Iizuka:2008eb} is significantly different from the IP model since the interaction preserves the adjoint number. See Appendix \ref{sec:AppIP}. The extra symmetry in the interaction restricts the Krylov subspace to a fixed adjoint number. As a result, in the IOP model, the dimension of the Krylov subspace is order one and the Krylov subspace cannot grow in time.

In the IP model, we have not imposed the singlet constraint. On the other hand, in a generic gauge/gravity setting, we have the singlet constraint. As discussed in detail in \cite{Aharony:2003sx}, in the large $N$ gauge theory where we have confinement/deconfinement phase transition at $T= T_c$, the eigenvalues of the Wilson-Polyakov line $U$ are uniformly distributed on the unit circle at $T < T_c$, but at $T \gg T_c$, they are concentrated around the identity. Since the thermal propagator for the adjoint contains an infinite sum over the thermal winding modes, these winding modes pick up a phase from the Wilson-Polyakov line $U$. Thus the thermal propagator for the adjoint behaves differently depending on the phase. At $T < T_c$,  the uniform distribution of the phase in $U$ cancel the winding contributions, and makes the thermal propagator effectively as zero temperature one in eq.~\eqref{kzero0}. On the other hand, at $T \gg T_c$, it reduces to the thermal one we used in eq.~\eqref{ktherm} \cite{Iizuka:2008hg}. Thus the exponential growth of the Krylov complexity at sufficiently high temperature is the feature appearing only at the deconfinement phase in the gauge theory. On the other hand, the feature that the Krylov complexity only oscillates and does not grow in time 
at zero temperature corresponds to the feature at the confinement phase. Since it is important to take a large $N$ limit in our analysis, 
we speculate that in $SU(N)$ gauge theories, only in the large $N$ limit and in the deconfinement phase, the Krylov complexity can show exponential growth at sufficiently high temperature.  

In interesting papers \cite{Avdoshkin:2022xuw, Kundu:2023hbk}, it was also shown that the Krylov complexity changes drastically by the temperature as our analysis: Their setting is 
two-dimensional CFT with large central charge, and 
in low temperature it oscillates and does not grow, but in high temperature it grows exponentially. These are essentially the same as our results. The difference between their set-ups and ours is that the IP model is a large $N$ matrix quantum mechanics. On the other hand, the analysis in \cite{Avdoshkin:2022xuw, Kundu:2023hbk} is 2D CFT and its dual AdS$_3$/BTZ.  As far as we are aware, our results of exponential growth in the Krylov complexity and Lanczos coefficients $b_n \sim n/\log n$ is the first example found in the large $N$ matrix quantum mechanics.


Finally, we discuss the future direction of our work. 
Our analysis shows that for the case of nonzero adjoint mass $m\neq 0$, the Krylov complexity does not grow at zero temperature. On the other hand, in the infinite temperature limit, the Krylov complexity grows exponentially in time as $\sim \exp({\order(\sqrt{t})})$.  Note that the spectral density keeps changing from discrete to continuous as seen in Figure \ref{fig:F(w)}. However, depending on the temperature, it differs a lot. In low temperature, it is gapped. As we increase the temperature, gaps begin to close, and finally, it becomes completely gapless. It is natural to expect that the Krylov complexity grows faster and faster as we increase the temperature, however how the Krylov complexity behaves as a function of time dependent on the temperature is an open issue, which must be dependent on the shape of the spectral density. We will report these in a separate paper \cite{WIP}. 

It is interesting that the Krylov complexity changes drastically as we increase the temperature. As we mention, since the zero temperature limit of the IP model corresponds to the confinement phase in the large $N$ gauge theory and nonzero temperature to deconfinement phase, the Krylov complexity might play the role of the order parameter to distinguish the confinement/deconfinement phase in the large $N$ limit. The deep connection between large $N$ deconfinement, continuous spectra and exponential decay of the correlator in time, and exponential growth of the Krylov complexity is quite interesting and is definitely worth further investigation.

It is also interesting to generalize the model. In the IOP model \cite{Iizuka:2008eb}, the interaction term is $\phi^\dagger A^\dagger A \phi$, therefore the adjoint number is preserved and this symmetry prohibits the Krylov complexity growth in time. However, once we include interaction terms such as $\phi^\dagger A^\dagger A^\dagger \phi+\phi^\dagger  A A \phi$, then there is no more adjoint number conservation, just like the IP model. We expect this model shows very similar properties of chaos as the IP model, though it is less tractable.

\acknowledgments
We are happy to thank Pratik Nandy for his helpful conversations at several stages and comments on the draft. We also thank him for his Mathematica code sharing with us.  
The work of NI was supported in part by JSPS KAKENHI Grant Number 18K03619 and also by MEXT KAKENHI Grant-in-Aid for Transformative Research Areas A ``Extreme Universe'' No.~21H05184. M.N.~was supported by the Basic Science Research Program through the National Research Foundation of Korea (NRF) funded by the Ministry of Education (NRF-2020R1I1A1A01072726). M.N.~carried out part of this work while visiting Osaka University and would like to thank Osaka University for its hospitality and support.

\appendix

\section{Overview of IP matrix-vector model}\label{sec:AppIP}

\subsection{A toy model of a vector $\phi_i$}
Let us consider a complex vector field, $\phi_i$, with $i = (1, \cdots, N)$, where the action is given by 
\be
\label{GUEmassmodel}
S = \int dt {\cal{L}}  = \frac{1}{g^2}\int dt \left(  ( \partial_t \phi_i^\dagger )  \left( \partial_t \phi_i  \right)  -   \phi_i^\dagger \left(  M^2  \delta_{ij} +  M M_{ij} \right) \phi_j \right) \,. 
\ee
Here we assume $M_{ij}$ to be a $N \times N$ random matrix, obeying Gaussian Unitary Ensemble (GUE) and $M$ is some mass scale and $g$ is a coupling constant. 
Square of the effective mass matrix $(M_{\rm eff})_{ij}$ for the vector $\phi_i$ is given by  
\be
\label{mi}
(M^2_{\rm eff})_{ij}  :=   M^2  \delta_{ij} +  M M_{ij}  \,.
\ee
By diagonalizing $M_{ij}$ as   $\lambda_i \delta_{ij}$, 
the model given by eq.~\eqref{GUEmassmodel} is made up by simply $N$ free harmonic oscillators. 
By the assumption that $M_{ij}$ obeys GUE, there is an eigenvalue repulsion for each $\lambda_i$ due to the Vandermonde determinant after the diagonalization.  
In fact, due to this eigenvalue $\lambda_i$ repulsion, the spectrum form factor (SFF)  of this model shows similar behavior to the RMT result seen in \cite{Cotler:2016fpe}.

After the diagonalization of $M_{ij}$, 
in the large $M$ limit, the effective mass matrix $(M_{\rm eff})_{ij}$ reduces to  
\be
(M_{\rm eff})_{ij}\to  \delta_{ij} \, \sqrt{M^2 +  M \lambda_i} = \delta_{ij} \left( M  + \frac{1}{2}  \lambda_i + \order\left({\frac{1}{M}}\right) \right)
\ee
and this shows that $M + \frac{1}{2}  \lambda_i $  
is the mass eigenvalue for $i$-th vector component $\phi_i$. 
Setting $M + \frac{1}{2}  \lambda_i  := \tilde{\lambda}_i$,  
the partition function of these $L$ free harmonic oscillators is\footnote{Here we neglect the ground state energy for each harmonic.}   
\begin{align}
&\hspace{-2mm} Z(\beta) = \Pi_{i=1}^{N}  \left( \sum_{n_i=0}^{\infty} \exp \left( - \beta n_i \tilde{\lambda}_i \right)  \right) 
=  \Pi_{i=1}^{N}  \left(\frac{1}{1 - \exp \left( - \beta \tilde{\lambda}_i \right) } \right) \,.
\end{align}
We consider the limit where the mass $M$ is much larger than the temperature, $\beta M \to \infty$. 
Thus  the mass eigenvalue $\tilde{\lambda}$  is large compared to the temperature, 
and then  
\begin{align}
&\hspace{-3mm}  
Z(\beta) \, {\approx}\, \, 1  + \sum_{i=1}^N  \exp \left( - \beta \tilde{\lambda}_i \right)
\approx 1 + N  e^{-\beta M} \int d \lambda\, \rho_N(\lambda) \, \exp \left( -  \frac{\beta}{2}  \lambda \right)  \,.
\end{align}
where we take large $N$ limit and $\rho_N(\lambda)$ is the density of eigenvalues in GUE, which is unit normalized, {\it i.e.,} $\int d \lambda  \rho_N(\lambda) = 1$. 

In the large $N$ limit, $\rho_N(\lambda)$ is given by famous Wigner semi-circle
\begin{align}
\lim_{N\to \infty}\rho_N(\lambda) = \frac{1}{2 \pi} \sqrt{4 - \lambda^2} \,.
\label{Wignersemicirclerho}
\end{align} 
Due to the eigenvalue repulsion of $\lambda$, SFF defined as 
\begin{align}
\frac{Z(\beta + i t) Z(\beta - i t)}{Z(\beta)^2}
\end{align}
shows slope, ramp, and plateau as a function of time as \cite{Cotler:2016fpe}. This indicates that the action \eqref{GUEmassmodel}  shows the nature of a chaotic system. Of course, this is simply because we have chosen the mass matrix $M_{ij}$ to obey the GUE by hand.   

In fact, the IP model \cite{Iizuka:2008hg} in the massless limit shows features similar to this model. The Schwinger-Dyson equation in $m=0$ gives the spectral density for the vector field $\phi_i$ as  eq.~\eqref{wignermassless} for zero temperature $(T=0)$ and eq.~\eqref{wignermassless2} for nonzero temperature $(T \neq 0)$. Both of these show Wigner semi-circle type for the spectrum of $\phi_i$.

\subsection{A toy model of a vector $\phi_i$ and a matrix $X_{ij}$}
By introducing matrix degrees of freedom, one can also construct a toy model of a vector with a matrix. 
Physical motivation to consider the model given by \eqref{GUEmassmodel} comes from D-brane dynamics. 
Consider $N$ D0 branes. Their dynamics is written by the so-called BFSS \cite{Banks:1996vh} matrix model, which is obtained by the dimensional reduction of ${\cal{N}} =4$ SYM to $0+1$-dimension  
\begin{align}
\label{BFSS}
& S =  \frac{1} {g_{YM}^2}\int dt \, 
\mbox{Tr}  \Biggl(
\sum_{A=1}^9 {1 \over 2} \left( D_t X^A \right)^{2}   +{1 \over 4} \sum_{A,B=1}^9 \left[X^A , X^B \right]^2  +  \cdots 
\Biggr)
\end{align}
where $X^A$ are adjoint $N \times N$ scalars representing the positions of D0's, and $\cdots$ are fermions. 
This model is conjectured to be 10-dimensional string theory or in the extremal IR, flat space 11-dimensional M-theory, 
\cite{Banks:1996vh, Itzhaki:1998dd, Polchinski:1999br}. 
If we add probe D0-brane in this background, the matrix $X^A$ will be enhanced to $(N+1) \times (N+1)$, where its 
$(N+1)$-th component represents the position of the probe D0-brane by the Higgs vacuum expectation value (vev), 
\be
X^A_{N+1, N+1} = M^A \,,
\ee
and $N \times N$ matrix $X_{background}$ represents the background $N$ D0's.
Then all are packaged into the following $(N+1) \times (N+1)$ matrix $X$, where
\be
X^A = \left(
\begin{array}{ccc:c}
&&&\\
&{\smash{\huge{X^A_{background}}}}&&  \phi^A_{i}\\
&&& \\ \hdashline
& \left( \phi^{A}_{i} \right)^\dagger && M^A
\end{array}
\right)
\label{N+1matrix}
\ee
where $N \times 1$  off-diagonal vector element, $\phi^A_i$, and  its conjugate represents the open strings between a prove D0 and background $N$ D0's. 
The dynamics of these off-diagonal vector elements are determined completely by the action obtained by expanding the $(N+1) \times (N+1)$ matrix model eq.~\eqref{BFSS} as 
\begin{align}
S &= S_{0} \left(X^A_{background} \right)  + \, S_{probe} \left(X^A_{background}, \phi^A_{i} \right) 
\end{align}
where $S_0$ is an action of the background D0-branes without the probe D0-brane determined only by the background $N \times N$ matrix $X^A_{background}$, giving $O(N^2)$ entropy. 
In large $N$ limit, the dynamics of $X_{background}$ is determined by $S_0(X_{background})$ since it is the dominant degree of freedom. Once $X_{background}$ is solved, then  using that knowledge, $ \phi_{i}$ is determined by $S_{probe}$. $S_{probe}$ takes schematically following form 
\begin{align}
\hspace{-1mm}S_{probe} 
&\sim   
\frac{1} {g_{YM}^2}
\int dt  \left(  ( \partial_t \phi_i^\dagger ) \left( \partial_t \phi_i \right)  -  M^2   \phi_i^{\dagger} \phi_i 
-  
\phi_i^{\dagger} B_{ij} \phi_j \right) \,, 
\label{SquadrapartB}
\end{align} 
where $B_{ij}$ is a matrix value, composed by $X^A_{background}$ and $M \delta_{ij}$, representing the interaction terms between a vector $\phi_i$ and the background matrix $X^A_{background}$.   \\

The full action $S_0$ and $S_{probe}$, obtained from \eqref{BFSS} using \eqref{N+1matrix} are very complicated.  
One can try solving it using numerical analysis under certain approximations like \cite{Iizuka:2001cw}. 
Instead, one can simplify it to some toy models. 
Both IP model \cite{Iizuka:2008hg} and IOP model \cite{Iizuka:2008eb} are examples of such simple toy models where we simplify both $S_0$ and $S_{probe}$ above in such a way that we have only one background matrix $X_{background}$, which we write simply $X$, and only one complex vector field $\phi_i$. 
Furthermore for both IP and IOP model, we choose $S_0$ such that $X$ is free, {\it i.e.,} 
\be
\label{backgroundXL}
S_{0} \left(X \right)  =  \frac{1}{g_{YM}^2} \int dt \mbox{Tr} \left[ \frac{1}{2}\left( \partial_t X \right) ^2  - \frac{m^2}{2} X^2 \right]   \,,
\ee
where one regard the effective mass $m$ is induced by the complicated dynamics of $X_{background}$.  
For the probe part $S_{probe}(\phi_i)$, we consider 
\begin{align}
\hspace{-1mm} S_{probe}(\phi_i)
&= S_{probe, kin}(\phi_i) + S_{probe, int}  (\phi_i)
\label{SquadrapartB}
\end{align} 
where we consider  
\begin{align}
S_{probe, kin}(\phi_i) &= \frac{1} {g_{YM}^2}
\int dt  \left(  ( \partial_t \phi_i^\dagger ) \left( \partial_t \phi_i \right)  -  M^2   \phi_i^{\dagger} \phi_i \right) \,, \\
S_{probe, int}(\phi_i)  &=- \frac{1} {g_{YM}^2}
\int dt  \left(   \phi_i^{\dagger} B_{ij} \phi_j \right) \,, \
\end{align}
The choice of $B_{ij}$ corresponds to the choice of the model we consider. 
For example, we can choose $B_{ij}$ in eq.~\eqref{SquadrapartB}, which characterizes the interaction between $X$ and $\phi$ as 
\begin{align}
\label{IPL}
S_{probe, int}(\phi_i)  &= -      \frac{ 1} {g_{YM}^2} M \int dt  \sum_{i, j}  
\phi_i^\dagger  X_{ij} \phi_j  \qquad \,  \, \mbox{(for the IP-model \cite{Iizuka:2008hg})} \, \\
\label{IOPL}
S_{probe, int}(\phi_i)  &= -   \frac{ 1} {g_{YM}^2}  \int dt  \sum_{i, j} 
\phi_i^\dagger  A^\dagger_{ik} A_{kj} \phi_j   \qquad \mbox{(for the IOP-model \cite{Iizuka:2008eb})} 
\end{align} 
Here, $A^\dagger$, $A$ are creation and annihilation operator for $X$, as 
\begin{align}
A_{ij} = \frac{1}{\sqrt{2m}} \left(\Pi_{ij} - i m X_{ij} \right) \,, \quad A^\dagger_{ij} = \frac{1}{\sqrt{2m}} \left(\Pi_{ij} + i m X_{ij} \right) \,.
\end{align}
Setting 
\begin{align}
 g := g_{YM} \,,
\end{align} 
then the IP model given by eq.~\eqref{IPL} has the same form given by the action eq.~\eqref{GUEmassmodel} if and only if $X$ obeys GUE. In the large $N$ and massless limit $m\to0$ in the IP model, the spectral density of the fundamental shows Wigner semi-circle behavior, which is the same as large $N$ limit of the GUE eq.~\eqref{Wignersemicirclerho}. 

Note also that in the IOP model \eqref{IOPL}, the interaction terms preserve the adjoint number,  $A^\dagger_{ik} A_{kj}$ is essentially the number operator. Due to this extra symmetry, the IOP model is more constrained and also more tractable than the IP model. In fact this results in the power law decay for the correlator in the IOP model at the planar limit, which is very different from the exponential decay of the correlator in the IP model. The IOP model allows various ways to solve it even including non-planar corrections \cite{Iizuka:2008eb}.

\section{Lanczos coefficients and Krylov complexity from the high frequency tail of the spectral density}\label{app1}
In this appendix, we review a relation between the high frequency tail of the spectral density $f(\omega)$ and asymptotic behaviors of Lanczos coefficients $b_n$ and Krylov complexity $K(t)$. 
Since the asymptotic behaviors of the Lanczos coefficients are sensitive to the high-frequency behavior of the spectral density, one can read off Lanczos coefficients $b_n$ from $f(\omega)$. 
Here we mainly focus on the case where $a_n = 0$.  

\subsection{From $f(\omega)$ to $b_n$}
\label{appftobn}
From the viewpoint of orthogonal polynomials, it was shown in \cite{Lubinsky:1988} that the asymptotic behavior of $b_n$, at large $n$, is associated with the spectrum density $f(\omega)$. Suppose its high frequency tail is given 
\begin{align}
f(\omega)\sim\exp\left[-2Q(\omega)\right] \, \quad \mbox{at large $\vert\omega\vert$} 
\end{align} 
where we implicitly assume that the leading exponential decay of $f(\omega)$ is represented by a continuous and even function $Q(\omega)$, {\it i.e.,} 
\begin{align}
f(\omega) \sim f(-\omega) \quad \mbox{at large $\vert\omega\vert$}
\end{align}
Then the asymptotic behaviors of $a_n$ and $b_n$ are determined by the following prescription:
\begin{align}
\lim_{n\to\infty}\frac{a_n}{B_n}=0, \;\;\; \lim_{n\to\infty}\frac{b_n}{B_n}=\frac{1}{2} \,,
\end{align}
where $B_n$ is a positive root of
\begin{align}
n=\frac{2}{\pi}\int_0^1 d\omega B_n \omega Q'(B_n\omega)(1-\omega^2)^{-1/2} \,.
\label{rootBn}
\end{align}
Several examples are 
\begin{enumerate}
\item For the case $2Q(\omega)=\vert\omega/\omega_0\vert^{1/p}$ with $p>0$. For $\omega > 0$, we have 
$Q'(\omega) = \frac{1}{2 p \omega}\vert \frac{\omega}{\omega_0}\vert^{1/p} $ and 
\begin{align}
n 
&= \frac{1}{p \pi}  \vert \frac{B_n }{ \omega_0}\vert^{1/p} \int_0^1  \omega^{1/p}  (1-\omega^2)^{-1/2} 
= \frac{1}{p \pi}  \vert \frac{B_n }{ \omega_0}\vert^{1/p}  \frac{\sqrt{\pi}\, \Gamma[\frac{1 + p}{2 p}]}{2 \Gamma[1 + \frac{1}{2p}]} \,,
\end{align}  
this gives 
\begin{align}
B_n=\omega_0 \left(\frac{\Gamma[1 + \frac{1}{2p}]}{ \Gamma[\frac{1 + p}{2 p}]}\right)^p (2 \sqrt{\pi}  p n )^p  \propto n^p \,,
\end{align}
In particular, for $p=1$ case, $B_n$  is given by
\begin{align}
B_n= \pi \omega_0 n \,.
\end{align}
therefore 
\begin{align}
\lim_{n\to\infty}{b_n} = \frac{\pi \omega_0}{2} n \,,
\end{align}
$b_n$ is $n$ linear. 
\item For the case $2Q(\omega)=\frac{\vert\omega\vert}{\omega_0}\log {\vert\omega\vert}$. 
This case is particularly interesting since 
\begin{align}
f(\omega)\sim\exp\left[-2Q(\omega)\right] =\vert\omega\vert^{- \frac{\vert\omega\vert}{\omega_0}} \, \quad \mbox{at large $\vert\omega\vert$} 
\end{align}
the spectrum density behaves asymptotically as we have seen in the infinite temperature limit of the IP model at eq.~\eqref{abIP}. 
For $\omega > 0$, we have 
$Q'(\omega)= \frac{1 + \log | \omega|}{2 \omega_0}$ and 
\begin{align}
n = \frac{2}{\pi} \int_0^1 d \omega B_n \omega \frac{1 + \log |B_n \omega|}{2 \omega_0}  (1-\omega^2)^{-1/2}  
= \frac{B_n \log | 2 B_n| }{\pi \omega_0}
\end{align}
this gives 
\begin{align}
B_n=\frac{ \pi \omega_0 n}{W(2 \pi \omega_0 n)} \,,
\end{align}
where $W(z)$ is the Lambert W function or the product-log function, which satisfies  
\begin{align}
W(z) e^{W(z)} = z \,.  
\end{align}
\end{enumerate}

\subsection{From $b_n$ to $K(t)$ (for $a_n = 0 $)}
If $b_n$ has a smooth asymptotic behavior with respect to $n$, then the asymptotic behavior of $K(t)$ can be derived by a continuous approximation with the following prescription \cite{Barbon:2019wsy}. Let us introduce a small cutoff $\varepsilon$ and define a new coordinate $x$ and a continuous function $v(x)$ as follows; 
\begin{align}
x: =\varepsilon n \,,  \quad v(x): =2\epsilon b_n\vert_{n=\frac{x}{\varepsilon}} \,, 
\end{align}
where we implicitly assume that $b_n$ can be approximated as a continuous function. 
Similarly assuming that $\varphi_n(t)$ can be approximated as a continuous function of two continuous parameters, $t$ and $x$ as 
\begin{align}
\varphi(x, t): =\varphi_n(t) 
\end{align}
Then for the case $a_n=0$, eq.~\eqref{recursionwf} becomes 
\begin{align}
\frac{\partial \varphi(x, t)}{\partial t} &=- \frac{1}{{2 \epsilon}} \left( {v\left(x+\epsilon\right)} \varphi \left(x+\epsilon , t \right) - {v \left(x \right)}   \varphi \left(x-\epsilon , t \right) \right) \nonumber \\
& = -v(x) \frac{\partial  \varphi \left(x, t \right) }{\partial x} - \frac{1}{2} \frac{\partial v(x)}{\partial x} \varphi \left(x , t \right) +O(\epsilon) \,.
\label{varphidiffeq}
\end{align}
Setting a new coordinate $y$ and rescaled function $\psi$ as 
\begin{align}
\partial_y &:= v(x)\partial_x \,, \quad \left(\mbox{or equivalently} \,, 
y := \int \frac{dx}{v(x)} \right) \,, \\ \psi(y, t) &:=\sqrt{v(x)} \, \varphi \left(x+\epsilon , t \right)
\end{align} 
Then eq.~\eqref{varphidiffeq} becomes in the $\epsilon \to 0$ limit 
\begin{align}
\left( \frac{\partial }{\partial t} + \frac{\partial }{\partial y} \right) \psi(y, t) = 0  \quad \Rightarrow \quad \psi(y, t)  =  \psi(y-t) \,.
\end{align}

Then the Krylov complexity is 
\begin{align}
K(t):=\sum_{n=1}^\infty n\vert\varphi_n(t)\vert^2  \sim \frac{1}{\epsilon^2} \int dy \, {|\psi(y-t)|^2} \, x(y)  
= \frac{1}{\epsilon^2} \int dy \, {|\psi(y)|^2} \, x(y+t)  
\end{align}
Thus, the $t$-dependence of $K(t)$ at late times can be extracted from the $t$-dependence of $x(y+t)$ at late times. \\

Several examples are
\begin{enumerate}
\item $b_n = \alpha n$ case: Then 
\begin{align}
v(x)=2\epsilon b_n\vert_{n=x/\varepsilon}=2\alpha x \,,
\end{align} 
and the $t$-dependence of $x(y+t)$ and $K(t)$ are
\begin{align}
x(y+t) \propto e^{2\alpha t} \,, \;\;\; K(t)\propto e^{2\alpha t} \,.
\end{align}
\item $b_n = \alpha n/\log n$ case: Then 
\begin{align}
v(x)=\frac{2\alpha x}{\log (x/\varepsilon)} \,,
\end{align} 
and  the $t$-dependence of $x(y+t)$ and $K(t)$ are
\begin{align}
x(y+t) \propto  e^{\sqrt{4\alpha t}} \,, \;\;\; K(t) \propto e^{\sqrt{4\alpha t}} \,.
\end{align}
\item $b_n = \alpha n^p$ case: Then 
\begin{align}
v(x)=2\alpha \epsilon \left(\frac{x}{\varepsilon} \right)^p \,,
\end{align} 
 the $t$-dependence of $x(y+t)$ and $K(t)$ are
\begin{align}
x(y+t) \propto (y+t)^{\frac{1}{1- p}}\,, \;\;\; K(t) \propto  t^{\frac{1}{1-p}} \,, 
\end{align}
where we implicitly assume that $ {|\psi(y)|^2}$ decays fast enough at large $y$.  
\end{enumerate}

\subsection{Short summary}
The above examples are classified into three classes dependent on the spectral density $f(\omega)$. Here we summarize the three classes with known literature. 
\begin{enumerate}
\item For $f(\omega)\sim\exp\left[-\vert\omega/\omega_0\vert\right]$, 
\begin{align}
b_n \sim \frac{\omega_0\pi}{2} n \,, \;\;\; K(t) \propto e^{\omega_0\pi t} \,.
\end{align}
The SYK model in the large $q$ limit corresponds to this class \cite{Parker:2018yvk}\footnote{As far as we are aware, there are no analytic results showing that in the SYK model for finite $q$, the Lanczos coefficients behave linearly in $n$, instead of $n/\log n$.}.

Furthermore, this is the maximal growth of the Lanczos coefficient in $n$ in the lattice models whose spatial dimension $d$ is greater than one, $d  > 1$  \cite{Parker:2018yvk}. These include $d=2$ square lattice model in the thermodynamic limit \cite{bouch2010complex}. 
\item For $f(\omega)\sim\exp\left[-\frac{\vert\omega\vert}{\omega_0}\log\vert\omega\vert\right]$, 
\begin{align}
b_n\sim  \frac{ \pi \omega_0 n}{2 W(2 \pi \omega_0 n)} \approx \frac{\pi \omega_0}{2} \frac{n}{\log n} \,,
 \;\;\; K(t) \propto e^{\sqrt{2 \pi \omega_0 t}} \,,\label{Kt1dchaos}
\end{align}
A 1d Ising model   
with both transverse and parallel magnetic fields \cite{Banulsetal} 
is known to belong to this second class \cite{Cao:2020zls}.  It is also known that in a 1d lattice model, this is the fastest growth for the Krylov complexity under the assumption that the correlator does not have a pole in the entire complex plane of $t$ \cite{Parker:2018yvk}. 
\item For $f(\omega)\sim \exp\left[-\vert\omega/\omega_0\vert^{1/p}\right]$, 
\begin{align}
b_n \propto n^p \,, \quad K(t) \propto t^{\frac{1}{1-p}} \, \;\;\; (0<p<1) \,.
\label{Ktintegrable}
\end{align}
This class contains various integrable systems. 
\end{enumerate}
The IP model in the infinite temperature limit \cite{Iizuka:2008hg}, which is the main focus of this paper, belongs to the second class.

\bibliography{Ref}
\bibliographystyle{JHEP}

\end{document}